\documentclass[amssymb,aps]{revtex4-2}
\usepackage{graphicx}
\usepackage{enumerate}
\usepackage{amssymb}
\usepackage{amsmath}
\usepackage{amsfonts}
\usepackage{color}
\usepackage{mathrsfs}
\usepackage{bm}
\usepackage{tikz}

\begin{document}
\title{Comprehensive quantum electrodynamics of fast electrons interacting with dissipative nanostructures and quantum light}

\author{A. Ciattoni}
\email{alessandro.ciattoni@spin.cnr.it}
\affiliation{Consiglio Nazionale delle Ricerche (CNR), SPIN Institute, c/o Dip.to di Scienze Fisiche e Chimiche, Via Vetoio, 67100 Coppito (L'Aquila), Italy}
\date{\today}

\begin{abstract}
Theoretical descriptions of the quantum electrodynamical interaction between fast electrons and nanostructures conventionally rely on standard macroscopic quantum electrodynamics (MQED), which, when applied to bounded objects in vacuum, formally requires an asymptotic limiting procedure to account for free-space radiation modes. To explicitly incorporate these modes from the outset, we employ the modified Langevin noise formalism (MLNF), which independently quantizes both the asymptotic scattering modes and the electromagnetic field generated by dissipative material sources. Utilizing an exact second-order Magnus expansion, we derive a closed-form, nonperturbative master formula for the transverse-momentum-resolved energy exchange probability. In the unilluminated regime, we demonstrate that at finite nanostructure temperatures, the vacuum scattering and thermal material fluctuations do not mathematically recombine into the total imaginary part of the dyadic Green's function ($\mathrm{Im}[\mathcal{G}]$). This breakdown does not violate the Fluctuation-Dissipation Theorem, because the explicit zero-temperature scattering field acts as an independent dynamical system that maintains the configuration out of global thermodynamic equilibrium. This fundamental separation reveals an intrinsic Stokes--Anti-Stokes asymmetry, yielding distinct transverse diffraction patterns for energy loss and thermal energy gain, and establishing a generalized detailed-balance relation that governs the radiative reduction of the apparent temperature in nanoscale electron thermometry. In the illuminated regime, our framework rigorously incorporates quantum-optical states into macroscopic dissipative electrodynamics, demonstrating that optically stimulated transitions are inevitably dressed and spatially decohered by the intrinsic spontaneous and thermal inelastic background that inherently governs the unilluminated regime. Evaluating multimode continuous-wave illuminations, we detail the breakdown of the semiclassical PINEM Dirac comb into continuous sidebands for coherent states, and the survival of phase-locked anomalous two-photon correlations allowing holographic reshaping of electron diffraction and active phase control of radiation-induced spatial decoherence for squeezed vacua. Finally, we identify the explicit approximations required to reduce our complete multipolariton framework to standard single-mode, lossless quantum PINEM models, thereby providing a rigorous first-principles derivation of the phenomenological single-photon coupling amplitude and demonstrating its complete dependence on the exact classical scattering kernel of the nanostructure.
\end{abstract}

\maketitle

\section{Introduction}

The interaction of fast free electrons with spatially confined electromagnetic fields near nanostructured materials provides a versatile platform for nanoscale spectroscopy and ultrafast quantum control \cite{DeAbajoRMP2010,Polman2019,Rivera2020,DeAbajoACSPhot2021,DeAbajoRoadmap}. In transmission electron microscopy, spontaneous inelastic scattering forms the basis of electron energy-loss spectroscopy (EELS) \cite{Egerton2008} and cathodoluminescence (CL) \cite{DeAbajoRMP2010}, whereas thermally activated energy gains enable nanoscale electron thermometry \cite{Idrobo2018,Lagos2018}. Beyond longitudinal energy-resolved measurements, monitoring the transverse momentum distribution of the post-interaction electron beam gives access to elastic diffraction induced by electromagnetic vacuum fluctuations \cite{Ford1993,Hsiang2006,Scheel2012,DiGiulio2020} as well as inelastic transverse decoherence, lateral momentum transfer, and entanglement with material excitations \cite{Schattschneider2005,Potapov2006,Mechel2021,CiattPRB2023,CiattPRApp2023,Adiv2023}. When the nanostructure is simultaneously irradiated by external optical fields, as in photon-induced near-field electron microscopy (PINEM) \cite{Barwick2009,DeAbajo2010,Park2010,Feist2015,Priebe2017,Morimoto2018,Vanacore2019,Wang2020,Kfir2020}, the evanescent near field mediates the stimulated exchange of multiple photon quanta, allowing coherent longitudinal and transverse shaping of the electron wave packet \cite{Kfir2019,DiGiulio2019,DiGiulio2020Optica,Reinhardt2020,BenHayun2021,Kfir2021,DiGiulioACSNano2021,Dahan2021,Feist2022}.

A self-consistent theoretical treatment of these phenomena relies on quantizing the macroscopic electromagnetic field in inhomogeneous, dispersive, and absorptive media. While normal-mode expansions suffice for transparent dielectrics \cite{Glauber1991,Dalton1996}, properly encoding material dispersion and dissipation requires coupling the field to a continuum of reservoir oscillators \cite{Fano1956,Hopfield1958,Huttner1992,Suttorp2004,Bhat2006,Philbin2010}. Following the canonical quantization of macroscopic electromagnetism (CQME) \cite{Philbin2010}, diagonalizing the corresponding field-matter Hamiltonian for an unbounded lossy medium directly yields the standard Langevin noise formalism (LNF), commonly known as macroscopic quantum electrodynamics (MQED) \cite{Gruner1996,Dung1998,Scheel1998,Scheel2008,Buhmann2012}. In its standard formulation, the LNF constructs the quantized electromagnetic field exclusively from the quantum fluctuating sources distributed within the material medium. Consequently, because the external radiation field is entirely omitted, this approach provides a rigorous description solely for unbounded dispersive environments. Moreover, this inherent restriction ensures that the LNF remains consistent with the Fluctuation-Dissipation Theorem: the asymptotic surface term in the fundamental integral relation of the dyadic Green's function---which physically represents the scattering modes---vanishes exclusively in unbounded lossy media, where the electromagnetic field exponentially decays at infinity \cite{Drezet2017b,Franke2020,Ciatt1}. Consequently, applying the standard LNF to finite objects in vacuum requires a singular limiting procedure where the surrounding space is treated as weakly absorptive and the limits ${\rm Im}(\varepsilon_\omega), {\rm Im}(\mu_\omega) \to 0^+$ are taken strictly at the end of the calculations \cite{Hanson2021}---an approach whose universal asymptotic validity has only recently been established \cite{CiattLimiting2026}. To bypass this singular limiting procedure, the scattering modes---which encapsulate both the incident free-space radiation and its exact scattering by the object---must be explicitly included in the macroscopic electric field operator alongside the medium-assisted field. This complete field decomposition constitutes the theoretical foundation of the modified Langevin noise formalism (MLNF). While initially introduced for planar and non-magnetic structures \cite{DiStefano2001,Drezet2017a,Dorier2019,Na2023}, the framework has recently been generalized to arbitrary finite magnetodielectric bodies \cite{Ciatt1} and directly derived from the first-quantized CQME Hamiltonian in the Schr\"odinger picture \cite{CiattDirect}. Providing the most complete macroscopic description, we adopt this fully general formulation of Ref. \cite{CiattDirect}, referring to it as the MLNF throughout this work. This formalism has recently proven essential for modeling quantum emitters in out-of-equilibrium dispersive environments \cite{Miano2025a,Miano2025b,Miano2026}, quantum-optical scattering and two-photon interference by lossy nanostructures \cite{Ciatt2,Ciatt3}, quantum optomechanics in structured squeezed vacua \cite{CiattOptomech2026}, and far-field spatial coherence unifying quantum-light scattering and thermal emission \cite{CiattCoherence2026}.

In previous theoretical investigations of fast-electron interactions with finite nanostructures in the absence of external illumination \cite{DiGiulio2020,Mechel2021,Kfir2021,CiattPRB2023,CiattPRApp2023,Adiv2023}, the standard LNF has commonly been applied directly to the bounded object without limiting procedures, thereby inherently omitting the scattering radiation. At zero temperature, these approaches recover the exact spontaneous EELS probability and vacuum phase shift by expressing the field correlators in terms of the complete imaginary part of the macroscopic Green's function \cite{DiGiulio2020,Mechel2021,CiattPRB2023,CiattPRApp2023}. As we show below, this identification holds at absolute zero because the quantum vacuum fluctuations of the scattering field and of the internal dissipative sources share identical zero-point statistical weights, allowing their sum to recombine into the full Green's tensor. However, a description based solely on the total $\mathrm{Im}[\mathcal{G}]$ cannot be directly applied when the two sectors require independent treatment. First, in the strictly lossless dielectric limit, the internal dissipative sources vanish, so that retaining the radiative response without asymptotic limiting procedures requires the explicit presence of the scattering modes. Second, for a heated nanostructure in a cold vacuum environment, weighting the full $\mathrm{Im}[\mathcal{G}]$ by a single Bose-Einstein factor effectively assigns the macroscopic temperature of the object to the surrounding radiation bath as well, corresponding to global thermodynamic equilibrium rather than an out-of-equilibrium configuration.

Regarding the illuminated regime, the existing literature on fast-electron interactions with external optical fields generally follows two distinct methodological paths. On the one hand, approaches incorporating the complex dispersive response of arbitrary nanostructures via the full macroscopic Green's tensor have been primarily restricted to classical illumination. These treatments introduce the external light either as unquantized fields in semiclassical photon-induced near-field electron microscopy (PINEM) models \cite{DeAbajo2010,Park2010,Rivera2020} or as deterministic classical source currents added to the Hamiltonian \cite{DiGiulioACSNano2021}, inherently precluding the description of nonclassical photon statistics or mixed incident states. On the other hand, theoretical investigations of quantum PINEM (QPINEM) with nonclassical light have typically relied on idealized single- or discrete-mode cavity models \cite{Kfir2019,DiGiulio2019,DiGiulio2020Optica,BenHayun2021,DiGiulioThesis}. While highly successful in describing coherent quantum effects, these lossless formulations are designed around optical states statically confined within a closed cavity. As such, they are naturally tailored for low-loss environments rather than metallic or plasmonic nanostructures featuring broad absorptive continua, and their focus remains separated from the concurrent interplay between optically stimulated transitions and environment-driven inelastic scattering.

In this paper, we develop the exact, nonperturbative theory of energy and transverse-momentum exchange for a fast aloof electron interacting with an arbitrary finite and dissipative magnetodielectric nanostructure, illuminated by any quantum state of light ranging from the vacuum to the most general mixed state. By explicitly incorporating the quantum scattering sector, our approach rigorously treats bounded geometries without invoking singular limiting procedures, while simultaneously bridging the divide between classical macroscopic treatments and quantum lossless-cavity models. Employing the MLNF to solve the interaction-picture dynamics, we derive a closed-form master formula for the joint energy-momentum probability distribution of the post-interaction electron. This fully unrestricted quantum description of both the field driven by localized material dissipation and the scattering field reveals fundamental physical phenomena inaccessible to standard approaches.

Specifically, in the unilluminated regime---where the electron dynamics are governed solely by the interplay between the thermal field of a heated nanostructure and the vacuum fluctuations of the scattering field---we demonstrate:
\begin{itemize}
\item The mathematical recombination of the material and scattering vacuum fluctuations into the imaginary part of the full macroscopic Green's function at absolute zero, which rigorously validates standard EELS descriptions, juxtaposed with the structural breakdown of this recombination at finite nanostructure temperatures.
\item The emergence of a fundamental Stokes--Anti-Stokes asymmetry, yielding distinct transverse diffraction patterns for energy loss versus thermal energy gain.
\item The establishment of a generalized detailed-balance relation that rules the impact-parameter-dependent radiative enhancement or reduction of the apparent temperature in nanoscale electron thermometry.
\end{itemize}

Furthermore, in the illuminated regime---where the electron dynamics are additionally driven by an externally prepared quantum state of light---we demonstrate:
\begin{itemize}
\item The essential convolution of the electron's optically stimulated transitions with the intrinsic spontaneous and thermal inelastic background that inherently governs the unilluminated regime, thereby inducing intrinsic spatial decoherence and establishing fundamental limits on coherent wave-packet shaping.
\item The emergence of distinct non-equilibrium phenomena under specific illumination states. Explicitly, we detail the exact physical signatures driven by ambient thermal radiation (revealing near-field-mediated energy gains from hot vacua), continuous-wave coherent fields (revealing the inelastic dressing of coherent sidebands, from which the standard semiclassical PINEM limit is rigorously recovered under appropriate approximations), and multimode squeezed vacua (revealing the phase-locked survival of anomalous two-photon correlations, the quantum holographic reshaping of electron diffraction, and the active phase manipulation of radiation-induced spatial decoherence).
\item The exact mathematical reduction of our complete asymptotic evolution operator to the standard scattering operator underlying all single-mode QPINEM models. By applying strict lossless-cavity approximations, this reduction inherently recovers established QPINEM phenomena while providing a first-principles electrodynamic expression for the phenomenological single-photon coupling amplitude, demonstrating its complete dependence on the exact classical scattering kernel of the nanostructure.
\end{itemize}

The paper is organized as follows. Section~II introduces the electron-field Hamiltonian and the MLNF framework. Section~III presents the exact closed-form master formula for the transverse-momentum-resolved energy exchange. Sections~IV and V detail the physical predictions of the theory in the unilluminated and quantum-light illumination regimes, respectively. Conclusions are drawn in Section~VI. The complete mathematical derivations of the evolution operator, the reduced density matrix, and the quantum-optical characteristic functionals are provided in the Appendices.

\begin{figure}
\centering
\includegraphics[width = 1\linewidth]{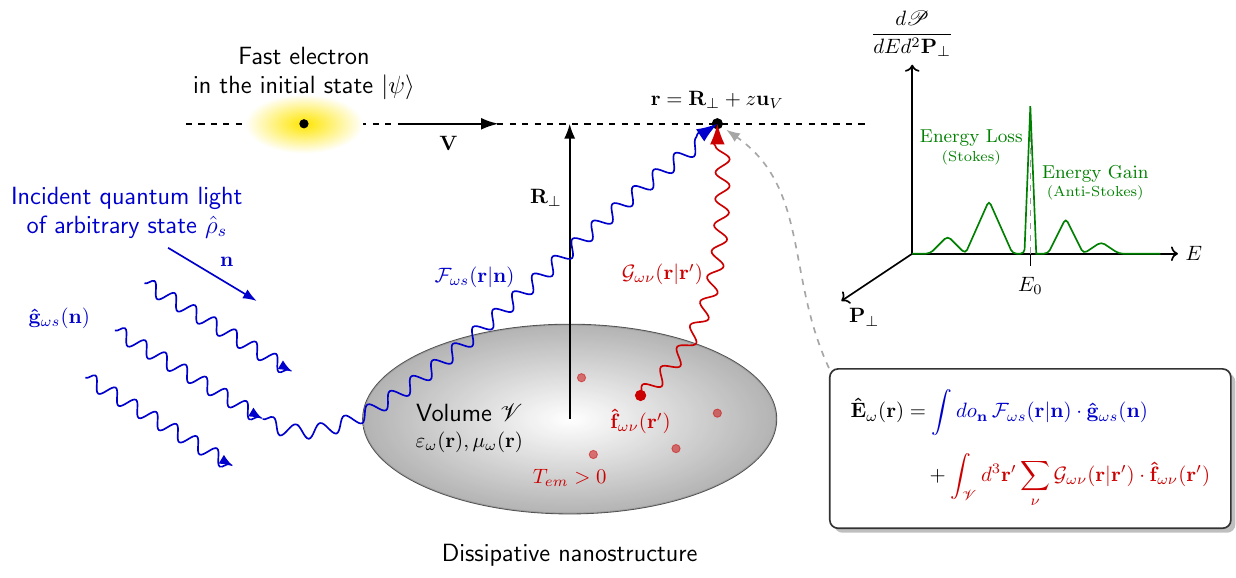}
\caption{\textbf{Schematic representation of the interaction between a fast electron, a dissipative nanostructure, and an incident quantum light field}. A fast electron, initially prepared in a paraxial state $\vert{}\psi\rangle$ whose transverse spatial profile spans the impact parameter coordinate $\mathbf{R}_\perp$, traverses the vicinity of a nanostructure of volume $\mathscr{V}$ along the longitudinal direction of velocity $\mathbf{V}$. The electron interacts with the total quantized electric field $\mathbf{\hat{E}}_\omega(\mathbf{r})$, which rigorously comprises two independent contributions originating from an incident quantum light field (blue, arbitrary state $\hat{\rho}_s$) and internal material polaritons (red, thermal equilibrium at $T_{em} > 0$). The total field operator is constructed by propagating the annihilation operators of the scattering modes, $\mathbf{\hat{g}}_{\omega s}(\mathbf{n})$, and the material modes, $\mathbf{\hat{f}}_{\omega \nu}(\mathbf{r}')$, through their respective dyadic kernels, $\mathcal{F}_{\omega s}(\mathbf{r}\vert{}\mathbf{n})$ and $\mathcal{G}_{\omega \nu}(\mathbf{r}\vert{}\mathbf{r}')$. This fully quantized interaction yields the multipolariton sideband spectrum (top-right panel), characterizing the energy loss (Stokes) and energy gain (Anti-Stokes) probability distribution, $\frac{d\mathscr{P}}{dE d^2\mathbf{P}_\perp}$, relative to the initial energy $E_0$.}
\label{Fig1}
\end{figure}

\section{Electron-field Hamiltonian and MLNF framework}

We consider a finite-size dispersive and absorptive nanostructure occupying a bounded spatial region $\mathscr{V}$ in vacuum, whose linear, inhomogeneous, isotropic, and causal magnetodielectric response at angular frequency $\omega > 0$ is described by the complex relative dielectric permittivity $\varepsilon_\omega({\bf{r}})$ and magnetic permeability $\mu_\omega({\bf{r}})$, with $\varepsilon_\omega({\bf{r}}) = \mu_\omega({\bf{r}}) = 1$ for ${\bf{r}} \notin \mathscr{V}$. A fast electron of mass $m$ and charge $-e < 0$ travels aloof past the nanostructure (see schematic in Fig.~\ref{Fig1}). Assuming the electron momentum distribution to remain tightly localized around a central relativistic momentum ${\bf{P}}_0$ throughout the interaction, the electron dynamics is governed by the paraxial nonrecoil approximation, in which the relativistic kinetic energy is linearized around ${\bf{P}}_0$ with constant central velocity ${\bf{V}} = c{\bf{P}}_0/\sqrt{m^2 c^2 + P_0^2} = V{\bf{u}}_V$, and spin-flip as well as ponderomotive ${\bf{\hat A}}^2$ terms are neglected. In the Coulomb gauge, the total Schr\"odinger-picture Hamiltonian of the coupled electron-field-matter system reads $\hat H =  \hat H_{el} + \hat H_{sem} + \hat H_{{\mathop{\rm int}}}$, where (see Appendix~A for complete mathematical details)
\begin{eqnarray} \label{H_MLNF_main}
\hat H_{el} &=& {\bf{V}} \cdot {\bf{\hat P}}, \nonumber \\
\hat H_{sem} &=& \int {d\omega } \;\hbar \omega \left[ {\int {do_{\bf{n}} } \;{\bf{\hat g}}_{\omega s}^\dag  \left( {\bf{n}} \right) \cdot {\bf{\hat g}}_{\omega s} \left( {\bf{n}} \right) + \int_\mathscr{V} {d^3 {\bf{r}}\,} \sum_\nu {{\bf{\hat f}}_{\omega \nu }^\dag  \left( {\bf{r}} \right) \cdot {\bf{\hat f}}_{\omega \nu } \left( {\bf{r}} \right)} } \right], \nonumber \\
\hat H_{{\mathop{\rm int}} } &=& \int {d\omega } \;\frac{e}{{i\omega }}{\bf{V}} \cdot \left[ {{\bf{\hat E}}_\omega  \left( {{\bf{\hat R}}} \right) - {\bf{\hat E}}_\omega ^\dag  \left( {{\bf{\hat R}}} \right)} \right].
\end{eqnarray}
Here, ${\bf{\hat R}}$ and ${\bf{\hat P}}$ are the canonical electron position and momentum operators satisfying $[{\bf{\hat R}}, {\bf{\hat P}}] = i\hbar {\cal I}$ (where ${\cal I}$ is the identity dyadic), while the positive-frequency spectral component of the electric field operator ${\bf{\hat E}}_\omega({\bf{r}})$ (linked to the total electric field via ${\bf{\hat E}}({\bf{r}}) = \int d\omega\,{\bf{\hat E}}_\omega({\bf{r}}) + {\rm H.c.}$) is partitioned in the MLNF into two commuting contributions belonging to distinct polariton sectors \cite{DiStefano2001,Drezet2017a,Dorier2019,Na2023,Ciatt1,CiattDirect} (depicted in the bottom-right panel of Fig.~\ref{Fig1}):
\begin{equation} \label{E_omega_main}
{\bf{\hat E}}_\omega  \left( {\bf{r}} \right) = \int {do_{\bf{n}} } \;{\cal F}_{\omega s} \left( {\left. {\bf{r}} \right|{\bf{n}}} \right) \cdot {\bf{\hat g}}_{\omega s} \left( {\bf{n}} \right) + \int_\mathscr{V} {d^3 {\bf{r}}'\,} \sum_\nu {{\cal G}_{\omega \nu } \left( {\left. {\bf{r}} \right|{\bf{r}}'} \right) \cdot {\bf{\hat f}}_{\omega \nu } \left( {{\bf{r}}'} \right)}.
\end{equation}
In Eqs.~(\ref{H_MLNF_main}) and (\ref{E_omega_main}), ${\bf{\hat g}}_{\omega s}({\bf{n}})$ are the bosonic annihilation operators for the \emph{scattering polaritons} whose asymptotic plane-wave component propagates along the unit vector ${\bf{n}}$ (with solid angle element $do_{\bf{n}}$ and transversality condition ${\bf{n}} \cdot {\bf{\hat g}}_{\omega s}({\bf{n}}) = 0$), whereas ${\bf{\hat f}}_{\omega e}({\bf{r}})$ and ${\bf{\hat f}}_{\omega m}({\bf{r}})$ are the bosonic annihilation operators for \emph{electric ($e$) and magnetic ($m$) polaritons} whose dipolar sources are localized within the object at point ${\bf{r}} \in \mathscr{V}$. These operators obey the exact macroscopic bosonic algebra
\begin{equation}
\left[ {{\bf{\hat g}}_{\omega s} \left( {\bf{n}} \right),{\bf{\hat g}}_{\omega 's}^\dag  \left( {{\bf{n}}'} \right)} \right] = \delta \left( {\omega  - \omega '} \right)\delta \left( {o_{\bf{n}}  - o_{{\bf{n}}'} } \right){\cal I}_{\bf{n}} \quad \quad \quad  \quad \left[ {{\bf{\hat f}}_{\omega \nu } \left( {\bf{r}} \right),{\bf{\hat f}}_{\omega '\nu '}^\dag  \left( {{\bf{r}}'} \right)} \right] = \delta \left( {\omega  - \omega '} \right)\delta _{\nu \nu '} \delta \left( {{\bf{r}} - {\bf{r}}'} \right){\cal I},
\end{equation}
where ${\cal I}_{\bf{n}} = {\cal I} - {\bf{n}}{\bf{n}}$ is the transverse dyadic projector, $\delta \left( {o_{\bf{n}}  - o_{{\bf{n}}'} } \right)$ is the Dirac delta function over the solid angle, and all other commutators vanish. The scattering and material dyadic kernels are defined as
\begin{equation}
{\cal F}_{\omega s} \left( {\left. {\bf{r}} \right|{\bf{n}}} \right) = {\cal F}_\omega  \left( {\left. {\bf{r}} \right|{\bf{n}}} \right)\sqrt {\frac{{\hbar k_\omega ^3 }}{{16\pi ^3 \varepsilon _0 }}} ,\quad \quad \begin{array}{*{20}c}
  {{\cal G}_{\omega e} \left( {\left. {\bf{r}} \right|{\bf{r}}'} \right) = {\cal G}_\omega  \left( {\left. {\bf{r}} \right|{\bf{r}}'} \right)i\sqrt {\frac{{\hbar k_\omega ^4 }}{{\pi \varepsilon _0 }}{\mathop{\rm Im}\nolimits} \left[ {\varepsilon _\omega  \left( {{\bf{r}}'} \right)} \right]} } \\
  {{\cal G}_{\omega m} \left( {\left. {\bf{r}} \right|{\bf{r}}'} \right) = {\cal G}_\omega  \left( {\left. {\bf{r}} \right|{\bf{r}}'} \right)\frac{{ \times \mathord{\buildrel{\lower3pt\hbox{$\scriptscriptstyle\leftarrow$}} 
\over \nabla } _{{\bf{r}}'} }}{{ik_\omega  }}\sqrt {\frac{{\hbar k_\omega ^4 }}{{\pi \varepsilon _0 }}{\mathop{\rm Im}\nolimits} \left[ {\frac{{ - 1}}{{\mu _\omega  \left( {{\bf{r}}'} \right)}}} \right]} } \\
\end{array}
\end{equation}
and are determined by the classical scattering modal dyadic ${\cal F}_\omega({\bf{r}}|{\bf{n}})$ \cite{Kristensson2016,Ciatt1} and the classical dyadic Green's function ${\cal G}_\omega({\bf{r}}|{\bf{r}}')$ defined in Appendix~A (with $k_\omega = \omega/c$). For any bounded object in vacuum, these kernels satisfy the fundamental integral relation:
\begin{equation} \label{Fund_Id_Comm_main}
\int {do_{\bf{n}} } \;{\cal F}_{\omega s} \left( {\left. {\bf{r}} \right|{\bf{n}}} \right) \cdot {\cal F}_{\omega s}^{T*} \left( {\left. {{\bf{r}}'} \right|{\bf{n}}} \right) + \int_\mathscr{V} {d^3 {\bf{s}}\,} \sum_\nu  {{\cal G}_{\omega \nu } \left( {\left. {\bf{r}} \right|{\bf{s}}} \right) \cdot {\cal G}_{\omega \nu }^{T*} \left( {\left. {{\bf{r}}'} \right|{\bf{s}}} \right)}  = \frac{{\hbar k_\omega ^2 }}{{\pi \varepsilon _0 }}{\mathop{\rm Im}\nolimits} \left[ {{\cal G}_\omega  \left( {\left. {\bf{r}} \right|{\bf{r}}'} \right)} \right],
\end{equation}
which is derived directly from the classical macroscopic Maxwell equations via Green's theorem, where the explicit scattering term rigorously arises from the non-vanishing boundary surface integral evaluated at spatial infinity \cite{Drezet2017b,Franke2020,Ciatt1,CiattDirect}. 

The standard LNF is mathematically recovered by setting the scattering kernel ${\cal F}_{\omega s} \equiv 0$ in the preceding equations, thereby omitting the  scattering modes from both the field expansion (\ref{E_omega_main}) and the fundamental integral identity (\ref{Fund_Id_Comm_main}). However, this condition is satisfied exclusively in an unbounded lossy medium, which inherently cannot support asymptotic free-space radiation—a physically intuitive fact that has been rigorously proven in Ref.~\cite{CiattLimiting2026}. Consequently, standard LNF descriptions applied to finite nanostructures strictly require singular vacuum limiting procedures to preserve the rigorous validity of the quantum framework \cite{Hanson2021,CiattLimiting2026}. In contrast, the MLNF explicitly incorporates the scattering modes—which physically do not vanish for any bounded object for which ${\cal F}_{\omega s} \neq 0$. Importantly, this inclusion guarantees that the formalism inherently complies with the fundamental integral relation: Eq.~(\ref{Fund_Id_Comm_main}) is strictly required to establish the MLNF exact electric field commutation relation $[ {{\bf{\hat E}}_\omega  \left( {\bf{r}} \right),{\bf{\hat E}}_{\omega '}^\dag  \left( {{\bf{r}}'} \right)} ] = \delta \left( {\omega  - \omega '} \right)\frac{{\hbar k_\omega ^2 }}{{\pi \varepsilon _0 }}{\mathop{\rm Im}\nolimits} \left[ {{\cal G}_\omega  \left( {\left. {\bf{r}} \right\vert{}{\bf{r}}'} \right)} \right]$, which ensures agreement with the zero-temperature fluctuation-dissipation theorem.

To analyze the quantum dynamics, we work in the interaction picture with respect to the free Hamiltonian $\hat H_0 = \hat H_{el} + \hat H_{sem}$, where the total density operator $\hat\rho_I(t) = e^{\frac{i}{\hbar}\hat H_0 t}\hat\rho(t)e^{-\frac{i}{\hbar}\hat H_0 t}$ evolves from the factorized asymptotic initial state at $t \to -\infty$,
\begin{equation} \label{Initial_State_main}
\hat \rho _I \left( { - \infty } \right) = \left| \psi  \right\rangle \left\langle \psi  \right|\hat \rho _s \hat \rho _{em}.
\end{equation}
Equation~(\ref{Initial_State_main}) establishes that the electron, the incident light, and the nanostructure are initially completely decoupled. First, the electron is prepared in a pure quantum state $\left|\psi\right\rangle$. In coordinate space, its wavefunction takes the form of a paraxial wave packet \cite{CiattPRB2023,CiattPRApp2023},
\begin{equation} \label{Initial_psi_coord}
\psi \left( {\bf{R}} \right) = H \left( {\frac{\ell }{2} - \left| {R_V } \right|} \right)\frac{1}{{\sqrt \ell  }}\;e^{\frac{i}{\hbar }P_0 R_V } \phi \left( {{\bf{R}}_\perp  } \right),
\end{equation}
which factors into a longitudinal wave train of spatial coherence length $\ell$ propagating along the trajectory coordinate $R_V = {\bf{u}}_V \cdot {\bf{R}}$ ($H(x)$ being the Heaviside step function) and a generic transverse spatial profile $\phi({\bf{R}}_\perp)$ (with ${\bf{R}}_\perp = ({\cal I} - {\bf{u}}_V{\bf{u}}_V)\cdot{\bf{R}}$) normalized as $\int d^2{\bf{R}}_\perp\,|\phi({\bf{R}}_\perp)|^2 = 1$. The theoretical limit $\ell \to \infty$ defines the monochromatic (monoenergetic) electron regime. Second, $\hat\rho_s$ (with ${\rm Tr}_s[\hat\rho_s] = 1$) is an arbitrary density operator within the scattering polariton sector, specifying the quantum state of the asymptotic incident external radiation. Third, 
\begin{equation} \label{Initial_rho_em}
\hat \rho _{em}  = \frac{{e^{ - \hat H_{em} /\left( {k_B T_{em} } \right)} }}{{{\rm Tr}_{em} \left[ {e^{ - \hat H_{em} /\left( {k_B T_{em} } \right)} } \right]}},
\end{equation}
is the thermal density operator within the material sector, where $\hat H_{em} = \int d\omega\,\hbar\omega \int_\mathscr{V} d^3{\bf{r}} \sum_\nu {\bf{\hat f}}_{\omega\nu}^\dag({\bf{r}})\cdot{\bf{\hat f}}_{\omega\nu}({\bf{r}})$ is the corresponding Hamiltonian operator, describing the internal material polaritons of the nanostructure in thermal equilibrium at macroscopic temperature $T_{em}$.

Ultimately, the joint probability distribution $\frac{d\mathscr{P}}{dE d^2{\bf{P}}_\perp}$ of finding the post-interaction electron ($t \to +\infty$) with final energy $E$ and transverse momentum ${\bf{P}}_\perp = ({\cal I} - {\bf{u}}_V{\bf{u}}_V)\cdot{\bf{P}}$ is determined by projecting the asymptotic electron reduced density operator $\hat\rho_{elI}(\infty) = {\rm Tr}_{sem}[\hat\rho_I(\infty)]$ onto the simultaneous eigenstates of the free-electron energy and transverse momentum, leading to the multipolariton sideband spectrum illustrated in the top-right panel of Fig.~\ref{Fig1}.

\section{Transverse-momentum-resolved energy exchange probability}

The evaluation of the asymptotic electron reduced density operator $\hat\rho_{elI}(\infty)$ and the resulting probability distribution $\frac{d\mathscr{P}}{dE d^2{\bf{P}}_\perp}$ can be carried out analytically without perturbative approximations in the electron-field coupling. As detailed in Appendix~B, the paraxial interaction Hamiltonian allows the exact second-order truncation of the Magnus expansion \cite{Magnus1954} for the time-evolution operator, as similarly exploited in standard LNF and cavity-based descriptions \cite{DiGiulio2020,Kfir2021,DiGiulioACSNano2021,CiattPRB2023,CiattPRApp2023}. By tracing out the macroscopic electromagnetic field---evaluating the thermal trace over the material polaritons via Wick's theorem and putting the scattering polariton trace in normal order \cite{Cahill1969}---we obtain a closed-form expression for the final electron state. Projecting the resulting asymptotic reduced density operator onto the simultaneous energy and transverse-momentum eigenstates $\left|E, {\bf{P}}_\perp\right\rangle$, we obtain the exact master formula for the transverse-momentum-resolved energy exchange probability:
\begin{eqnarray} \label{dP_dE_dPt_main}
\frac{{d\mathscr{P}}}{{dEd^2 {\bf{P}}_\perp  }} &=& \frac{1}{{\left( {2\pi \hbar } \right)^3 V}}\int {d^2 {\bf{R}}_{1\perp} } \int {d^2 {\bf{R}}_{2\perp} } \; e^{ - \frac{i}{\hbar }{\bf{P}}_\perp   \cdot \left( {{\bf{R}}_{1\perp}  - {\bf{R}}_{2\perp} } \right)} \widetilde \phi \left( {{\bf{R}}_{1\perp} } \right)\widetilde \phi ^* \left( {{\bf{R}}_{2\perp} } \right)\nonumber \\
&& \times \int_{ - \ell }^\ell  {dZ} \;e^{ - \frac{i}{{\hbar V}}\left( {E - E_0 } \right)Z} e^{K\left( {{\bf{R}}_{1\perp} ,{\bf{R}}_{2\perp} ;Z} \right)} \frac{1}{\ell }\int_{ - \frac{\left( {\ell  - \left| Z \right|} \right)}{2}}^{\frac{\left( {\ell  - \left| Z \right|} \right)}{2}} {d\bar R_V } \; \left[\Upsilon _s \left( {{\bf{R}}_1 ,{\bf{R}}_2 } \right) \right]_{\substack{ {\bf{R}}_1 = {\bf{R}}_{1\perp} + (\bar R_V + Z/2){\bf{u}}_V \\ {\bf{R}}_2 = {\bf{R}}_{2\perp} + (\bar R_V - Z/2){\bf{u}}_V }} ,
\end{eqnarray}
where $E_0 = V P_0$ is the central paraxial energy of the electron and $\widetilde\phi({\bf{R}}_\perp)$ is the dressed transverse wave function defined as
\begin{equation} \label{dressed_phi_main}
\widetilde \phi \left( {{\bf{R}}_\perp  } \right) \equiv \phi \left( {{\bf{R}}_\perp  } \right)e^{i\Phi \left( {{\bf{R}}_\perp  } \right) }
e^{-W\left( {{\bf{R}}_\perp  } \right)}.
\end{equation}
All constituent quantities in Eqs.~(\ref{dP_dE_dPt_main}) and (\ref{dressed_phi_main}) have a direct physical interpretation and are entirely determined by the classical electrodynamic kernels ${\cal F}_{\omega s}$ and ${\cal G}_{\omega\nu}$, the incident quantum light state $\hat\rho_s$, and the macroscopic nanostructure temperature $T_{em}$---which enters strictly through the Bose-Einstein thermal occupation factor
\begin{equation}
n_\omega(T_{em}) =\frac{1}{e^{\hbar\omega/(k_B T_{em})} - 1}.
\end{equation}
Specifically, the dynamical building blocks governing the electron-matter interaction are defined as follows:

\begin{itemize}
\item \textbf{Conservative vacuum phase shift:}
\begin{equation} \label{Phi_Rt_main}
\Phi \left( {{\bf{R}}_\perp  } \right) = \frac{{e^2 }}{{\pi \varepsilon _0 \hbar c^2 }}\int_{ - \infty }^{ + \infty } {dz} \int_{ - \infty }^z {dz'} \int {d\omega } \;\sin \left[ {\frac{\omega }{V}\left( {z - z'} \right)} \right]{\bf{u}}_V  \cdot {\mathop{\rm Im}\nolimits} \left[ {{\cal G}_\omega  \left( {\left. {{\bf{R}}_\perp   + z{\bf{u}}_V } \right|{\bf{R}}_\perp   + z'{\bf{u}}_V } \right)} \right] \cdot {\bf{u}}_V .
\end{equation}
The purely real phase $\Phi$ encodes the causal, conservative interaction between the electron and the zero-point electromagnetic fluctuations modified by the presence of the nanostructure \cite{Ford1993,Hsiang2006,Scheel2012,DiGiulio2020,CiattPRB2023,CiattPRApp2023}. 
Acting formally as a quantum image-potential \cite{DiGiulio2020,CiattPRB2023}, the spatial dependence of the Green's function on the transverse coordinate ${\bf{R}}_\perp$ across the electron transverse profile imparts a local phase gradient that induces purely elastic transverse diffraction of the electron beam, without the exchange of real energy quanta with the environment.
\item \textbf{Transverse correlation kernels:}
\begin{eqnarray} \label{Gamma_kernels_main}
\Gamma _{\omega s} \left( {{\bf{R}}_{1\perp} ,{\bf{R}}_{2\perp} } \right) &=& \left( {\frac{e}{{\hbar \omega }}} \right)^2 \int_{ - \infty }^{ + \infty } {dz} \int_{ - \infty }^{ + \infty } {dz'} \;e^{ - i\frac{\omega }{V}\left( {z - z'} \right)} \nonumber \\
&& \times {\bf{u}}_V  \cdot \left[ {\int {do_{\bf{n}} } \;{\cal F}_{\omega s} \left( {\left. {{\bf{R}}_{1\perp}  + z{\bf{u}}_V } \right|{\bf{n}}} \right) \cdot {\cal F}_{\omega s}^{T*} \left( {\left. {{\bf{R}}_{2\perp}  + z'{\bf{u}}_V } \right|{\bf{n}}} \right)} \right] \cdot {\bf{u}}_V , \nonumber \\
\Gamma _{\omega em} \left( {{\bf{R}}_{1\perp} ,{\bf{R}}_{2\perp} } \right) &=& \left( {\frac{e}{{\hbar \omega }}} \right)^2 \int_{ - \infty }^{ + \infty } {dz} \int_{ - \infty }^{ + \infty } {dz'} \;e^{ - i\frac{\omega }{V}\left( {z - z'} \right)} \nonumber \\
&& \times {\bf{u}}_V  \cdot \left[ {\int_\mathscr{V} {d^3 {\bf{r}}\,} \sum\limits_\nu  {{\cal G}_{\omega \nu } \left( {\left. {{\bf{R}}_{1\perp}  + z{\bf{u}}_V } \right|{\bf{r}}} \right) \cdot {\cal G}_{\omega \nu }^{T*} \left( {\left. {{\bf{R}}_{2\perp}  + z'{\bf{u}}_V } \right|{\bf{r}}} \right)} } \right] \cdot {\bf{u}}_V .
\end{eqnarray}
These Hermitian non-local kernels represent the spatial Fourier transforms of the macroscopic dyadic fields along the paraxial electron paths. Their diagonal elements evaluated at ${\bf{R}}_{1\perp} = {\bf{R}}_{2\perp} = {\bf{R}}_\perp$ represent the fundamental spectral rates of excitation of the scattering polaritons associated with radiative emission into the free-space continuum ($\Gamma_{\omega s}$), and of the material polaritons governing dipolar absorption within the nanostructure volume ($\Gamma_{\omega em}$), providing the fast-electron counterpart of the radiative and nonradiative spectral densities established in the MLNF for localized quantum emitters \cite{Miano2025a,Miano2025b,Miano2026}. By virtue of the fundamental integral relation [Eq.~(\ref{Fund_Id_Comm_main})], their sum exactly reconstructs the total macroscopic correlation kernel governed by the imaginary part of the full dyadic Green's function (as explicitly derived in Appendix~B):
\begin{eqnarray} \label{Gamma_sum_Green_main}
\Gamma _{\omega s} \left( {{\bf{R}}_{1\perp} ,{\bf{R}}_{2\perp} } \right) + \Gamma _{\omega em} \left( {{\bf{R}}_{1\perp} ,{\bf{R}}_{2\perp} } \right) &=& \frac{{e^2 }}{{\pi \varepsilon _0 \hbar c^2 }}\int_{ - \infty }^{ + \infty } {dz} \int_{ - \infty }^{ + \infty } {dz'} \;e^{ - i\frac{\omega }{V}\left( {z - z'} \right)} \nonumber \\
&\times& {\bf{u}}_V  \cdot {\mathop{\rm Im}\nolimits} \left[ {{\cal G}_\omega  \left( {\left. {{\bf{R}}_{1\perp}  + z{\bf{u}}_V } \right|{\bf{R}}_{2\perp}  + z'{\bf{u}}_V } \right)} \right] \cdot {\bf{u}}_V .
\end{eqnarray}
This expression corresponds precisely to the non-local two-path inelastic correlation kernel governing transverse electron coherence in standard macroscopic electrodynamic treatments \cite{Schattschneider2005,Potapov2006,DiGiulio2020,Mechel2021,CiattPRB2023,CiattPRApp2023}, reducing on the diagonal (${\bf{R}}_{1\perp} = {\bf{R}}_{2\perp}$) to the canonical zero-temperature spontaneous EELS probability kernel \cite{Ritchie1988,Egerton2008,DeAbajoRMP2010,DeAbajoACSPhot2021}. 
\item \textbf{Elastic depletion optical depth:}
\begin{equation} \label{W_Rt_main}
W\left( {{\bf{R}}_\perp  } \right) = \int {d\omega } \;\left\{ {\frac{1}{2}\Gamma _{\omega s} \left( {{\bf{R}}_\perp  ,{\bf{R}}_\perp  } \right) + \left[ {n_\omega  \left( {T_{em} } \right) + \frac{1}{2}} \right]\Gamma _{\omega em} \left( {{\bf{R}}_\perp  ,{\bf{R}}_\perp  } \right)} \right\}.
\end{equation}
The non-negative exponent $W$ acts as a wave-function amplitude attenuation factor $e^{-W}$ characterizing the depletion of the electron from the strictly elastic zero-loss channel due to inelastic scattering. Crucially, by formally decoupling the scattering and material sectors, Eq.~(\ref{W_Rt_main}) explicitly sums the zero-point vacuum fluctuations (the $1/2$ factors) associated with both the radiative ($\Gamma_{\omega s}$) and material ($\Gamma_{\omega em}$) channels, while amplifying solely the internal material dissipation $\Gamma_{\omega em}$ by the thermal population $n_\omega(T_{em})$. Notably, any depletion arising from stimulated transitions driven by external incident light is strictly excluded from $W$; by virtue of normal ordering \cite{Cahill1969}, the non-vacuum contributions of the scattering sector ($\hat\rho_s$) are entirely encapsulated within the characteristic functional $\Upsilon_s$ (see below), leaving $\Gamma_{\omega s}$ in Eq.~(\ref{W_Rt_main}) to govern solely the fundamental zero-point radiative decoherence.
\item \textbf{Non-local cumulant for spontaneous and thermal polaritonic processes:}
\begin{eqnarray} \label{K_cumulant_main}
K\left( {{\bf{R}}_{1\perp} ,{\bf{R}}_{2\perp} ;Z} \right) &=& \int {d\omega } \;\Bigg\{ e^{ - i\frac{\omega }{V}Z} \Big[ {\Gamma _{\omega s}^* \left( {{\bf{R}}_{1\perp} ,{\bf{R}}_{2\perp} } \right) + \left( {n_\omega  \left( {T_{em} } \right) + 1} \right)\Gamma _{\omega em}^* \left( {{\bf{R}}_{1\perp} ,{\bf{R}}_{2\perp} } \right)} \Big] \nonumber \\
&& \quad \quad \quad \; + e^{i\frac{\omega }{V}Z} n_\omega  \left( {T_{em} } \right)\Gamma _{\omega em} \left( {{\bf{R}}_{1\perp} ,{\bf{R}}_{2\perp} } \right) \Bigg\}.
\end{eqnarray}
The non-local two-path cumulant function $K$ nonperturbatively generates (via the generating functional $e^K$) all multipolariton inelastic transitions in the absence of external light. The first term ($\propto e^{-i\frac{\omega}{V}Z}$) drives electron energy losses (Stokes transitions, $E < E_0$) through the spontaneous emission of scattering polaritons ($\Gamma_{\omega s}^*$), alongside the spontaneous and thermally stimulated emission of material polaritons ($\Gamma_{\omega em}^*$). Conversely, the second term ($\propto e^{i\frac{\omega}{V}Z}$) drives thermally activated electron energy gains (Anti-Stokes transitions, $E > E_0$) purely via the absorption of material polaritons ($\Gamma_{\omega em}$). Crucially, the mathematical structure of $K$ enforces a fundamental thermodynamic asymmetry between the two branches: while the Stokes channel allows energy dissipation into both the radiative continuum and the material, energy extraction from the cold vacuum scattering modes is strictly forbidden on the Anti-Stokes branch. Consequently, electron energy gains are restricted solely to the thermally populated material channels ($\Gamma_{\omega em}$), physically preventing a cold nanostructure from heating the electron beam.
\item \textbf{Normal-ordered characteristic functional of incident light:}
\begin{eqnarray} \label{Upsilon_functional_main}
\Upsilon _s \left( {{\bf{R}}_1 ,{\bf{R}}_2 } \right) &=& {\rm Tr}_s \left\{ \hat \rho _s \exp \left[ {\int {d\omega } \int {do_{\bf{n}} } \;{\bf{C}}_\omega ^* \left( {{\bf{R}}_1 ,{\bf{R}}_2 \left| {\bf{n}} \right.} \right) \cdot {\bf{\hat g}}_{\omega s}^\dag  \left( {\bf{n}} \right)} \right] \right. \nonumber \\
&& \quad \quad \times \left. \exp \left[ { - \int {d\omega } \int {do_{\bf{n}} } \;{\bf{C}}_\omega  \left( {{\bf{R}}_1 ,{\bf{R}}_2 \left| {\bf{n}} \right.} \right) \cdot {\bf{\hat g}}_{\omega s} \left( {\bf{n}} \right)} \right] \right\},
\end{eqnarray}
governed by the geometric two-path scattering difference vector:
\begin{eqnarray} \label{C_vector_main}
{\bf{C}}_\omega  \left( {{\bf{R}}_1 ,{\bf{R}}_2 \left| {\bf{n}} \right.} \right) &=& \frac{e}{{\hbar \omega }}\int_{ - \infty }^{ + \infty } {dz} \;e^{ - i\frac{\omega }{V}z} {\bf{u}}_V  \cdot \left[ {e^{i\frac{\omega }{V}R_{1V} } {\cal F}_{\omega s} \left( {\left. {{\bf{R}}_{1\perp}  + z{\bf{u}}_V } \right|{\bf{n}}} \right)}  { - e^{i\frac{\omega }{V}R_{2V} } {\cal F}_{\omega s} \left( {\left. {{\bf{R}}_{2\perp}  + z{\bf{u}}_V } \right|{\bf{n}}} \right)} \right],
\end{eqnarray}
evaluated in Eq.~(\ref{dP_dE_dPt_main}) over the 3D trajectory points ${\bf{R}}_j = {\bf{R}}_{j\perp} + R_{jV}{\bf{u}}_V$ with $R_{1V,2V} = \bar{R}_V \pm Z/2$ ($j=1,2$). The normal-ordered characteristic functional $\Upsilon_s$ \cite{Cahill1969} acts as the universal generating kernel for all energy and momentum exchange processes driven by the state $\hat\rho_s$ of the impinging quantum light. Its structure bridges the macroscopic electrodynamic and quantum-optical domains \cite{Ciatt2,CiattOptomech2026,CiattCoherence2026}: the classical scattering dyadic ${\cal F}_{\omega s}$ maps the incident fields onto the exact macroscopic near-field modes dressing the nanostructure, while the density operator $\hat\rho_s$ dictates the underlying photon statistics. Factoring out the center-of-mass coordinate $\bar{R}_V$ to yield $\mathbf{C}_\omega \propto e^{i\frac{\omega}{V}\bar{R}_V}$, the functional explicitly separates the exchange channels: the term governed by $\mathbf{C}_\omega^*$ couples to the creation operators $\hat{\mathbf{g}}_{\omega s}^\dag$ (carrying the macroscopic phase $\propto e^{-i\frac{\omega}{V}\bar R_V}$) to drive electron energy losses via the stimulated emission of scattering polaritons, whereas the term governed by $\mathbf{C}_\omega$ couples to the annihilation operators $\hat{\mathbf{g}}_{\omega s}$ (carrying the phase $\propto e^{i\frac{\omega}{V}\bar R_V}$) to drive electron energy gains via scattering polariton absorption. Finally, the explicit normal ordering guarantees that $\Upsilon_s \equiv 1$ whenever the incident radiation is in the pure vacuum state $\hat\rho_s = \left|0_s\right\rangle\left\langle 0_s\right|$, ensuring that zero-point vacuum fluctuations are completely isolated from the externally driven dynamics and strictly governed by $\Phi, W$, and $K$.
\end{itemize}

\section{Unilluminated regime: Interplay of scattering zero-point fluctuations and thermal material excitations}

We first examine the unilluminated regime ($\hat\rho_s = \left|0_s\right\rangle\left\langle 0_s\right|$, $\Upsilon_s \equiv 1$). Although external impinging photons are absent, the asymptotic scattering sector remains dynamically active through its zero-point vacuum fluctuations. As demonstrated below, their interplay with localized material polaritons yields striking non-equilibrium features purely driven by the nanostructure finite temperature $T_{em} > 0$.

For $\Upsilon_s \equiv 1$, the integrand in Eq.~(\ref{dP_dE_dPt_main}) is independent of the center-of-mass coordinate $\bar R_V$, and the $\bar R_V$ integral evaluates directly to $1 - |Z|/\ell$. Expanding the multipolariton generating functional $e^{K} = \sum_{N=0}^\infty \frac{1}{N!}K^N$, the distribution (\ref{dP_dE_dPt_main}) decomposes into an $N$-polariton exchange series:
\begin{eqnarray} \label{dP_no_light_series}
\frac{{d\mathscr{P}}}{{dEd^2 {\bf{P}}_\perp  }}  &=& \frac{1}{{\left( {2\pi \hbar } \right)^3 V}}\int {d^2 {\bf{R}}_{1\perp} } \int {d^2 {\bf{R}}_{2\perp} } \;e^{ - \frac{i}{\hbar }{\bf{P}}_\perp   \cdot \left( {{\bf{R}}_{1\perp}  - {\bf{R}}_{2\perp} } \right)}  \widetilde \phi \left( {{\bf{R}}_{1\perp} } \right)\widetilde \phi ^* \left( {{\bf{R}}_{2\perp} } \right)\nonumber \\
&& \times \int_{ - \ell }^\ell  {dZ} \;\left( {1 - \frac{{\left| Z \right|}}{\ell }} \right)e^{ - i\frac{{E - E_0 }}{{\hbar V}}Z} e^{K\left( {{\bf{R}}_{1\perp} ,{\bf{R}}_{2\perp} ;Z} \right)}  = \sum\limits_{N = 0}^\infty  {\left( {\frac{{d\mathscr{P}}}{{dEd^2 {\bf{P}}_\perp  }}} \right)^{\left( N \right)} }.
\end{eqnarray}

Under strong coupling (where the cumulant magnitude $\vert{}K\vert{} \sim 1$, e.g., in grazing-incidence or phase-matched 2D Cherenkov configurations \cite{Adiv2023}), multipolariton terms ($N \ge 2$) become significant. They broaden the single-loss profile into a continuum and, at $T_{em} > 0$, capture mixed thermal events comprising sequential polariton absorption and emission. Total probability across all channels remains automatically conserved, self-consistently normalized by the elastic attenuation $e^{-W}$ encoded within $\widetilde\phi$.

We explicitly evaluate the individual $N$-terms using the longitudinal Fourier identity (Appendix~A):
\begin{equation} \label{Z_Fourier_identity}
\frac{1}{{2\pi \hbar V}}\int_{ - \ell }^\ell  {dZ} \;\left( {1 - \frac{{\left| Z \right|}}{\ell }} \right)e^{ - i\frac{U}{{\hbar V}}Z}    = \delta ^{\left( {\Delta E} \right)} \left( U \right),
\end{equation}
where $\Delta E = 2\hbar V / \ell$. In the monoenergetic limit ($\ell \to \infty$, $\Delta E \to 0^+$), this regularized sequence rigorously converges to the exact Dirac delta function $\delta(U)$.

\subsection{Zero-polariton channel: Elastic zero-loss peak}

\begin{figure*}[t]
\centering
\includegraphics[width = 1\linewidth]{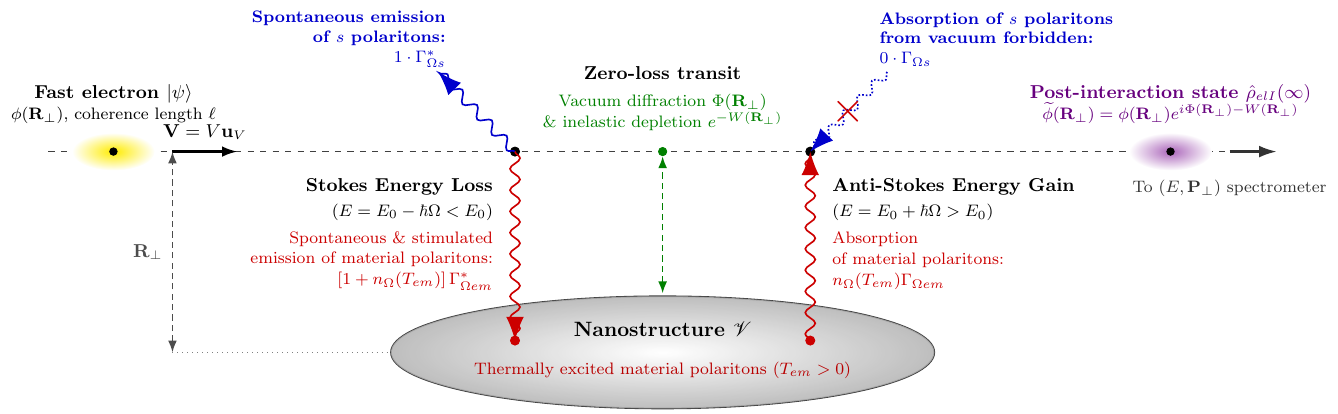}
\caption{\textbf{Physical channels of fast-electron energy exchange in the unilluminated regime.} Schematic of zero- and single-polariton interactions for an electron traversing a nanostructure $\mathscr{V}$ at finite temperature $T_{em} > 0$ in a cold vacuum. The elastic zero-loss channel ($N=0$, center) is governed by the conservative vacuum phase $\Phi(\mathbf{R}_\perp)$ and the inelastic depletion optical depth $W(\mathbf{R}_\perp)$. For single-polariton exchanges ($N=1$), the Stokes energy-loss branch (left) is driven by the spontaneous and thermally stimulated emission of material polaritons (red) alongside the spontaneous emission of free-space scattering polaritons (blue). Conversely, the Anti-Stokes energy-gain branch (right) proceeds exclusively via the absorption of thermal material polaritons confined to $\mathscr{V}$.}
\label{Fig2}
\end{figure*}

The $N = 0$ term in Eq.~(\ref{dP_no_light_series}), obtained by substituting $e^K \to 1$ and evaluating the integral via Eq.~(\ref{Z_Fourier_identity}) with $U = E - E_0$, describes the elastic zero-loss peak (ZLP) [see central panel of Fig.~\ref{Fig2}]. In the ideal monoenergetic limit, this term is strictly governed by the Dirac delta $\delta(E - E_0)$, corresponding to electrons undergoing no net energy exchange:
\begin{equation} \label{ZLP_main}
\left( {\frac{{d\mathscr{P}}}{{dEd^2 {\bf{P}}_\perp  }}} \right)^{\left( 0 \right)}  = \delta \left( {E - E_0 } \right)\left| {\frac{1}{{2\pi \hbar }}\int {d^2 {\bf{R}}_\perp  } \; e^{ - \frac{i}{\hbar }{\bf{P}}_\perp   \cdot {\bf{R}}_\perp  }  \phi \left( {{\bf{R}}_\perp  } \right)e^{i\Phi \left( {{\bf{R}}_\perp  } \right) - W\left( {{\bf{R}}_\perp  } \right)} } \right|^2 .
\end{equation}
Equation~(\ref{ZLP_main}) demonstrates that the elastic transverse momentum distribution is proportional to the squared modulus of the two-dimensional Fourier transform of the dressed wave function $\widetilde\phi({\bf{R}}_\perp) = \phi({\bf{R}}_\perp)e^{i\Phi({\bf{R}}_\perp) - W({\bf{R}}_\perp)}$. 

Crucially, the structure of $W({\bf{R}}_\perp)$ captures the physical recombination of scattering and material vacuum fluctuations into ${\rm Im}[{\cal G}_\omega]$ at $T_{em} = 0$, as well as its fundamental breakdown at $T_{em} > 0$. At zero nanostructure temperature ($T_{em} = 0$, $n_\omega(0) = 0$), the scattering ($\Gamma_{\omega s}$) and material ($\Gamma_{\omega em}$) rates in Eq.~(\ref{W_Rt_main}) share the identical $1/2$ zero-point statistical weight. Consequently, their unweighted sum exactly recombines via Eq.~(\ref{Gamma_sum_Green_main}) into ${\rm Im}[{\cal G}_\omega]$, yielding the well-known zero-temperature elastic depletion optical depth \cite{DiGiulio2020,CiattPRApp2023}:
\begin{equation} \label{W_LNF_T0}
W\left( {{\bf{R}}_\perp  } \right)  = \frac{{e^2 }}{{2\pi \varepsilon _0 \hbar c^2 }}\int {d\omega } \int_{ - \infty }^{ + \infty } {dz} \int_{ - \infty }^{ + \infty } {dz'} \;e^{ - i\frac{\omega }{V}\left( {z - z'} \right)} {\bf{u}}_V  \cdot {\mathop{\rm Im}\nolimits} \left[ {{\cal G}_\omega  \left( {\left. {{\bf{R}}_\perp   + z{\bf{u}}_V } \right|{\bf{R}}_\perp   + z'{\bf{u}}_V } \right)} \right] \cdot {\bf{u}}_V .
\end{equation}
This recombination explains why direct applications of the standard LNF to bounded objects reproduce Eq.~(\ref{W_LNF_T0}) at $T_{em} = 0$: when both the field expansion [Eq.~(\ref{E_omega_main})] and the integral relation [Eq.~(\ref{Fund_Id_Comm_main})] are restricted to the internal volume, replacing the material correlator with the full ${\rm Im}[{\cal G}_\omega]$ implicitly restores the zero-point radiative contribution. However, at $T_{em} > 0$, Eq.~(\ref{W_Rt_main}) dictates that the thermal factor $[n_\omega(T_{em}) + 1/2]$ amplifies exclusively the localized material rate $\Gamma_{\omega em}$, leaving the radiative rate $\Gamma_{\omega s}$ weighted by $1/2$. Because the system is out of global equilibrium with the zero-temperature surrounding vacuum, these rates no longer synthesize into ${\rm Im}[{\cal G}_\omega]$. Applying a single thermal factor to the full ${\rm Im}[{\cal G}_\omega]$ at $T_{em} > 0$ would instead describe a nanostructure in thermal equilibrium with a heated blackbody environment, thereby overestimating the ZLP depletion in a cold vacuum for radiatively bright modes.

\subsection{Single-polariton transitions: Stokes and Anti-Stokes branches}

The $N = 1$ polariton contribution in Eq.~(\ref{dP_no_light_series}), linear in $K({\bf{R}}_{1\perp}, {\bf{R}}_{2\perp}; Z)$, describes the inelastic scattering of the electron via single-polariton exchange, whose Stokes and Anti-Stokes branches are illustrated in the left and right panels of Fig.~\ref{Fig2}. Substituting Eq.~(\ref{K_cumulant_main}) and applying the identity (\ref{Z_Fourier_identity}) with $U = E - E_0 \pm \hbar\omega$, the longitudinal integral over $Z$ yields the regularized distributions $\delta^{(\Delta E)}(E - E_0 \pm \hbar\omega)$. In the monoenergetic limit ($\Delta E \to 0^+$), both regularized distributions converge to the exact Dirac delta $\frac{1}{\hbar}\delta(\omega - \Omega)$, where $\hbar\Omega = \vert{}E - E_0\vert{} > 0$ is the absolute transferred energy. The subsequent frequency integral $\int d\omega$ inherently present in the definition of $K$ then trivially absorbs the deltas, directly yielding the Stokes ($E = E_0 - \hbar\Omega < E_0$) and Anti-Stokes ($E = E_0 + \hbar\Omega > E_0$) energy-momentum distributions:
\begin{eqnarray} \label{N1_Stokes_AntiStokes_main}
\left. {\left( {\frac{{d\mathscr{P}}}{{dEd^2 {\bf{P}}_\perp  }}} \right)^{\left( 1 \right)} } \right|_{E = E_0 - \hbar \Omega  < E_0 }  &=& \frac{1}{{\hbar \left( {2\pi \hbar } \right)^2 }}\int {d^2 {\bf{R}}_{1\perp} } \int {d^2 {\bf{R}}_{2\perp} } \; e^{ - \frac{i}{\hbar }{\bf{P}}_\perp   \cdot \left( {{\bf{R}}_{1\perp}  - {\bf{R}}_{2\perp} } \right)} \widetilde \phi \left( {{\bf{R}}_{1\perp} } \right)\widetilde \phi ^* \left( {{\bf{R}}_{2\perp} } \right) \nonumber \\
&& \times \left\{ {\Gamma _{\Omega s}^* \left( {{\bf{R}}_{1\perp} ,{\bf{R}}_{2\perp} } \right) + \left[ {n_\Omega  \left( {T_{em} } \right) + 1} \right]\Gamma _{\Omega em}^* \left( {{\bf{R}}_{1\perp} ,{\bf{R}}_{2\perp} } \right)} \right\}, \nonumber \\
\left. {\left( {\frac{{d\mathscr{P}}}{{dEd^2 {\bf{P}}_\perp  }}} \right)^{\left( 1 \right)} } \right|_{E = E_0  + \hbar \Omega  > E_0 }  &=& \frac{1}{{\hbar \left( {2\pi \hbar } \right)^2 }}\int {d^2 {\bf{R}}_{1\perp} } \int {d^2 {\bf{R}}_{2\perp} } \;e^{ - \frac{i}{\hbar }{\bf{P}}_\perp   \cdot \left( {{\bf{R}}_{1\perp}  - {\bf{R}}_{2\perp} } \right)} \widetilde \phi \left( {{\bf{R}}_{1\perp} } \right)\widetilde \phi ^* \left( {{\bf{R}}_{2\perp} } \right) \nonumber \\
&& \times n_\Omega  \left( {T_{em} } \right)\Gamma _{\Omega em} \left( {{\bf{R}}_{1\perp} ,{\bf{R}}_{2\perp} } \right).
\end{eqnarray}
Note that the necessary reality and non-negativity of the probability distributions in Eq.~(\ref{N1_Stokes_AntiStokes_main}) are explicitly guaranteed. Indeed, by inserting the classical kernels $\Gamma_{\Omega s}$ and $\Gamma_{\Omega em}$ from Eq.~(\ref{Gamma_kernels_main}), they can be equivalently rewritten as sums of integrals of squared two-dimensional transverse Fourier amplitudes, scaled by strictly positive vacuum and thermal weights (i.e., $1$, $n_\Omega(T_{em}) + 1$, and $n_\Omega(T_{em})$).

\begin{figure*}[t]
\centering
\includegraphics[width = 1\linewidth]{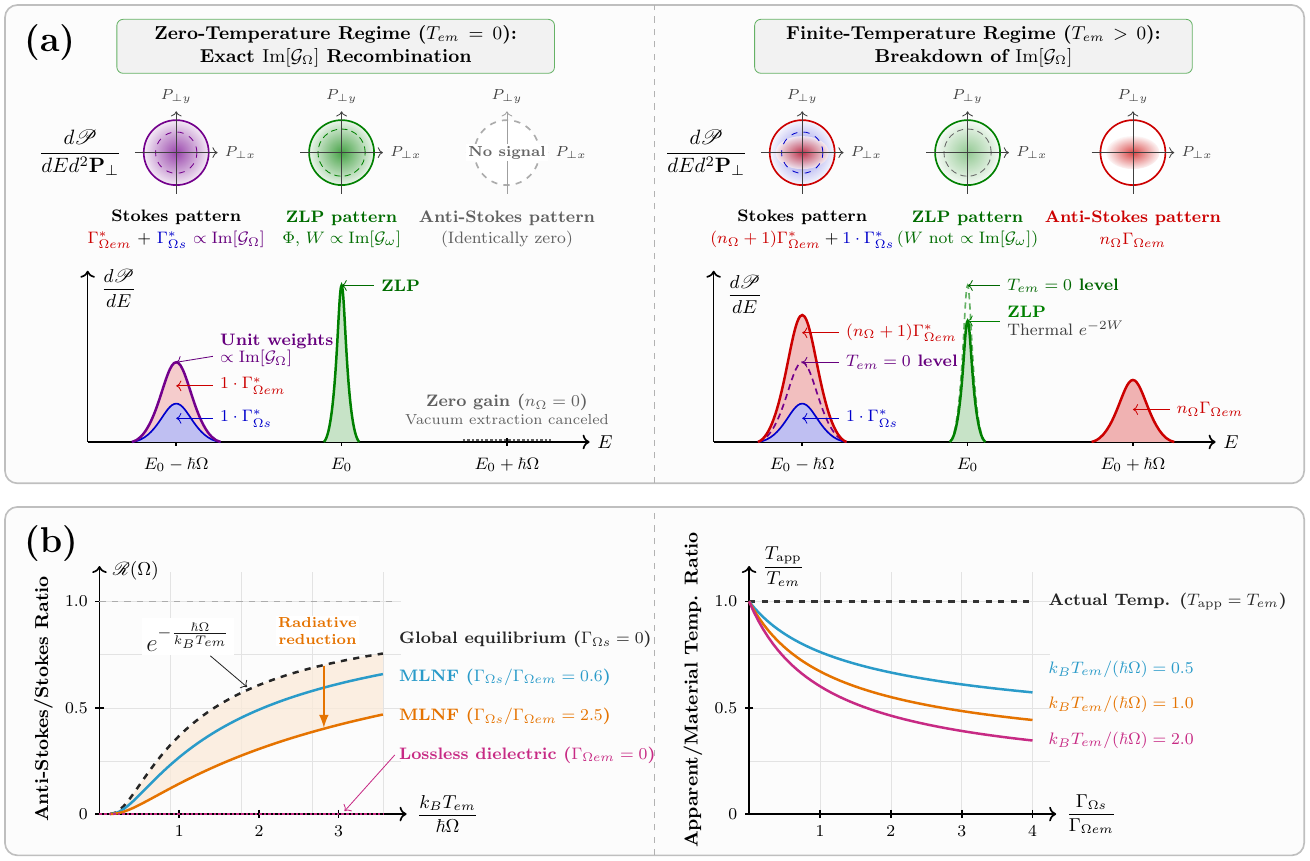}
\caption{\textbf{Stokes–Anti-Stokes asymmetry in single-polariton electron diffraction and non-equilibrium thermometry.} 
(a) Transverse diffraction patterns $d^2\mathscr{P}/(dE d^2\mathbf{P}_\perp)$ and marginal spectra $d\mathscr{P}/dE$. At $T_{em} = 0$ (left), unit statistical weights enable exact $\mathrm{Im}[\mathcal{G}_\Omega]$ recombination, yielding a unified Stokes pattern and vanishing Anti-Stokes. At $T_{em} > 0$ (right), distinct thermal weights break this recombination: the Stokes pattern mixes delocalized scattering and localized material modes, diverging qualitatively from the purely material-driven Anti-Stokes pattern. 
(b) Left: Spontaneous emission of scattering polaritons ($\Gamma_{\Omega s} > 0$) suppresses the Anti-Stokes-to-Stokes probability ratio $\mathscr{R}(\Omega)$ [see Eq.(\ref{Thermometry_Ratio_main})] below the global-equilibrium Boltzmann limit (dashed). Right: Consequently, standard Boltzmann inversion yields an apparent temperature $T_{\mathrm{app}}$ that systematically underestimates the true material temperature $T_{em}$, governed by the branching ratio $\Gamma_{\Omega s}/\Gamma_{\Omega em}$.}
\label{Fig3}
\end{figure*}

\subsection{Single-polariton transitions at zero temperature: Stokes recombination versus lossless dielectric limit}

In the strict zero-temperature regime ($T_{em} = 0$, $n_\Omega(0) = 0$), the initial absence of thermally excited material polaritons strictly forbids electron energy gains. The Anti-Stokes branch in the second of Eqs.~(\ref{N1_Stokes_AntiStokes_main}) thus vanishes identically. Under this condition, the Stokes branch in the first of Eqs.~(\ref{N1_Stokes_AntiStokes_main}) establishes two fundamental results:

\begin{itemize}
\item {\bf Zero-temperature Stokes recombination into ${\rm Im}[{\cal G}_\Omega]$ and derivation of the standard EELS formula:}
At $T_{em} = 0$, the statistical weight $[n_\Omega(0) + 1] = 1$ of the internal material polaritons ($\Gamma_{\Omega em}^*$) coincides with the unit statistical weight of the asymptotic scattering polaritons ($\Gamma_{\Omega s}^*$). Consequently, the two channels enter the first of Eqs.~(\ref{N1_Stokes_AntiStokes_main}) through their unweighted sum $\Gamma_{\Omega s}^*({\bf{R}}_{1\perp}, {\bf{R}}_{2\perp}) + \Gamma_{\Omega em}^*({\bf{R}}_{1\perp}, {\bf{R}}_{2\perp})$. By virtue of the fundamental integral identity [Eq.~(\ref{Fund_Id_Comm_main})], this sum rigorously recombines to reconstruct the total macroscopic correlation kernel governed by the imaginary part of the classical dyadic Green's function [as explicitly shown in Eq.~(\ref{Gamma_sum_Green_main})]—mirroring the mathematical recombination previously discussed for the elastic depletion optical depth $W$:
\begin{eqnarray} \label{Stokes_T0_Green_main}
\left. {\left( {\frac{{d\mathscr{P}}}{{dEd^2 {\bf{P}}_\perp  }}} \right)^{\left( 1 \right)} } \right|_{\substack{E = E_0 - \hbar \Omega  \\ T_{em}  = 0}} &=& \frac{{e^2 }}{{\pi \varepsilon _0 \hbar ^2 c^2 \left( {2\pi \hbar } \right)^2 }}\int {d^2 {\bf{R}}_{1\perp} } \int {d^2 {\bf{R}}_{2\perp} } \;e^{ - \frac{i}{\hbar }{\bf{P}}_\perp   \cdot \left( {{\bf{R}}_{1\perp}  - {\bf{R}}_{2\perp} } \right)}  \widetilde \phi \left( {{\bf{R}}_{1\perp} } \right)\widetilde \phi ^* \left( {{\bf{R}}_{2\perp} } \right)\nonumber \\
&& \times \int_{ - \infty }^{ + \infty } {dz} \int_{ - \infty }^{ + \infty } {dz'} \;e^{i\frac{\Omega }{V}\left( {z - z'} \right)} {\bf{u}}_V  \cdot {\mathop{\rm Im}\nolimits} \left[ {{\cal G}_\Omega  \left( {\left. {{\bf{R}}_{1\perp}  + z{\bf{u}}_V } \right|{\bf{R}}_{2\perp}  + z'{\bf{u}}_V } \right)} \right] \cdot {\bf{u}}_V .
\end{eqnarray}
In the strictly perturbative weak-coupling regime ($W({\bf{R}}_\perp) \ll 1$ and $\Phi({\bf{R}}_\perp) \ll 1$), Eq.~(\ref{Stokes_T0_Green_main}) recovers the first-order perturbative transverse coherence expression of Ref.~\cite{Mechel2021}, and its integral over $d^2{\bf{P}}_\perp$ reproduces the standard Ritchie--Howie EELS probability \cite{DeAbajoRMP2010,Ritchie1988}:
\begin{equation} \label{Ritchie_Howie_EELS}
\frac{{d\mathscr{P}}}{{dE}} = \frac{{e^2 }}{{\pi \varepsilon _0 \hbar ^2 c^2 }}\int {d^2 {\bf{R}}_\perp  } \;\left| {\phi \left( {{\bf{R}}_\perp  } \right)} \right|^2 \int_{ - \infty }^{ + \infty } {dz} \int_{ - \infty }^{ + \infty } {dz'} \;e^{i\frac{\Omega }{V}\left( {z - z'} \right)} {\bf{u}}_V  \cdot {\mathop{\rm Im}\nolimits} \left[ {{\cal G}_\Omega  \left( {\left. {{\bf{R}}_\perp   + z{\bf{u}}_V } \right|{\bf{R}}_\perp   + z'{\bf{u}}_V } \right)} \right] \cdot {\bf{u}}_V .
\end{equation}
As noted for the elastic depletion $W$, this derivation clarifies why direct LNF treatments of finite objects \cite{DiGiulio2020,Mechel2021,CiattPRB2023,CiattPRApp2023} yield the exact EELS probability at $T_{em} = 0$: because the scattering and material vacuum channels carry equal statistical weights, identifying the internal source integral with the full ${\rm Im}[{\cal G}_\Omega]$ automatically accounts for both radiative and nonradiative energy losses.

\item {\bf Rigorous lossless dielectric limit:}
The physical necessity of the scattering sector ${\bf{\hat g}}_{\omega s}({\bf{n}})$ at $T_{em} = 0$ is manifest for lossless dielectrics (e.g., transparent waveguides \cite{Kfir2020,Kfir2021}, where ${\rm Im}[\varepsilon_\omega({\bf{r}})] = {\rm Im}[\mu_\omega({\bf{r}})] = 0$). Here, the material kernels identically vanish (${\cal G}_{\omega\nu} \equiv 0$, $\Gamma_{\Omega em} \equiv 0$). Consequently, the standard LNF—which constructs the field exclusively from sources within $\mathscr{V}$—yields a vanishing field operator unless singular vacuum limiting procedures are applied \cite{Hanson2021,CiattLimiting2026}. Although Eq.~(\ref{Stokes_T0_Green_main}) remains well-defined across the lossless limit—since ${\mathop{\rm Im}\nolimits}[{\cal G}_\Omega]$ inherently captures radiative losses—this formulation can only be established from first principles via the MLNF. Without the explicit scattering kernel $\Gamma_{\Omega s}^*({\bf{R}}_{1\perp}, {\bf{R}}_{2\perp})$ in the first of Eqs.~(\ref{N1_Stokes_AntiStokes_main}), the integral identity required to reconstruct ${\mathop{\rm Im}\nolimits}[{\cal G}_\Omega]$ breaks down. By retaining $\Gamma_{\Omega s}^*$, the MLNF natively accounts for radiative EELS, Cherenkov, and transition radiation without singular mathematical procedures.
\end{itemize}

\subsection{Single-polariton transitions at $T_{em} > 0$: Non-equilibrium Stokes--Anti-Stokes asymmetry}

At a finite nanostructure temperature $T_{em} > 0$, the traversing electron can absorb available thermal material polaritons. Because the material statistical weight $[n_\Omega(T_{em}) + 1]$ on the Stokes branch no longer matches the unit weight of vacuum scattering polaritons ($\Gamma_{\Omega s}^*$), the two kernels in the first of Eqs.~(\ref{N1_Stokes_AntiStokes_main}) cannot recombine into ${\rm Im}[{\cal G}_\Omega]$. Furthermore, the Anti-Stokes energy-gain branch [second of Eqs.~(\ref{N1_Stokes_AntiStokes_main})] lacks the scattering channel entirely, being strictly driven by thermal material absorption via $n_\Omega(T_{em})\Gamma_{\Omega em}$.

This structural Stokes--Anti-Stokes asymmetry yields two major physical predictions stemming from the explicit and independent MLNF quantization of the scattering and material polariton sectors:
\begin{itemize}
\item {\bf Distinct transverse diffraction patterns for energy loss and thermal gain:}
At $T_{em} > 0$, electrons losing energy $\hbar\Omega$ diffract according to the combined kernel $\Gamma_{\Omega s}^*({\bf{R}}_{1\perp}, {\bf{R}}_{2\perp}) + [n_\Omega(T_{em}) + 1]\Gamma_{\Omega em}^*({\bf{R}}_{1\perp}, {\bf{R}}_{2\perp})$, whereas electrons gaining energy diffract solely via the material kernel $n_\Omega(T_{em})\Gamma_{\Omega em}({\bf{R}}_{1\perp}, {\bf{R}}_{2\perp})$. Because $\Gamma_{\Omega s}$ (propagating radiation) and $\Gamma_{\Omega em}$ (localized thermal near-fields) possess fundamentally different spatial profiles, their Stokes superposition yields a normalized angular diffraction pattern qualitatively distinct from the purely material Anti-Stokes pattern [see Fig.~\ref{Fig3}(a)], provided $\Gamma_{\Omega s} \neq 0$. While standard EELS thermometry relies purely on spectral intensity ratios \cite{Idrobo2018,Lagos2018}, this spatial asymmetry demonstrates that momentum-space signatures of loss and gain are intrinsically uncoupled, establishing a novel observable for angle-resolved electron spectroscopy.

\item {\bf Generalized non-equilibrium detailed-balance relation:}
Integrating Eqs.~(\ref{N1_Stokes_AntiStokes_main}) over ${\bf{P}}_\perp$ yields the marginal single-polariton energy spectra:
\begin{eqnarray} \label{Marginal_Stokes_AntiStokes_main}
\left. {\left( {\frac{{d\mathscr{P}}}{{dE}}} \right)^{\left( 1 \right)} } \right|_{E = E_0  - \hbar \Omega  < E_0 }  &=& \frac{1}{\hbar }\int {d^2 {\bf{R}}_\perp  } \;\left| {\phi \left( {{\bf{R}}_\perp  } \right)} \right|^2 e^{ - 2W\left( {{\bf{R}}_\perp  } \right)} \left\{ {\Gamma _{\Omega s} \left( {{\bf{R}}_\perp  ,{\bf{R}}_\perp  } \right) + \left[ {n_\Omega  \left( {T_{em} } \right) + 1} \right]\Gamma _{\Omega em} \left( {{\bf{R}}_\perp  ,{\bf{R}}_\perp  } \right)} \right\}, \nonumber \\
\left. {\left( {\frac{{d\mathscr{P}}}{{dE}}} \right)^{\left( 1 \right)} } \right|_{E = E_0  + \hbar \Omega  > E_0 }  &=& \frac{1}{\hbar }\int {d^2 {\bf{R}}_\perp  } \;\left| {\phi \left( {{\bf{R}}_\perp  } \right)} \right|^2 e^{ - 2W\left( {{\bf{R}}_\perp  } \right)} n_\Omega  \left( {T_{em} } \right)\Gamma _{\Omega em} \left( {{\bf{R}}_\perp  ,{\bf{R}}_\perp  } \right).
\end{eqnarray}
For a tightly focused electron beam at impact parameter ${\bf{R}}_\perp$, the Anti-Stokes-to-Stokes probability ratio $\mathscr{R}(\Omega)$ becomes: 
\begin{eqnarray} \label{Thermometry_Ratio_main}
\mathscr{R}\left( \Omega  \right) &\equiv& \frac{{\left( {d\mathscr{P}/dE} \right)_{E_0  + \hbar \Omega }^{\left( 1 \right)} }}{{\left( {d\mathscr{P}/dE} \right)_{E_0  - \hbar \Omega }^{\left( 1 \right)} }} = \frac{{n_\Omega  \left( {T_{em} } \right)\Gamma _{\Omega em} \left( {{\bf{R}}_\perp  ,{\bf{R}}_\perp  } \right)}}{{\left[ {n_\Omega  \left( {T_{em} } \right) + 1} \right]\Gamma _{\Omega em} \left( {{\bf{R}}_\perp  ,{\bf{R}}_\perp  } \right) + \Gamma _{\Omega s} \left( {{\bf{R}}_\perp  ,{\bf{R}}_\perp  } \right)}} \nonumber \\
&=& \frac{{e^{ - \frac{{\hbar \Omega }}{{k_B T_{em} }}} }}{{1 + \left( {1 - e^{ - \frac{{\hbar \Omega }}{{k_B T_{em} }}} } \right)\frac{{\Gamma _{\Omega s} \left( {{\bf{R}}_\perp  ,{\bf{R}}_\perp  } \right)}}{{\Gamma _{\Omega em} \left( {{\bf{R}}_\perp  ,{\bf{R}}_\perp  } \right)}}}} \le e^{ - \frac{{\hbar \Omega }}{{k_B T_{em} }}} .
\end{eqnarray}
Evaluating the finite-temperature response purely via ${\rm Im}[{\cal G}_\Omega]$ inherently forces both material and radiative channels to share the thermal weight $n_\Omega(T_{em})$, mathematically simulating global equilibrium with a hot blackbody vacuum (Appendix~C\,1) and recovering the standard Boltzmann ratio $e^{-\hbar\Omega/(k_B T_{em})}$ \cite{Idrobo2018,Lagos2018}. In contrast, Eq.~(\ref{Thermometry_Ratio_main}) proves that for a heated nanostructure in a cold vacuum, the spontaneous emission of scattering polaritons ($\Gamma_{\Omega s} > 0$) strictly reduces the gain-to-loss ratio below the Boltzmann limit [Fig.~\ref{Fig3}(b), left]. Furthermore, Eq.~(\ref{Thermometry_Ratio_main}) rigorously vanishes in the lossless dielectric limit ($\Gamma_{\Omega em} \to 0$); this complies with Kirchhoff's law of thermal radiation, preventing electron energy gains from objects unable to absorb. Conventionally, electron thermometry defines an apparent temperature $T_{\mathrm{app}}$ by inverting the pure Boltzmann law, $\mathscr{R}(\Omega) \equiv e^{-\hbar\Omega/(k_B T_{\mathrm{app}})}$. Consequently, neglecting the radiative branching ratio $\Gamma_{\Omega s}/\Gamma_{\Omega em}$ in Eq.~(\ref{Thermometry_Ratio_main}) for radiatively bright modes yields an apparent temperature $T_{\mathrm{app}}$ that systematically underestimates the true macroscopic temperature $T_{em}$ [Fig.~\ref{Fig3}(b), right].
\end{itemize}

\section{Quantum-light illumination regime: Multimode continuous-variable fields and the standard PINEM and QPINEM limits}

We now examine the general scenario in which the nanostructure is externally illuminated by an arbitrary quantum state of light described by the density operator $\hat\rho_s \neq \left|0_s\right\rangle\left\langle 0_s\right|$ such that $\Upsilon_s({\bf{R}}_1, {\bf{R}}_2) \neq 1$. Before analyzing specific states of incident light, we highlight two structural properties of the master formula (\ref{dP_dE_dPt_main})--(\ref{C_vector_main}), which stem directly from the MLNF partitioning of the polariton Fock space into distinct material and scattering sectors.

\paragraph{\bf Unified encoding of arbitrary quantum states and complex nanostructures:}
Equations~(\ref{dP_dE_dPt_main})--(\ref{C_vector_main}) bridge macroscopic classical-illumination models and discrete lossless-cavity approximations by merging quantum optics with dissipative macroscopic electrodynamics. Any pure or mixed incident state is strictly accommodated by the density operator $\hat\rho_s$ defined over the asymptotic scattering modes. Concurrently, the geometry, dispersion, and dissipation of the nanostructure $\mathscr{V}$ are encoded via two distinct classical kernels: the scattering kernel ${\cal F}_{\omega s}({\bf{r}}|{\bf{n}})$ within $\Upsilon_s$ maps the incident quantum radiation onto the macroscopic scattered fields, while the material kernels ${\cal G}_{\omega\nu}({\bf{r}}|{\bf{r}}')$ within the macroscopic exponents ($\Phi$, $W$, and $K$) independently govern spontaneous and thermal transitions.

\paragraph{\bf Dressing of light-stimulated transitions by the inelastic background of the unilluminated regime:}
In Eq.~(\ref{dP_dE_dPt_main}), the longitudinal average of $\Upsilon_s({\bf{R}}_1, {\bf{R}}_2)$ over $\bar R_V$ acts multiplicatively on the dressed transverse wave functions $\widetilde \phi \left( {{\bf{R}}_{1\perp} } \right)\widetilde \phi ^* \left( {{\bf{R}}_{2\perp} } \right)$ and the cumulant $e^{K\left( {{\bf{R}}_{1\perp} ,{\bf{R}}_{2\perp} ;Z} \right)}$. By the Fourier convolution theorem, the total distribution $\frac{d\mathscr{P}}{dE d^2{\bf{P}}_\perp}$ constitutes the exact energy-momentum convolution of the optically stimulated profile with the unilluminated multipolariton distribution (\ref{dP_no_light_series}). Physically, this dictates that any optically driven transition is inherently accompanied by elastic attenuation $e^{-W({\bf{R}}_\perp)}$, conservative vacuum diffraction $\Phi({\bf{R}}_\perp)$, and environment-driven inelastic scattering. Consequently, every quantum sideband is intrinsically dressed and broadened by the spontaneous loss continuum and, at finite nanostructure temperatures, by thermal loss and gain shoulders.

We now apply this framework to evaluate the exact electron dynamics under continuous-variable illuminations (thermal, coherent, and squeezed states) and derive the rigorous limits to standard semiclassical PINEM and single-mode QPINEM models.

\subsection{Ambient thermal radiation}

We first consider incident free-space radiation in thermal equilibrium at temperature $T_s$, such that the state of the asymptotic scattering sector ($s$) is described by the canonical Gibbs density operator [Eq.~(\ref{AppC_rho_s_thermal})]
\begin{equation} \label{rho_s_thermal_main}
\hat \rho _s  = \frac{{e^{ - \hat H_s /\left( {k_B T_s } \right)} }}{{{\rm Tr}_s \left[ {e^{ - \hat H_s /\left( {k_B T_s } \right)} } \right]}},
\end{equation}
where $\hat H_s = \int d\omega\,\hbar\omega \int do_{\bf{n}}\,{\bf{\hat g}}_{\omega s}^\dag({\bf{n}})\cdot{\bf{\hat g}}_{\omega s}({\bf{n}})$ is the free-field Hamiltonian of the MLNF scattering sector, and $n_\omega(T_s) = [e^{\hbar\omega/(k_B T_s)} - 1]^{-1}$ is the corresponding Bose-Einstein occupation. Evaluating the normal-ordered functional $\Upsilon_s({\bf{R}}_1, {\bf{R}}_2)$ via Wick's theorem [Appendix~C\,1, Eq.~(\ref{AppC_Upsilon_T_explicit})] yields the exact transverse-momentum-resolved energy exchange probability in the monoenergetic limit ($\ell \to \infty$) [Eqs.~(\ref{AppC_dP_T_final}) and (\ref{AppC_W_K_T_defs})]:
\begin{eqnarray} \label{dP_T_main}
\frac{{d\mathscr{P}}}{{dEd^2 {\bf{P}}_\perp  }} &=& \frac{1}{{\left( {2\pi \hbar } \right)^3 V}}\int {d^2 {\bf{R}}_{1\perp} } \int {d^2 {\bf{R}}_{2\perp} } \;e^{ - \frac{i}{\hbar }{\bf{P}}_\perp   \cdot \left( {{\bf{R}}_{1\perp}  - {\bf{R}}_{2\perp} } \right)} \widetilde \phi _T \left( {{\bf{R}}_{1\perp} } \right)\widetilde \phi _T^* \left( {{\bf{R}}_{2\perp} } \right) \nonumber \\
&& \times \int_{ - \infty }^{ + \infty } {dZ} \;e^{ - i\frac{{E - E_0 }}{{\hbar V}}Z} e^{K_T \left( {{\bf{R}}_{1\perp} ,{\bf{R}}_{2\perp} ;Z} \right)} ,
\end{eqnarray}
where $\widetilde\phi_T({\bf{R}}_\perp) = \phi({\bf{R}}_\perp)e^{i\Phi({\bf{R}}_\perp) - W_T({\bf{R}}_\perp)}$, and the total optical depth $W_T({\bf{R}}_\perp)$ and two-path cumulant $K_T({\bf{R}}_{1\perp}, {\bf{R}}_{2\perp}; Z)$ read
\begin{eqnarray} \label{W_K_T_main}
W_T \left( {{\bf{R}}_\perp  } \right) &=& \int {d\omega } \;\left\{ {\left[ {n_\omega  \left( {T_s } \right) + \frac{1}{2}} \right]\Gamma _{\omega s} \left( {{\bf{R}}_\perp  ,{\bf{R}}_\perp  } \right) + \left[ {n_\omega  \left( {T_{em} } \right) + \frac{1}{2}} \right]\Gamma _{\omega em} \left( {{\bf{R}}_\perp  ,{\bf{R}}_\perp  } \right)} \right\}, \nonumber \\
K_T \left( {{\bf{R}}_{1\perp} ,{\bf{R}}_{2\perp} ;Z} \right) &=& \int {d\omega } \;\Bigg\{ e^{ - i\frac{\omega }{V}Z} \Big[ {\left( {n_\omega  \left( {T_s } \right) + 1} \right)\Gamma _{\omega s}^* \left( {{\bf{R}}_{1\perp} ,{\bf{R}}_{2\perp} } \right) + \left( {n_\omega  \left( {T_{em} } \right) + 1} \right)\Gamma _{\omega em}^* \left( {{\bf{R}}_{1\perp} ,{\bf{R}}_{2\perp} } \right)} \Big] \nonumber \\
&& \quad \quad \; + e^{i\frac{\omega }{V}Z} \Big[ {n_\omega  \left( {T_s } \right)\Gamma _{\omega s} \left( {{\bf{R}}_{1\perp} ,{\bf{R}}_{2\perp} } \right) + n_\omega  \left( {T_{em} } \right)\Gamma _{\omega em} \left( {{\bf{R}}_{1\perp} ,{\bf{R}}_{2\perp} } \right)} \Big] \Bigg\}.
\end{eqnarray}

Equations~(\ref{dP_T_main}) and (\ref{W_K_T_main}) treat the scattering and material sectors symmetrically at their respective temperatures $T_s$ and $T_{em}$, establishing two main physical results:

\begin{itemize}
\item {\bf Non-equilibrium interplay and near-field-mediated thermal gain:}
Equations~(\ref{dP_T_main}) and (\ref{W_K_T_main}) exactly describe an electron probing a nanostructure at $T_{em}$ immersed in blackbody radiation at $T_s$. In contrast to the unilluminated regime---where isolated material heating induces a fundamental Stokes--Anti-Stokes asymmetry---here the localized material kernel ($\Gamma_{\omega em}$) and the delocalized scattering kernel ($\Gamma_{\omega s}$) enter the dynamics as perfectly symmetric counterparts, strictly governed by their respective macroscopic temperatures. When the system is in global thermal equilibrium ($T_s = T_{em}$), the identical statistical weights allow the radiative and material contributions to exactly recombine into the full ${\rm Im}[{\cal G}_\omega]$. In this limit alone, the inelastic dynamics are completely determined by the standard fluctuation-dissipation theorem, rigorously recovering established theoretical models based exclusively on the full ${\rm Im}[{\cal G}_\omega]$. Conversely, in out-of-equilibrium scenarios ($T_s \neq T_{em}$), this structural recombination strictly breaks down. Crucially, even for a zero-temperature ($T_{em} = 0$) or lossless ($\Gamma_{\omega em}=0$) object, a hot ambient vacuum ($T_s > 0$) drives Anti-Stokes energy gains. In a purely thermal analog to the stimulated PINEM interaction, the nanostructure acts as a fundamental kinematic mediator: it geometrically scatters the free-space thermal radiation, generating the high-momentum evanescent near-fields necessary to bridge the kinematic gap and allow the traversing electron to extract energy from the ambient thermal bath.

\item {\bf Generalized detailed balance and nanoscale thermometry out of equilibrium:}
For any $(T_s, T_{em})$, the Anti-Stokes-to-Stokes ratio at frequency $\Omega$ for a focused beam at ${\bf{R}}_\perp$ obeys the generalized non-equilibrium detailed-balance law:
\begin{equation} \label{Two_Temp_Ratio_main}
\mathscr{R}_T \left( \Omega  \right) = \frac{{n_\Omega  \left( {T_s } \right)\Gamma _{\Omega s} \left( {{\bf{R}}_\perp  ,{\bf{R}}_\perp  } \right) + n_\Omega  \left( {T_{em} } \right)\Gamma _{\Omega em} \left( {{\bf{R}}_\perp  ,{\bf{R}}_\perp  } \right)}}{{\left[ {n_\Omega  \left( {T_s } \right) + 1} \right]\Gamma _{\Omega s} \left( {{\bf{R}}_\perp  ,{\bf{R}}_\perp  } \right) + \left[ {n_\Omega  \left( {T_{em} } \right) + 1} \right]\Gamma _{\Omega em} \left( {{\bf{R}}_\perp  ,{\bf{R}}_\perp  } \right)}}.
\end{equation}
This ratio is strictly governed by the radiative-to-dissipative branching ratio $\Gamma_{\Omega s}/\Gamma_{\Omega em}$, natively reducing to Eq.~(\ref{Thermometry_Ratio_main}) in a cold vacuum ($T_s = 0$). Standard electron thermometry extracts an apparent temperature $T_{\text{app}}$ by formally imposing $\mathscr{R}_T(\Omega) \equiv e^{-\hbar\Omega/(k_B T_{\text{app}})}$. Consequently, out of equilibrium ($T_s \neq T_{em}$), spontaneous radiative losses ($T_s < T_{em}$) systematically reduce $T_{\text{app}}$ below $T_{em}$, whereas near-field-mediated thermal absorption ($T_s > T_{em}$) radiatively enhances it. Crucially, because the radiative and dissipative kernels exhibit distinct spatial decay lengths, the extracted $T_{\text{app}}$ acquires an explicit dependence on the impact parameter ${\bf{R}}_\perp$, dynamically mapping the thermodynamic transition from the material surface to the free-space vacuum. The standard spatially uniform Boltzmann factor $\mathscr{R}_T(\Omega) = e^{-\hbar\Omega/(k_B T)}$ is exclusively recovered in global equilibrium ($T_s = T_{em} \equiv T$), confirming that formulations expressed solely in terms of the total ${\rm Im}[{\cal G}_\omega]$ describe global equilibrium and do not resolve out-of-equilibrium radiative corrections.
\end{itemize}

\subsection{Multimode coherent illumination and the semiclassical PINEM limit}

We now consider continuous-wave multimode coherent illumination at frequency $\bar\omega$, prepared in the pure state $\hat\rho_s = \left|\psi_s\right\rangle\left\langle\psi_s\right|$ defined in Eq.~(\ref{AppC_coh_state}):
\begin{equation} \label{coh_state_main}
\left| {\psi _s } \right\rangle  = \exp \left\{ {\int {do_{\bf{n}} } \;\left[ {{\bm{\alpha }}\left( {\bf{n}} \right) \cdot {\bf{\hat g}}_{\bar \omega s}^\dag  \left( {\bf{n}} \right) - {\bm{\alpha }}^* \left( {\bf{n}} \right) \cdot {\bf{\hat g}}_{\bar \omega s} \left( {\bf{n}} \right)} \right]} \right\}\left| {0_s } \right\rangle ,
\end{equation}
where ${\bm{\alpha}}({\bf{n}})$ is the transverse angular amplitude [${\bf{n}}\cdot{\bm{\alpha}}({\bf{n}}) = 0$]. The expectation value of the field operator (\ref{E_omega_main}) yields the total classical electric field---comprising both the incident radiation and its exact scattering by the nanostructure---given by ${\bf{E}}_{\bar\omega}^{{\rm cla}}({\bf{r}}) = \int do_{\bf{n}}\,{\cal F}_{\bar\omega s}({\bf{r}}|{\bf{n}})\cdot{\bm{\alpha}}({\bf{n}})$ [Eq.~(\ref{AppC_E_cla})]. Integrating this macroscopic field along the electron trajectory directly provides the general multimode dimensionless PINEM coupling parameter [Eq.~(\ref{AppC_beta_def})]:
\begin{equation}
\beta({\bf{R}}_\perp) = \frac{e}{\hbar\bar\omega}\int_{-\infty}^{+\infty} dz\,e^{-i\frac{\bar\omega}{V}z}{\bf{u}}_V\cdot{\bf{E}}_{\bar\omega}^{{\rm cla}}({\bf{R}}_\perp + z{\bf{u}}_V).
\end{equation}
Expanding the normal-ordered functional $\Upsilon_s({\bf{R}}_1, {\bf{R}}_2)$ [Eq.~(\ref{AppC_Upsilon_factor})] into Bessel functions of the first kind $J_m$ and performing the longitudinal $\bar R_V$ integration in the monoenergetic limit ($\ell \to \infty$) strictly enforces macroscopic phase matching, yielding the exact discrete sideband distribution [Eqs.~(\ref{AppC_dP_coh_diag}) and (\ref{AppC_phi_m_def})]:
\begin{eqnarray} \label{dP_coh_main}
\frac{{d\mathscr{P}}}{{dEd^2 {\bf{P}}_\perp  }} &=& \frac{1}{{\left( {2\pi \hbar } \right)^3 V}}\sum\limits_{m =  - \infty }^{ + \infty } {\int {d^2 {\bf{R}}_{1\perp} } \int {d^2 {\bf{R}}_{2\perp} } \;e^{ - \frac{i}{\hbar }{\bf{P}}_\perp   \cdot \left( {{\bf{R}}_{1\perp}  - {\bf{R}}_{2\perp} } \right)} \widetilde \phi _m \left( {{\bf{R}}_{1\perp} } \right)\widetilde \phi _m^* \left( {{\bf{R}}_{2\perp} } \right) } \nonumber \\
&& \times \int_{ - \infty }^{ + \infty } {dZ} \;e^{ - i\frac{{E - E_0 - m\hbar \bar \omega }}{{\hbar V}}Z} e^{K\left( {{\bf{R}}_{1\perp} ,{\bf{R}}_{2\perp} ;Z} \right)} ,
\end{eqnarray}
where the dressed transverse amplitude of the $m$-th order is:
\begin{equation} \label{phi_m_main}
\widetilde \phi _m \left( {{\bf{R}}_\perp  } \right) = \phi \left( {{\bf{R}}_\perp  } \right)e^{i\Phi \left( {{\bf{R}}_\perp  } \right) - W\left( {{\bf{R}}_\perp  } \right)} J_m \left( {2\left| {\beta \left( {{\bf{R}}_\perp  } \right)} \right|} \right)e^{im\arg \left[ {\beta \left( {{\bf{R}}_\perp  } \right)} \right]} .
\end{equation}
By strictly discarding all quantum vacuum and thermal excitations---setting the conservative phase shift $\Phi({\bf{R}}_\perp) = 0$, the inelastic depletion optical depth $W({\bf{R}}_\perp) = 0$, and the multipolariton cumulant $K({\bf{R}}_{1\perp}, {\bf{R}}_{2\perp}; Z) = 0$---Eqs.~(\ref{dP_coh_main}) and (\ref{phi_m_main}) directly reduce to the conventional semiclassical PINEM theory \cite{Barwick2009,DeAbajo2010,Park2010,Feist2015,DiGiulio2019}. Under this approximation, the transverse amplitude trivially factorizes as $\widetilde\phi_m({\bf{R}}_\perp) \to \phi({\bf{R}}_\perp)J_m(2|\beta({\bf{R}}_\perp)|)e^{im\arg[\beta({\bf{R}}_\perp)]}$, while the longitudinal integral over $Z$ in Eq.~(\ref{dP_coh_main}) collapses into the Dirac delta function $2\pi\hbar V\delta(E - E_0 - m\hbar\bar\omega)$, yielding the standard discrete energy-momentum spectrum
\begin{equation} \label{dP_semiclassical_PINEM}
\left( {\frac{{d\mathscr{P}}}{{dEd^2 {\bf{P}}_\perp  }}} \right)_{{\rm semicla}} = \sum\limits_{m =  - \infty }^{ + \infty } {\delta \left( {E - E_0  - m\hbar \bar \omega } \right)\left| {\frac{1}{{2\pi \hbar }}\int {d^2 {\bf{R}}_\perp  } \;e^{ - \frac{i}{\hbar }{\bf{P}}_\perp   \cdot {\bf{R}}_\perp  } \phi \left( {{\bf{R}}_\perp  } \right)J_m \left( {2\left| {\beta \left( {{\bf{R}}_\perp  } \right)} \right|} \right)e^{im\arg \left[ {\beta \left( {{\bf{R}}_\perp  } \right)} \right]} } \right|^2 } .
\end{equation}
Comparing the exact distribution (\ref{dP_coh_main}) with its semiclassical limit (\ref{dP_semiclassical_PINEM}) reveals three key quantum phenomena:

\begin{figure}
\centering
\includegraphics[width = 1\linewidth]{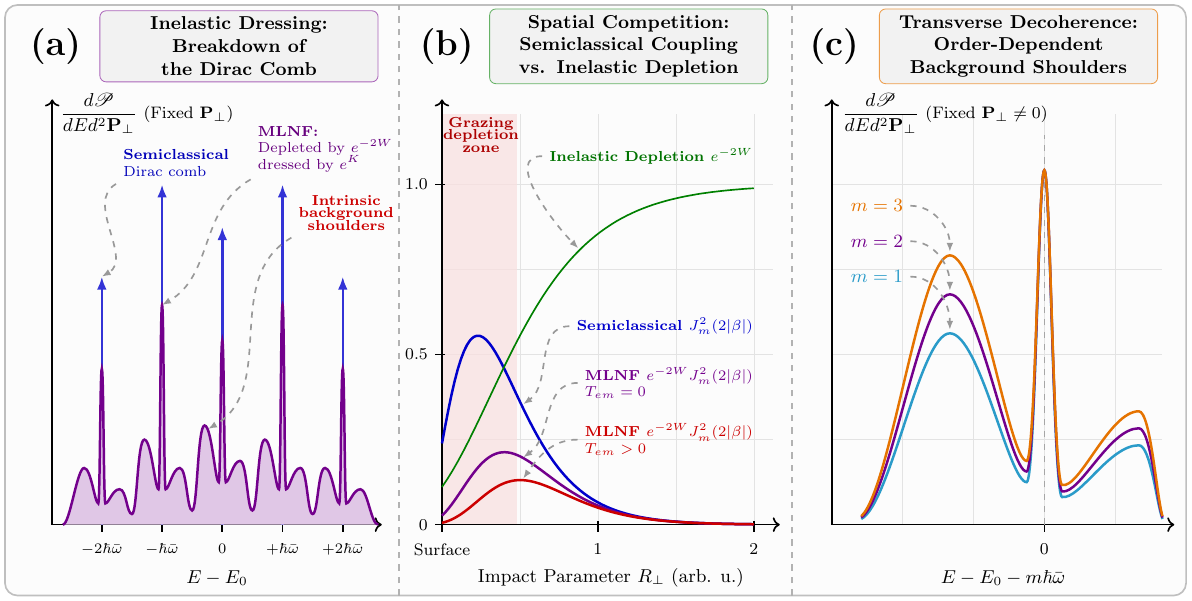}
\caption{\textbf{Quantum-electrodynamical PINEM: Inelastic dressing, spatial competition, and transverse decoherence.} (a) Energy spectrum at fixed transverse momentum. Unlike the semiclassical Dirac comb, exact MLNF sidebands [Eq.~(\ref{dP_coh_main})] are exponentially depleted by the elastic optical depth $e^{-2W}$ and dynamically dressed by the multipolariton cumulant $e^K$, yielding intrinsic background shoulders. (b) Spatial competition between semiclassical near-field coupling $J_m^2(2|\beta|)$ and inelastic depletion $e^{-2W}$. The optimal interaction distance is pushed outward from the surface, and is further suppressed at finite temperatures ($T_{em}>0$). (c) Transverse decoherence. At finite transverse momentum transfers ($\mathbf{P}_\perp \neq 0$), off-diagonal spatial sampling (${\bf{R}}_{1\perp} \neq {\bf{R}}_{2\perp}$) of the non-local cumulant $e^K$ prevents the spatial factorization of the transition probability. Consequently, the inelastic shoulders acquire an explicit dependence on the multiphoton order $m$.}
\label{Fig4}
\end{figure}

\begin{itemize}
\item {\bf Breakdown of the semiclassical Dirac comb:}
Whereas the semiclassical PINEM limit in Eq.~(\ref{dP_semiclassical_PINEM}) predicts a ladder of discrete Dirac-delta peaks, the exact distribution in Eq.~(\ref{dP_coh_main}) precludes a purely discrete spectrum. Because each $m$-photon transition is dynamically dressed by the non-local cumulant $e^K$, the sharp energy comb is intrinsically replaced by a sequence of continuous multipolariton sidebands [see Fig.~\ref{Fig4}(a)]. In the strong-coupling regime ($K \gtrsim 1$)---typically achieved for grazing trajectories, extended phase-matched interactions, or highly dissipative nanostructures at elevated temperatures---the spontaneous and thermal inelastic broadening causes adjacent sidebands to overlap. This spectral overlap inherently erases the discrete PINEM structure, establishing a fundamental quantum-electrodynamical limit on the energy resolution of coherent electron manipulation.

\item {\bf QED self-diffraction and spatial depletion of individual spectral orders:}
Even in regimes where the inelastic spectral broadening is negligible ($e^K \to 1$), the dressed amplitude $\widetilde\phi_m({\bf{R}}_\perp)$ in Eq.~(\ref{phi_m_main}) inherently retains the conservative vacuum phase shift $\Phi({\bf{R}}_\perp)$ and the inelastic optical depth $W({\bf{R}}_\perp)$. While the structured laser phase $m\arg[\beta({\bf{R}}_\perp)]$ dictates the primary transverse momentum transfer of the stimulated transition, the conservative shift $\Phi({\bf{R}}_\perp)$ superimposes a purely macroscopic QED vacuum-induced self-diffraction onto every individual spectral order. Simultaneously, $e^{-W({\bf{R}}_\perp)}$ exponentially suppresses the probability amplitude of purely coherent multiphoton exchanges to preserve unitarity.  Because the semiclassical multiphoton transition probability $J_m^2(2|\beta({\bf{R}}_\perp)|)$ and the inelastic loss rate $W({\bf{R}}_\perp)$ both peak significantly for grazing trajectories, sideband generation is governed by a fundamental spatial competition---which strictly factorizes as $e^{-2W({\bf{R}}_\perp)}J_m^2(2|\beta({\bf{R}}_\perp)|)$ for localized beams or marginal energy spectra [see Fig.~\ref{Fig4}(b)].  

\item {\bf Transverse spatial decoherence and order-dependent sideband dressing:}
Semiclassical PINEM models [Eq.~(\ref{dP_semiclassical_PINEM})] predict that upon exchanging $m$ photons, the initial transverse electron wave function $\phi({\bf{R}}_\perp)$ is simply multiplied by a local complex amplitude [Eq.~(\ref{phi_m_main})], implying that energy-filtering a specific sideband yields a pure transverse quantum state. Equation~(\ref{dP_coh_main}) generalizes this idealized limit: the resulting spatial decoherence is exclusively driven by the entanglement of the electron with the internal material polaritons, while the electron remains completely unentangled from the scattering polaritons. This fundamental asymmetry derives from the strict factorizability of the coherent-state characteristic functional $\Upsilon_s$ into independent single-coordinate amplitudes [see Eq.~(\ref{AppC_Upsilon_factor})]: because the optical field acts purely as a deterministic classical drive, it introduces no cross-path quantum correlations. Conversely, upon tracing out the electromagnetic field, the active entanglement with the material polaritons manifests as the non-local, cross-path dependence of the cumulant $e^{K({\bf{R}}_{1\perp}, {\bf{R}}_{2\perp}; Z)}$. Because this cumulant cannot factorize, the energy-filtered electron necessarily emerges in a mixed state, setting an intrinsic quantum-electrodynamical limit on the purity of the shaped wave packet. Furthermore, at finite transverse momentum transfers (${\bf{P}}_\perp \neq 0$), the non-local integration explicitly samples the off-diagonal terms (${\bf{R}}_{1\perp} \neq {\bf{R}}_{2\perp}$). Since the spatial profile of the coherent excitation $J_m$ varies with the index $m$, its spatial overlap with the non-local cumulant $e^K$ fundamentally differs across multiphoton channels. Consequently, in the momentum-resolved energy spectrum, the spontaneous and thermal inelastic shoulders acquire a strict dependence on the sideband order $m$, as illustrated in Fig.~\ref{Fig4}(c).
\end{itemize}

\subsection{Multimode squeezed vacuum illumination}

As the primary non-classical case, we consider illumination by a quasi-monochromatic multimode squeezed vacuum state centered at frequency $\bar\omega$ within a narrow spectral bandwidth $\Delta\omega \ll \bar\omega$. The incident pure state $\hat\rho_s = \left|\psi_s\right\rangle\left\langle\psi_s\right|$ is generated by the unitary squeezing operator acting on the scattering vacuum [Eq.~(\ref{AppC_sq_state})]:
\begin{equation} \label{sq_state_main}
\left| {\psi _s } \right\rangle   = \exp \left\{ {\frac{1}{2}\int_{\Delta \omega } {d\omega } \int {do_{\bf{n}} } \;\left[ {\zeta ^* \left( {\bf{n}} \right){\bf{\hat g}}_{\omega s} \left( {\bf{n}} \right) \cdot {\bf{\hat g}}_{\omega s} \left( {\bf{n}} \right) - \zeta \left( {\bf{n}} \right){\bf{\hat g}}_{\omega s}^\dag  \left( {\bf{n}} \right) \cdot {\bf{\hat g}}_{\omega s}^\dag  \left( {\bf{n}} \right)} \right]} \right\}\left| {0_s } \right\rangle ,
\end{equation}
where $\zeta({\bf{n}}) = r({\bf{n}})e^{i\theta({\bf{n}})}$ is the complex directional squeezing parameter. As detailed in Appendix~C\,3, evaluating the Gaussian characteristic functional $\Upsilon_s({\bf{R}}_1, {\bf{R}}_2)$ inside the master formula (\ref{dP_dE_dPt_main}) in the monoenergetic electron regime ($\ell \to \infty$) involves expanding $\Upsilon_s$ into modified Bessel functions of the first kind $I_n$. This strictly enforces center-of-mass phase matching, yielding the exact discrete sideband representation: 
\begin{eqnarray} \label{dP_sq_mono_main}
\frac{{d\mathscr{P}}}{{dEd^2 {\bf{P}}_\perp  }} &=& \frac{1}{{\left( {2\pi \hbar } \right)^3 V}}\sum\limits_{m =  - \infty }^{ + \infty } {\int {d^2 {\bf{R}}_{1\perp} } \int {d^2 {\bf{R}}_{2\perp} } \; e^{ - \frac{i}{\hbar }{\bf{P}}_\perp   \cdot \left( {{\bf{R}}_{1\perp}  - {\bf{R}}_{2\perp} } \right)} \widetilde \phi _{sq} \left( {{\bf{R}}_{1\perp} } \right)\widetilde \phi _{sq}^* \left( {{\bf{R}}_{2\perp} } \right)M_m \left( {{\bf{R}}_{1\perp} ,{\bf{R}}_{2\perp} } \right) } \nonumber \\
&& \times \int_{ - \infty }^{ + \infty } {dZ} \;e^{ - i\frac{{E - E_0 - m\hbar \bar \omega }}{{\hbar V}}Z} e^{K\left( {{\bf{R}}_{1\perp} ,{\bf{R}}_{2\perp} ;Z} \right)} ,
\end{eqnarray}
where the squeeze-dressed transverse amplitude becomes
\begin{equation} 
\widetilde\phi_{sq}({\bf{R}}_{i\perp}) = \phi({\bf{R}}_{i\perp})e^{i\Phi({\bf{R}}_{i\perp}) - W({\bf{R}}_{i\perp}) - \Lambda_{ii}^N}
\end{equation}
and the two-point sideband correlation weight $M_m({\bf{R}}_{1\perp}, {\bf{R}}_{2\perp})$ reads:
\begin{eqnarray} \label{Mm_sq_main}
M_m \left( {{\bf{R}}_{1\perp} ,{\bf{R}}_{2\perp} } \right) &=& \sum\limits_{k_{11}  =  - \infty }^{ + \infty } {\sum\limits_{k_{22}  =  - \infty }^{ + \infty } {\left( { - 1} \right)^{k_{11}  + k_{22} } } } I_{m - k_{11}  + k_{22} } \left( {2\left| {\Lambda _{12}^N } \right|} \right)I_{k_{11} } \left( {\left| {\Lambda _{11}^A } \right|} \right)I_{k_{22} } \left( {\left| {\Lambda _{22}^A } \right|} \right)I_{k_{11}  + k_{22} } \left( {2\left| {\Lambda _{12}^A } \right|} \right) \nonumber \\
&& \times e^{i\left\{ {\left( {m - k_{11}  + k_{22} } \right)\arg \left[ {\Lambda _{12}^N } \right] + k_{11} \arg \left[ {\Lambda _{11}^A } \right] + k_{22} \arg \left[ {\Lambda _{22}^A } \right] - \left( {k_{11}  + k_{22} } \right)\arg \left[ {\Lambda _{12}^A } \right]} \right\}} .
\end{eqnarray}
Here, the angular integration over the incident squeezed field and its scattering by the macroscopic object are entirely encapsulated by two dimensionless correlation kernels (with indices $i,j \in \{1,2\}$):
\begin{eqnarray} \label{Lambda_NA_main}
\Lambda _{ij}^N  &=& \int {do_{\bf{n}} } \;\sinh ^2 \left[ {r\left( {\bf{n}} \right)} \right]{\bm{\gamma }}\left( {\left. {{\bf{R}}_{i\perp} } \right|{\bf{n}}} \right) \cdot {\bm{\gamma }}^ *  \left( {\left. {{\bf{R}}_{j\perp} } \right|{\bf{n}}} \right), \nonumber \\
\Lambda _{ij}^A  &=& \int {do_{\bf{n}} } \;\sinh \left[ {r\left( {\bf{n}} \right)} \right]\cosh \left[ {r\left( {\bf{n}} \right)} \right]e^{ - i\theta \left( {\bf{n}} \right)} {\bm{\gamma }}\left( {\left. {{\bf{R}}_{i\perp} } \right|{\bf{n}}} \right) \cdot {\bm{\gamma }}\left( {\left. {{\bf{R}}_{j\perp} } \right|{\bf{n}}} \right),
\end{eqnarray}
which are driven by the dimensionless directional near-field coupling vector 
\begin{equation}
{\bm{\gamma}}({\bf{R}}_\perp|{\bf{n}}) = \sqrt{\Delta\omega}\,\frac{e}{\hbar\bar\omega}\int_{-\infty}^{+\infty} dz\,e^{-i\frac{\bar\omega}{V}z}{\bf{u}}_V\cdot{\cal F}_{\bar\omega s}({\bf{R}}_\perp + z{\bf{u}}_V|{\bf{n}}).
\end{equation}
Physically, the Hermitian normal kernel $\Lambda_{ij}^N$ governs the interaction driven by the mean photon-number intensity of the squeezed vacuum, strictly proportional to the mode population $\sinh^2[r({\bf{n}})]$. Conversely, the complex-symmetric anomalous kernel $\Lambda_{ij}^A$ encodes the quantum two-photon correlations, scaled by the pairing amplitude $\sinh[r({\bf{n}})]\cosh[r({\bf{n}})]$. This anomalous term directly couples the macroscopic spatial phase of the incident light $\theta({\bf{n}})$ to the transverse spatial degrees of freedom of the electron.

\begin{figure}
\centering
\includegraphics[width = 1\linewidth]{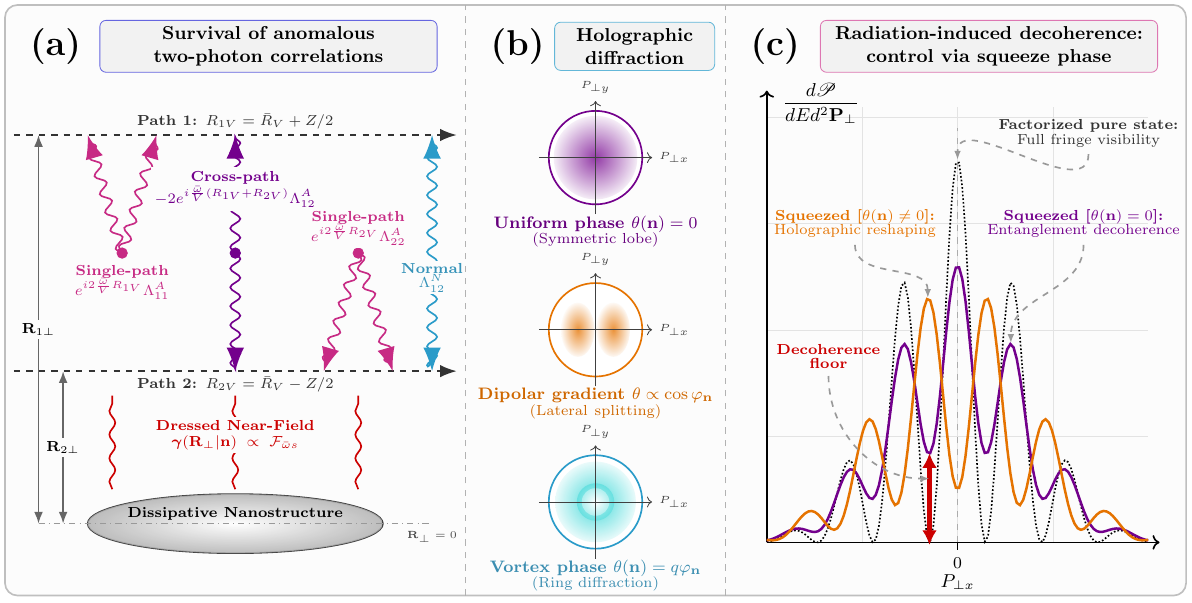}
\caption{\textbf{Transverse electron diffraction and spatial decoherence under multimode squeezed vacuum illumination.} 
(a) Two-path electron interaction with the nanostructure-dressed squeezed vacuum. A strict macroscopic selection rule phase-locks the anomalous cross-path ($\Lambda_{12}^A$) and single-path ($\Lambda_{11}^A$, $\Lambda_{22}^A$) two-photon exchanges, allowing their survival and intrinsic mixing with normal intensity-driven processes ($\Lambda_{12}^N$). (b) Quantum holographic reshaping of the 2D diffraction profile. Structuring the incident squeeze phase $\theta(\mathbf{n})$ directly modulates the anomalous correlation kernels, reshaping the momentum distribution (shown for uniform, dipolar, and vortex phases). (c) 1D diffraction cross-sections demonstrating radiation-induced spatial decoherence and its active phase control. In contrast to an idealized factorized pure state exhibiting full fringe visibility (dotted line), the squeezed vacuum interaction manifests an intrinsic decoherence floor (loss of perfect destructive interference) that can be actively manipulated via the squeeze phase.}
\label{Fig5}
\end{figure}

Equation~(\ref{dP_sq_mono_main}) reveals three distinctive quantum-optical features of squeezed-vacuum electron-matter interactions:

\begin{itemize}
\item {\bf Survival of anomalous two-photon correlations and generation of sidebands:}
Because the photon pairing encoded in the anomalous kernel $\Lambda_{ij}^A$ lacks the optical phase cancellation inherent to intensity-driven processes, these interactions introduce macroscopic spatial oscillations at frequency $2\bar\omega/V$. Naively, these oscillations would average to zero over a macroscopic electron wave train, seemingly eliminating the effect of $\Lambda_{ij}^A$. However, the master formula [Eq.~(\ref{dP_dE_dPt_main})] evaluates the interaction along two independent spatial paths, parameterized by the relative separation $Z$ and center-of-mass coordinate $\bar R_V$ via $\mathbf{R}_{1,2} = \mathbf{R}_{1,2\perp} + (\bar R_V \pm Z/2)\mathbf{u}_V$. As shown in Appendix~C\,3, expanding $\Upsilon_s(\mathbf{R}_1, \mathbf{R}_2)$ into modified Bessel functions yields a multi-harmonic series where each term carries the longitudinal phase factor $\exp [ i(n_{12} + k_{11} - k_{22})\frac{\bar \omega}{V}Z ] \exp [ i2(k_{11} + k_{22} + k_{12})\frac{\bar \omega}{V}\bar R_V ]$. Here, $n_{12}$ counts the normal cross-path exchanges, while $k_{11}$, $k_{22}$, and $k_{12}$ denote correlated photon pairs exchanged via the anomalous single- and cross-path channels. Crucially, the macroscopic integration over $\bar R_V$ acts as a strict selection rule, enforcing the phase-locking condition $k_{12} = -(k_{11} + k_{22})$ [see the two-path schematic in Fig.~\ref{Fig5}(a)]. Physically, this dictates that the independent exchange of correlated photon pairs along the single paths ($\mathbf{R}_1$ and $\mathbf{R}_2$) dynamically balances the pairs shared across both paths. This synchronous interference exactly cancels the macroscopic phase mismatch, preventing the associated modified Bessel functions in Eq.~(\ref{Mm_sq_main}) from reducing to unity. As a result, the surviving anomalous correlations intrinsically mix with the normal processes to define the net sideband order $m = n_{12} + k_{11} - k_{22}$. This fundamental mixing mechanism ensures that squeezed light populates a complete frequency comb of both even and odd sidebands, contrary to the strictly even-order transitions expected from isolated photon-pair exchanges.

\item {\bf Quantum holographic reshaping of transverse diffraction:}
The functional dependence on the squeezing phase $\theta({\bf{n}})$ establishes a fundamental mechanism to coherently control the transverse spatial degrees of freedom of the electron. Probing finite transverse momentum transfers (${\bf{P}}_\perp \neq 0$) intrinsically samples the off-diagonal spatial separations (${\bf{R}}_{1\perp} \neq {\bf{R}}_{2\perp}$) of the paraxial electron wave packet. In this regime, the sideband correlation weight $M_m({\bf{R}}_{1\perp}, {\bf{R}}_{2\perp})$ in Eq.~(\ref{Mm_sq_main}) depends critically on the macroscopic phase mismatches between the cross-path anomalous kernel ($\Lambda_{12}^A$) and the single-path ones ($\Lambda_{jj}^A$). Physically, these off-diagonal terms govern the spatial interference of correlated photon pairs exchanged across distinct trajectories of the electron. Because the anomalous kernel explicitly maps the macroscopic optical phase $\theta({\bf{n}})$ onto these biphoton exchanges, tailoring the angular profile of the incident squeezed light dynamically reshapes the spatial phase landscape of $M_m({\bf{R}}_{1\perp}, {\bf{R}}_{2\perp})$. This dependence provides a purely quantum-optical holographic knob: rather than relying on deterministic classical field gradients, it utilizes the programmable spatial distribution of quantum vacuum correlations to reshape the electron transverse momentum diffraction---as visualized for various angularly structured phase profiles in Fig.~\ref{Fig5}(b).

\item {\bf Radiation-induced electron transverse decoherence and its active manipulation by squeeze phase:}
Under exact quantum interaction with a coherent field, the sole source of transverse decoherence is the background cumulant $e^K$, because the characteristic functional $\Upsilon_s({\bf{R}}_1, {\bf{R}}_2)$ governing the stimulated multiphoton exchange rigorously factorizes into independent single-coordinate amplitudes, yielding the term $\widetilde\phi_m({\bf{R}}_{1\perp})\widetilde\phi_m^*({\bf{R}}_{2\perp})$ in Eq.~(\ref{dP_coh_main}). Conversely, because the squeezed vacuum inherently comprises quantum-correlated photon pairs, the multiphoton exchange actively entangles the electron with the scattering polaritons. Upon tracing out this photonic environment, this entanglement manifests as the strictly non-separable sideband correlation weight $M_m({\bf{R}}_{1\perp}, {\bf{R}}_{2\perp})$ in Eq.~(\ref{Mm_sq_main}), making the illumination itself a fundamental source of transverse decoherence. This mathematical non-separability occurs because the cross-path correlation kernels, $\Lambda_{12}^N$ and $\Lambda_{12}^A$, are nonlinearly embedded within the modified Bessel functions. As a direct consequence, the explicit dependence of the anomalous kernel $\Lambda_{12}^A$ on the squeeze phase $\theta({\bf{n}})$ enables the active manipulation of this radiation-induced spatial decoherence by tuning $\theta({\bf{n}})$, as illustrated in Fig.~\ref{Fig5}(c).
\end{itemize}

\subsection{The lossless single-mode QPINEM limit}

In the existing literature, quantum PINEM (QPINEM) encompasses phenomena extending well beyond asymptotic energy-exchange spectroscopy, including attosecond wave-packet shaping \cite{DiGiulio2020Optica,Dahan2021}, electron-photon entanglement \cite{Kfir2019,Kfir2021}, and the active engineering of photonic states \cite{BenHayun2021}. Because these applications frequently rely on joint electron-photon observables, postselected projections, or subsequent free-space propagation rather than the asymptotic energy-momentum distribution $\frac{d\mathscr{P}}{dE d^2{\bf{P}}_\perp}$, the most direct route to connect our formalism with standard QPINEM models is through the asymptotic time-evolution operator $\hat U_I(\infty)$. Indeed, conventional QPINEM descriptions \cite{Kfir2019,DiGiulio2019,DiGiulio2020Optica,BenHayun2021,Dahan2021,DiGiulioThesis} are ultimately governed by a common single-mode scattering operator $\hat S$.

Within the MLNF framework, as derived in Appendix~B, the exact nonperturbative post-interaction evolution operator of the coupled electron-field-matter system is given by
\begin{equation} \label{UI_exact_QPINEM_sec}
\hat U_I \left( \infty  \right) = \int {d^3 {\bf{R}}} \;\left| {\bf{R}} \right\rangle \left\langle {\bf{R}} \right|e^{i\Phi \left( {{\bf{R}}_\perp  } \right)} e^{\hat K_s \left( {\bf{R}} \right)} e^{\hat K_{em} \left( {\bf{R}} \right)} ,
\end{equation}
where $\left|{\bf{R}}\right\rangle$ are the electron position eigenstates, $\Phi({\bf{R}}_\perp)$ is the conservative vacuum phase shift (\ref{Phi_Rt_main}), and $\hat K_s({\bf{R}})$ and $\hat K_{em}({\bf{R}})$ are the commuting anti-Hermitian generators for the asymptotic scattering ($s$) and internal material ($em$) polariton sectors, explicitly given by
\begin{eqnarray} \label{Ks_Kem_main_text}
\hat K_s \left( {\bf{R}} \right) &=& \int {d\omega } \;\frac{e}{{\hbar \omega }}\int {do_{\bf{n}} } \;\int_{ - \infty }^{ + \infty } {dz} \; {\bf{u}}_V  \cdot  \left[ {e^{ - i\frac{\omega }{V}\left( {R_V  - z} \right)} {\cal F}_{\omega s}^* \left( {\left. {{\bf{R}}_\perp  + z{\bf{u}}_V } \right|{\bf{n}}} \right) \cdot {\bf{\hat g}}_{\omega s}^\dag  \left( {\bf{n}} \right)} \right. \nonumber \\
&& \quad \quad \quad \quad \quad \quad \quad \quad \quad \quad \;\; \left. { - e^{i\frac{\omega }{V}\left( {R_V  - z} \right)} {\cal F}_{\omega s} \left( {\left. {{\bf{R}}_\perp  + z{\bf{u}}_V } \right|{\bf{n}}} \right) \cdot {\bf{\hat g}}_{\omega s} \left( {\bf{n}} \right)} \right], \nonumber \\
\hat K_{em} \left( {\bf{R}} \right) &=& \int {d\omega } \;\frac{e}{{\hbar \omega }}\int_\mathscr{V} {d^3 {\bf{r}}\,} \sum\limits_\nu  {\int_{ - \infty }^{ + \infty } {dz} \;{\bf{u}}_V  \cdot \left[ {e^{ - i\frac{\omega }{V}\left( {R_V  - z} \right)} {\cal G}_{\omega \nu }^* \left( {\left. {{\bf{R}}_\perp  + z{\bf{u}}_V } \right|{\bf{r}}} \right) \cdot {\bf{\hat f}}_{\omega \nu }^\dag  \left( {\bf{r}} \right)} \right.} \nonumber \\
&& \quad \quad \quad \quad \quad \quad \quad \quad \quad \quad \quad \quad \; \left. { - e^{i\frac{\omega }{V}\left( {R_V  - z} \right)} {\cal G}_{\omega \nu } \left( {\left. {{\bf{R}}_\perp  + z{\bf{u}}_V } \right|{\bf{r}}} \right) \cdot {\bf{\hat f}}_{\omega \nu } \left( {\bf{r}} \right)} \right].
\end{eqnarray}
Standard QPINEM models are recovered from the exact operator (\ref{UI_exact_QPINEM_sec}) by introducing three sequential physical approximations:

\begin{enumerate}[(i)]
\item \emph{Neglect of material absorption and thermal fluctuations.} The nanostructure is idealized as perfectly lossless, setting ${\rm Im}[\varepsilon_\omega({\bf{r}})] = {\rm Im}[-1/\mu_\omega({\bf{r}})] = 0$ throughout $\mathscr{V}$. Consequently, the material dyadic kernels vanish identically (${\cal G}_{\omega\nu}({\bf{r}}|{\bf{r}}') \equiv 0$), which eliminates the material generator, $\hat K_{em}({\bf{R}}) \equiv 0$, and reduces the material displacement operator to the identity, $e^{\hat K_{em}({\bf{R}})} = \hat I_{em}$.

\item \emph{Single-mode truncation of the radiative continuum.} The external electromagnetic field is assumed to interact with the electron exclusively through a single quasi-monochromatic optical mode of central frequency $\bar\omega$ (within a narrow spectral bandwidth $\Delta\omega \ll \bar\omega$) characterized by a normalized transverse angular profile ${\bf{u}}_0({\bf{n}})$ [satisfying ${\bf{n}}\cdot{\bf{u}}_0({\bf{n}}) = 0$ and $\int do_{\bf{n}}\,|{\bf{u}}_0({\bf{n}})|^2 = 1$]. Introducing the corresponding discrete bosonic annihilation operator
\begin{equation} \label{a_single_mode_def}
\hat a \equiv \frac{1}{{\sqrt {\Delta \omega } }}\int_{\Delta \omega } {d\omega } \int {do_{\bf{n}} } \;{\bf{u}}_0^* \left( {\bf{n}} \right) \cdot {\bf{\hat g}}_{\omega s} \left( {\bf{n}} \right),
\end{equation}
which obeys the canonical commutation relation $[\hat a, \hat a^\dag] = 1$, and discarding the coupling to all orthogonal spectral and angular scattering modes via the single-mode projection ${\bf{\hat g}}_{\omega s}({\bf{n}}) \to \frac{1}{\sqrt{\Delta\omega}}{\bf{u}}_0({\bf{n}})\hat a$ for $\omega \in \Delta\omega$, the continuum generator $\hat K_s({\bf{R}})$ in the first of Eqs.~(\ref{Ks_Kem_main_text}) collapses into
\begin{equation} \label{Ks_single_mode_red}
\hat K_s \left( {\bf{R}} \right) \to g_0^* \left( {{\bf{R}}_\perp  } \right)e^{ - i\frac{{\bar \omega }}{V}R_V } \hat a^\dag   - g_0 \left( {{\bf{R}}_\perp  } \right)e^{i\frac{{\bar \omega }}{V}R_V } \hat a.
\end{equation}
Here, $g_0({\bf{R}}_\perp)$ is the dimensionless single-photon near-field coupling amplitude at transverse impact parameter ${\bf{R}}_\perp$:
\begin{equation} \label{g0_def}
g_0 \left( {{\bf{R}}_\perp  } \right) \equiv \sqrt {\Delta \omega } \,\frac{e}{{\hbar \bar \omega }}\int_{ - \infty }^{ + \infty } {dz} \;e^{ - i\frac{{\bar \omega }}{V}z} {\bf{u}}_V  \cdot \left[ {\int {do_{\bf{n}} } \;{\cal F}_{\bar \omega s} \left( {\left. {{\bf{R}}_\perp   + z{\bf{u}}_V } \right|{\bf{n}}} \right) \cdot {\bf{u}}_0 \left( {\bf{n}} \right)} \right],
\end{equation}
which corresponds to the single-photon coupling parameter $\beta_0$ (or $-g_{{\rm Qu}}^*$) in the QPINEM literature \cite{Kfir2019,DiGiulio2019,DiGiulio2020Optica,BenHayun2021}.

\item \emph{Parametric transverse treatment and electron energy-ladder operators.} Factorizing the three-dimensional electron coordinate projector into transverse and longitudinal components, $\int d^3{\bf{R}}\,\left|{\bf{R}}\right\rangle\left\langle{\bf{R}}\right| = \int d^2{\bf{R}}_\perp\,\left|{\bf{R}}_\perp\right\rangle\left\langle{\bf{R}}_\perp\right| \otimes \int_{-\infty}^{+\infty} dR_V\,\left|R_V\right\rangle\left\langle R_V\right|$, the longitudinal spatial phase factors in Eq.~(\ref{Ks_single_mode_red}) directly define the unitary electron energy-ladder operators \cite{Kfir2019,BenHayun2021,DiGiulioThesis}
\begin{align} \label{b_ladder_def}
\hat b &\equiv \int_{ - \infty }^{ + \infty } {dR_V } \;\left| {R_V } \right\rangle \left\langle {R_V } \right|e^{ - i\frac{{\bar \omega }}{V}R_V }  = e^{ - i\frac{{\bar \omega }}{V}\hat R_V } , &
\hat b^\dag   &= e^{i\frac{{\bar \omega }}{V}\hat R_V } .
\end{align}
Because $\hat b$ and $\hat b^\dag$ depend solely on the longitudinal position operator $\hat R_V = {\bf{u}}_V\cdot{\bf{\hat R}}$, they mutually commute ($[\hat b, \hat b^\dag] = 0$) and act on the energy eigenstates (\ref{AppA_E_Pt_basis}) by shifting the longitudinal electron energy by discrete photon quanta, $\hat b\left|E, {\bf{P}}_\perp\right\rangle = \left|E - \hbar\bar\omega, {\bf{P}}_\perp\right\rangle$ and $\hat b^\dag\left|E, {\bf{P}}_\perp\right\rangle = \left|E + \hbar\bar\omega, {\bf{P}}_\perp\right\rangle$.
\end{enumerate}
Substituting Eqs.~(\ref{Ks_single_mode_red}) and (\ref{b_ladder_def}) into Eq.~(\ref{UI_exact_QPINEM_sec}), the reduced evolution operator reads $\hat U_I(\infty) \to \int d^2{\bf{R}}_\perp\,\left|{\bf{R}}_\perp\right\rangle\left\langle{\bf{R}}_\perp\right|\hat S({\bf{R}}_\perp)$, where the operator acting on the longitudinal electron and single-mode photon subspaces at a given impact parameter ${\bf{R}}_\perp$ is precisely the canonical QPINEM scattering operator \cite{Kfir2019,DiGiulio2019,DiGiulio2020Optica,BenHayun2021,DiGiulioThesis}:
\begin{equation} \label{S_QPINEM_final}
\hat S\left( {{\bf{R}}_\perp  } \right) = e^{i\Phi \left( {{\bf{R}}_\perp  } \right)} \exp \left[ {g_0^* \left( {{\bf{R}}_\perp  } \right)\hat b\hat a^\dag   - g_0 \left( {{\bf{R}}_\perp  } \right)\hat b^\dag  \hat a} \right].
\end{equation}
In standard 1D QPINEM treatments, the electron beam is assumed to be tightly focused at a fixed impact parameter ${\bf{R}}_\perp$, so that the conservative phase $e^{i\Phi({\bf{R}}_\perp)}$ reduces to a global phase factor $e^{i\chi}$ that drops out of longitudinal and optical observables \cite{DiGiulio2019,DiGiulio2020Optica,BenHayun2021}. 

Equation~(\ref{S_QPINEM_final}) coincides with the canonical scattering operator underlying standard QPINEM and discrete-cavity models \cite{Kfir2019,DiGiulio2019,DiGiulio2020Optica,BenHayun2021,DiGiulioThesis}. Consequently, established results---such as the Laguerre-polynomial transition amplitudes for Fock-state illumination and their semiclassical Bessel-function limit---follow directly from this operator. Notably, even within this strict single-mode approximation, the MLNF provides a first-principles derivation for the single-photon coupling amplitude $g_0$ [Eq.~(\ref{g0_def})]. Whereas standard models often introduce this parameter phenomenologically or assume idealized closed cavities, our framework rigorously determines $g_0$ through the exact classical scattering kernel of the macroscopic object.

\section{Conclusions}

In this paper, we have developed a nonperturbative quantum electrodynamical theory of the energy and transverse-momentum exchange between a fast aloof electron, an arbitrary finite dispersive and dissipative magnetodielectric nanostructure of volume $\mathscr{V}$ held at temperature $T_{em}$ in vacuum, and an arbitrary incident quantum state of light $\hat{\rho}_s$. To rigorously capture this complete interplay, we employ the modified Langevin noise formalism (MLNF), a formulation of macroscopic quantum electrodynamics that explicitly accounts for the quantum nature of the external radiation field associated with scattering phenomena, rather than solely the fields generated inside the medium. A theoretical treatment of this scope is challenging within standard macroscopic quantum electrodynamics—or Langevin noise formalism (LNF)—since this framework structurally omits scattering modes. While these modes can formally be recovered via singular vacuum limiting procedures, adopting the MLNF completely bypasses this mathematical complication, enabling the native inclusion of arbitrary incident radiation. Here, describing the electron kinematics within the paraxial nonrecoil approximation, the exact second-order truncation of the Magnus expansion yields a closed-form master formula for the transverse-momentum-resolved energy exchange distribution $\frac{d\mathscr{P}}{dE d^2{\bf{P}}_\perp}$ [Eq.~(\ref{dP_dE_dPt_main})]. This formulation captures physical phenomena stemming directly from the unrestricted quantum description of the scattering field, which remain inaccessible to standard macroscopic quantum electrodynamics applied without limiting procedures.

In the unilluminated regime, the absence of externally prepared photons (scattering polaritons) does not render the scattering sector inactive; rather, its zero-point vacuum fluctuations remain a fundamental dynamical driver. The explicit inclusion of this cold scattering vacuum alongside the internal material polaritons at temperature $T_{em}$ reveals precisely where and how these zero-point fields manifest in the electron dynamics, yielding the following key results:
\begin{itemize}
\item Recombination of scattering and material vacuum fluctuations into ${\rm Im}[{\cal G}_\omega]$ at $T_{em} = 0$---rigorously recovering previous LNF-based expressions for both the elastic depletion optical depth $W$ and the single-polariton Stokes loss distribution (by showing how the identical zero-point weights of the two sectors allow their exact synthesis into the full Green's tensor)---and the breakdown of this recombination at $T_{em} > 0$; 
\item Rigorous lossless dielectric limit (${\rm Im}[\varepsilon_\omega] = {\rm Im}[\mu_\omega] = 0$) for radiative EELS, Cherenkov, and transition radiation without singular limiting procedures;
\item Qualitatively distinct transverse diffraction patterns for Stokes energy loss and Anti-Stokes thermal energy gain at $T_{em} > 0$;
\item Generalized non-equilibrium detailed-balance relation and systematic radiative reduction of the apparent temperature in nanoscale EELS thermometry.
\end{itemize}

In the quantum-light illumination regime, the MLNF framework leverages the normal-ordered characteristic functional $\Upsilon_s$ to describe the external radiation field in full quantum-optical generality. Accommodating both arbitrary nonclassical incident light and the exact macroscopic response of dissipative nanostructures, this approach resolves the dichotomy between classical-field models and idealized lossless cavities. Consequently, it demonstrates that light-stimulated transitions are inherently dressed in energy and transverse momentum by the spontaneous and thermal inelastic background. The explicit evaluation of the closed-form master formula uncovers specific quantum-electrodynamical features under three paradigmatic classes of externally prepared radiation:

\begin{itemize}
\item \emph{Ambient thermal radiation at temperature $T_s$:} Non-equilibrium two-temperature ($T_s, T_{em}$) energy-momentum exchange; near-field-mediated thermal energy gain from ambient blackbody fields; generalized detailed-balance relation governing the radiative enhancement or reduction of the apparent temperature in nanoscale electron thermometry;
\item \emph{Multimode coherent illumination:} Breakdown of the semiclassical PINEM Dirac comb into continuous multipolariton sidebands; QED self-diffraction and spatial inelastic depletion of individual spectral orders; transverse spatial decoherence resulting from electron entanglement strictly with material polaritons, alongside order-dependent sideband dressing;
\item \emph{Multimode squeezed vacuum illumination:} Phase-locked survival of anomalous two-photon correlations populating both even and odd sidebands; quantum holographic reshaping of transverse electron diffraction; radiation-induced transverse spatial decoherence driven by the inherent entanglement between the electron and the scattering polaritons, actively controlled by the squeeze phase.
\end{itemize}
Furthermore, by applying sequential lossless, single-mode, and parametric transverse approximations to our exact asymptotic evolution operator $\hat U_I(\infty)$, we have rigorously recovered the canonical single-mode scattering operator $\hat S$ underlying standard quantum PINEM (QPINEM) models, which describe coherent electron wave-packet shaping, electron-photon entanglement, and quantum photonic state engineering in idealized lossless cavities. Crucially, beyond recovering this established limit, our framework provides a first-principles electrodynamic expression for the single-photon coupling amplitude $g_0$, explicitly determining it in terms of the exact geometry and electromagnetic response of the nanostructure through its classical scattering kernel.

While this work has focused on the asymptotic energy and transverse-momentum exchange of the fast electron, our derivation of the exact, nonperturbative electron-radiation evolution operator paves the way for exploring more highly articulated phenomena. In the electron sector, this complete theoretical framework provides the necessary first-principles tools to investigate advanced wave-packet shaping, complex spatial decoherence, and electron-photon entanglement, all underpinned by a fully quantum description of both the free electron and the macroscopic electromagnetic field. As a natural development of the present theory, accessing the corresponding electromagnetic observables will enable a rigorous description of light-emission processes---such as stimulated cathodoluminescence, non-equilibrium thermal luminescence, and far-field quantum photon statistics---generated by fast aloof electrons interacting with arbitrary dissipative nanostructures.

\appendix

\section{MLNF quantization framework and fast-electron quantum kinematics}

In this Appendix, we summarize the constitutive equations of the modified Langevin noise formalism (MLNF) for a bounded magnetodielectric object in vacuum \cite{Ciatt1,CiattDirect} together with the kinematic representation of the fast electron in the paraxial nonrecoil regime.

Let $\mathscr{V}$ denote the finite spatial region occupied by an inhomogeneous, isotropic, dispersive, and absorptive magnetodielectric nanostructure surrounded by vacuum. In the frequency domain ($e^{-i\omega t}$ with $\omega > 0$), the global relative dielectric permittivity $\varepsilon_\omega({\bf{r}})$ and magnetic permeability $\mu_\omega({\bf{r}})$ are defined piecewise as
\begin{align} \label{AppA_eps_mu}
\varepsilon _\omega  \left( {\bf{r}} \right) &= \left\{ {\begin{array}{*{20}l}
   {\varepsilon _\omega ^\mathscr{V} \left( {\bf{r}} \right),} & {{\bf{r}} \in \mathscr{V},}  \\
   {1,} & {{\bf{r}} \notin \mathscr{V},}  \\
\end{array}} \right. &
\mu _\omega  \left( {\bf{r}} \right) &= \left\{ {\begin{array}{*{20}l}
   {\mu _\omega ^\mathscr{V} \left( {\bf{r}} \right),} & {{\bf{r}} \in \mathscr{V},}  \\
   {1,} & {{\bf{r}} \notin \mathscr{V},}  \\
\end{array}} \right.
\end{align}
where $\varepsilon_\omega^\mathscr{V}({\bf{r}})$ and $\mu_\omega^\mathscr{V}({\bf{r}})$ are holomorphic functions in the upper complex frequency half-plane as required by causality. In the MLNF, the field-matter system is exactly diagonalized into a scattering ($s$) sector and a material ($em$) sector. Consequently, the total Hamiltonian $\hat H_{sem} = \hat H_s + \hat H_{em}$ is strictly decoupled into the free-energy operators for the corresponding bosonic polaritons: 
\begin{align} \label{AppA_H_sectors}
\hat H_s &= \int {d\omega } \;\hbar \omega \int {do_{\bf{n}} } \;{\bf{\hat g}}_{\omega s}^\dag  \left( {\bf{n}} \right) \cdot {\bf{\hat g}}_{\omega s} \left( {\bf{n}} \right), &
\hat H_{em} &= \int  {d\omega } \;\hbar \omega \int_{\mathscr{V}} {d^3 {\bf{r}}\,} \sum_\nu  {{\bf{\hat f}}_{\omega \nu }^\dag  \left( {\bf{r}} \right) \cdot {\bf{\hat f}}_{\omega \nu } \left( {\bf{r}} \right)} .
\end{align}
The scattering polariton operators ${\bf{\hat g}}_{\omega s}({\bf{n}})$ are parameterized by the unit propagation vector ${\bf{n}} = \sin\theta_{\bf{n}}(\cos\varphi_{\bf{n}}{\bf{u}}_x + \sin\varphi_{\bf{n}}{\bf{u}}_y) + \cos\theta_{\bf{n}}{\bf{u}}_z$ and satisfy the transversality constraint ${\bf{n}} \cdot {\bf{\hat g}}_{\omega s}({\bf{n}}) = 0$. Conversely, the electric ($e$) and magnetic ($m$) material polariton operators ${\bf{\hat f}}_{\omega\nu}({\bf{r}})$ ($\nu = e,m$) are defined strictly inside the object volume (${\bf{r}} \in \mathscr{V}$) and vanish identically for ${\bf{r}} \notin \mathscr{V}$. They satisfy the bosonic commutation relations
\begin{align} \label{AppA_boson_comm}
\left[ {{\bf{\hat g}}_{\omega s} \left( {\bf{n}} \right),{\bf{\hat g}}_{\omega 's}^\dag  \left( {{\bf{n}}'} \right)} \right] &= \delta \left( {\omega  - \omega '} \right)\delta \left( {o_{\bf{n}}  - o_{{\bf{n}}'} } \right){\cal I}_{\bf{n}} , &
\left[ {{\bf{\hat f}}_{\omega \nu } \left( {\bf{r}} \right),{\bf{\hat f}}_{\omega '\nu '}^\dag  \left( {{\bf{r}}'} \right)} \right] &= \delta \left( {\omega  - \omega '} \right)\delta _{\nu \nu '} \delta \left( {{\bf{r}} - {\bf{r}}'} \right){\cal I},
\end{align}
with all other commutators vanishing, where $\delta(o_{\bf{n}} - o_{{\bf{n}}'}) = \delta(\theta_{\bf{n}} - \theta_{{\bf{n}}'})\delta(\varphi_{\bf{n}} - \varphi_{{\bf{n}}'})/\sin\theta_{\bf{n}}$ is the solid-angle Dirac delta function, $do_{\bf{n}} = \sin\theta_{\bf{n}}d\theta_{\bf{n}}d\varphi_{\bf{n}}$, ${\cal I}$ is the identity dyadic, and ${\cal I}_{\bf{n}} = {\cal I} - {\bf{n}}{\bf{n}}$ is the projector onto the plane orthogonal to ${\bf{n}}$. The electric field operator ${\bf{\hat E}}({\bf{r}}) = \int d\omega\,{\bf{\hat E}}_\omega({\bf{r}}) + {\rm H.c.}$ is fully determined by its positive-frequency spectral component ${\bf{\hat E}}_\omega({\bf{r}})$, which is explicitly expanded in terms of both sets of polariton operators as
\begin{equation} \label{AppA_E_omega_recalled}
{\bf{\hat E}}_\omega  \left( {\bf{r}} \right) = \int {do_{\bf{n}} } \;{\cal F}_{\omega s} \left( {\left. {\bf{r}} \right|{\bf{n}}} \right) \cdot {\bf{\hat g}}_{\omega s} \left( {\bf{n}} \right) + \int_\mathscr{V} {d^3 {\bf{r}}'\,} \sum_\nu {{\cal G}_{\omega \nu } \left( {\left. {\bf{r}} \right|{\bf{r}}'} \right) \cdot {\bf{\hat f}}_{\omega \nu } \left( {{\bf{r}}'} \right)}.
\end{equation}
The scattering dyadic kernel ${\cal F}_{\omega s}({\bf{r}}|{\bf{n}})$ and the electric and magnetic material dyadic kernels ${\cal G}_{\omega\nu}({\bf{r}}|{\bf{r}}')$ ($\nu = e,m$) appearing in Eq.~(\ref{AppA_E_omega_recalled}) are given by
\begin{equation} \label{AppA_Kernels}
{\cal F}_{\omega s} \left( {\left. {\bf{r}} \right|{\bf{n}}} \right) = {\cal F}_\omega  \left( {\left. {\bf{r}} \right|{\bf{n}}} \right)\sqrt {\frac{{\hbar k_\omega ^3 }}{{16\pi ^3 \varepsilon _0 }}} , \quad \quad \begin{array}{*{20}l}
   {{\cal G}_{\omega e} \left( {\left. {\bf{r}} \right|{\bf{r}}'} \right) = {\cal G}_\omega  \left( {\left. {\bf{r}} \right|{\bf{r}}'} \right)i\sqrt {\displaystyle\frac{{\hbar k_\omega ^4 }}{{\pi \varepsilon _0 }}{\mathop{\rm Im}\nolimits} \left[ {\varepsilon _\omega  \left( {{\bf{r}}'} \right)} \right]} ,}  \\
   {{\cal G}_{\omega m} \left( {\left. {\bf{r}} \right|{\bf{r}}'} \right) = {\cal G}_\omega  \left( {\left. {\bf{r}} \right|{\bf{r}}'} \right)\displaystyle\frac{{ \times \overleftarrow{\nabla }_{{\bf{r}}'} }}{{ik_\omega  }}\sqrt {\frac{{\hbar k_\omega ^4 }}{{\pi \varepsilon _0 }}{\mathop{\rm Im}\nolimits} \left[ {\frac{{ - 1}}{{\mu _\omega  \left( {{\bf{r}}'} \right)}}} \right]} ,}  \\
\end{array}
\end{equation}
where $k_\omega = \omega/c$ and the right curl operator acts as $({\cal Q} \times \overleftarrow{\nabla}_{{\bf{r}}'})_{ij} = \varepsilon_{jpk}\partial'_k Q_{ip}$. In Eq.~(\ref{AppA_Kernels}), the classical modal dyadic ${\cal F}_\omega({\bf{r}}|{\bf{n}})$ and the classical dyadic Green's function ${\cal G}_\omega({\bf{r}}|{\bf{r}}')$ are uniquely defined by the Helmholtz boundary-value problems and asymptotic radiation conditions at $r \to +\infty$ (with ${\bf{r}} = r{\bf{m}}$ and unit radial vector ${\bf{m}}$):
\begin{equation} \label{AppA_Helmholtz_BVP}
\begin{aligned}[t]
& \left[ {\nabla _{\bf{r}}  \times \frac{1}{{\mu _\omega  \left( {\bf{r}} \right)}}\nabla _{\bf{r}}  \times  - k_\omega ^2 \varepsilon _\omega  \left( {\bf{r}} \right)} \right]{\cal F}_\omega  \left( {\left. {\bf{r}} \right|{\bf{n}}} \right) = 0, \\
& {\cal F}_\omega  \left( {\left. {r{\bf{m}}} \right|{\bf{n}}} \right) \mathop  \approx \limits_{r \to  + \infty } e^{ik_\omega  {\bf{n}} \cdot \left( {r{\bf{m}}} \right)} {\cal I}_{\bf{n}}  + \frac{{e^{ik_\omega  r} }}{r}{\cal S}_\omega  \left( {\left. {\bf{m}} \right|{\bf{n}}} \right),
\end{aligned}
\quad \quad \quad
\begin{aligned}[t]
& \left[ {\nabla _{\bf{r}}  \times \frac{1}{{\mu _\omega  \left( {\bf{r}} \right)}}\nabla _{\bf{r}}  \times  - k_\omega ^2 \varepsilon _\omega  \left( {\bf{r}} \right)} \right]{\cal G}_\omega  \left( {\left. {\bf{r}} \right|{\bf{r}}'} \right) = \delta \left( {{\bf{r}} - {\bf{r}}'} \right){\cal I}, \\
& {\cal G}_\omega  \left( {\left. {r{\bf{m}}} \right|{\bf{r}}'} \right) \mathop  \approx \limits_{r \to  + \infty } \frac{{e^{ik_\omega  r} }}{r}{\cal W}_\omega  \left( {\left. {\bf{m}} \right|{\bf{r}}'} \right),
\end{aligned}
\end{equation}
where ${\cal S}_\omega({\bf{m}}|{\bf{n}})$ is the classical scattering dyadic \cite{Kristensson2016} (satisfying ${\bf{m}}\cdot{\cal S}_\omega({\bf{m}}|{\bf{n}}) = 0$ and ${\cal S}_\omega({\bf{m}}|{\bf{n}})\cdot{\bf{n}} = 0$, which entails ${\cal F}_\omega({\bf{r}}|{\bf{n}})\cdot{\bf{n}} = 0$) and ${\cal W}_\omega({\bf{m}}|{\bf{r}}')$ is the far-field Green's function amplitude (satisfying ${\bf{m}}\cdot{\cal W}_\omega({\bf{m}}|{\bf{r}}') = 0$). Physically, for any polarization unit vector ${\bf{e}}_{{\bf{n}}\lambda}$ orthogonal to ${\bf{n}}$, the vector field ${\bf{F}}_{\omega{\bf{n}}\lambda}({\bf{r}}) = {\cal F}_\omega({\bf{r}}|{\bf{n}})\cdot{\bf{e}}_{{\bf{n}}\lambda}$ represents the complete classical electric field (incident plane wave plus scattered near and far fields) generated when the nanostructure is illuminated by an asymptotic unit-amplitude plane wave of frequency $\omega$, wave vector $k_\omega{\bf{n}}$, and polarization ${\bf{e}}_{{\bf{n}}\lambda}$. By virtue of Lorentz reciprocity, the classical dyadic Green's function inherently satisfies spatial symmetry, ${\cal G}_\omega^T({\bf{r}}|{\bf{r}}') = {\cal G}_\omega({\bf{r}}'|{\bf{r}})$. Furthermore, as demonstrated in Ref.~\cite{Ciatt1}, its far-field amplitude is rigorously linked to the transposed modal dyadic through the relation
\begin{equation} \label{AppA_W_F_relation}
{\cal W}_\omega({\bf{m}}|{\bf{r}}) = \frac{1}{4\pi}{\cal F}_\omega^T({\bf{r}}| - {\bf{m}}).
\end{equation}
Applying the divergence theorem to the inhomogeneous Helmholtz equation for ${\cal G}_\omega({\bf{r}}|{\bf{r}}')$ over the entire space and evaluating the resulting surface term over the sphere at infinity $S_\infty$ via the far-field relation (\ref{AppA_W_F_relation}) yields the fundamental integral identity (\ref{Fund_Id_Comm_main}) (see Ref.~\cite{CiattDirect}):
\begin{equation} \label{AppA_Fund_Id}
\int {do_{\bf{n}} } \;{\cal F}_{\omega s} \left( {\left. {\bf{r}} \right|{\bf{n}}} \right) \cdot {\cal F}_{\omega s}^{T*} \left( {\left. {{\bf{r}}'} \right|{\bf{n}}} \right) + \int_\mathscr{V} {d^3 {\bf{s}}\,} \sum_\nu  {{\cal G}_{\omega \nu } \left( {\left. {\bf{r}} \right|{\bf{s}}} \right) \cdot {\cal G}_{\omega \nu }^{T*} \left( {\left. {{\bf{r}}'} \right|{\bf{s}}} \right)}  = \frac{{\hbar k_\omega ^2 }}{{\pi \varepsilon _0 }}{\mathop{\rm Im}\nolimits} \left[ {{\cal G}_\omega  \left( {\left. {\bf{r}} \right|{\bf{r}}'} \right)} \right],
\end{equation}
where the spatial integral of the material kernels is restricted to $\mathscr{V}$ because ${\rm Im}[\varepsilon_\omega({\bf{s}})] = {\rm Im}[-1/\mu_\omega({\bf{s}})] = 0$ for ${\bf{s}} \notin \mathscr{V}$. Combining Eqs.~(\ref{AppA_E_omega_recalled}), (\ref{AppA_boson_comm}), and (\ref{AppA_Fund_Id}) directly yields the spectral electric field commutators
\begin{align} \label{AppA_E_comm}
\left[ {{\bf{\hat E}}_\omega  \left( {\bf{r}} \right),{\bf{\hat E}}_{\omega '}^\dag  \left( {{\bf{r}}'} \right)} \right] &= \delta \left( {\omega  - \omega '} \right)\frac{{\hbar k_\omega ^2 }}{{\pi \varepsilon _0 }}{\mathop{\rm Im}\nolimits} \left[ {{\cal G}_\omega  \left( {\left. {\bf{r}} \right|{\bf{r}}'} \right)} \right], &
\left[ {{\bf{\hat E}}_\omega  \left( {\bf{r}} \right),{\bf{\hat E}}_{\omega '} \left( {{\bf{r}}'} \right)} \right] &= 0.
\end{align}
We emphasize that the inclusion of all three polariton sectors---asymptotic scattering, internal electric, and internal magnetic---is strictly necessary to exactly reconstruct ${\mathop{\rm Im}\nolimits} [{\cal G}_\omega]$ for a bounded object in vacuum. Furthermore, while the first of  Eqs.~(\ref{AppA_E_comm}) is a universal operator identity independent of the system state, evaluating its expectation value over the global vacuum state correctly reproduces the standard zero-temperature Fluctuation-Dissipation Theorem.

In the initial state (\ref{Initial_State_main}), the material polaritons of the nanostructure are in the canonical thermal Gibbs state at temperature $T_{em}$,
\begin{equation} \label{AppA_rho_em}
\hat \rho _{em}  = \frac{{e^{ - \hat H_{em} /\left( {k_B T_{em} } \right)} }}{{{\rm Tr}_{em} \left[ {e^{ - \hat H_{em} /\left( {k_B T_{em} } \right)} } \right]}},
\end{equation}
whose one- and two-point thermal expectation values read
\begin{eqnarray} \label{AppA_thermal_averages}
{\rm Tr}_{em} \left[ {\hat \rho _{em} {\bf{\hat f}}_{\omega \nu } \left( {\bf{r}} \right)} \right] &=& {\rm Tr}_{em} \left[ {\hat \rho _{em} {\bf{\hat f}}_{\omega \nu }^\dag  \left( {\bf{r}} \right)} \right] = 0, \nonumber \\
{\rm Tr}_{em} \left[ {\hat \rho _{em} {\bf{\hat f}}_{\omega \nu }^\dag  \left( {\bf{r}} \right){\bf{\hat f}}_{\omega '\nu '} \left( {{\bf{r}}'} \right)} \right] &=& n_\omega  \left( {T_{em} } \right)\delta \left( {\omega  - \omega '} \right)\delta _{\nu \nu '} \delta \left( {{\bf{r}} - {\bf{r}}'} \right){\cal I}, \nonumber \\
{\rm Tr}_{em} \left[ {\hat \rho _{em} {\bf{\hat f}}_{\omega \nu } \left( {\bf{r}} \right){\bf{\hat f}}_{\omega '\nu '}^\dag  \left( {{\bf{r}}'} \right)} \right] &=& \left[ {n_\omega  \left( {T_{em} } \right) + 1} \right]\delta \left( {\omega  - \omega '} \right)\delta _{\nu \nu '} \delta \left( {{\bf{r}} - {\bf{r}}'} \right){\cal I}, \nonumber \\
{\rm Tr}_{em} \left[ {\hat \rho _{em} {\bf{\hat f}}_{\omega \nu } \left( {\bf{r}} \right){\bf{\hat f}}_{\omega '\nu '} \left( {{\bf{r}}'} \right)} \right] &=& {\rm Tr}_{em} \left[ {\hat \rho _{em} {\bf{\hat f}}_{\omega \nu }^\dag  \left( {\bf{r}} \right){\bf{\hat f}}_{\omega '\nu '}^\dag  \left( {{\bf{r}}'} \right)} \right] = 0,
\end{eqnarray}
where $n_\omega(T) = \left[e^{\hbar\omega/(k_B T)} - 1\right]^{-1}$ is the Bose-Einstein thermal distribution.

Regarding the fast electron of mass $m$ and charge $-e < 0$, the paraxial nonrecoil approximation relies on the assumption that its momentum distribution remains tightly localized around the central relativistic momentum ${\bf{P}}_0$ before, during, and after the interaction, as typical in transmission electron microscopy where the exchanged photon energies and momenta are several orders of magnitude smaller than the central kinetic energy and momentum \cite{DiGiulio2019,DiGiulio2020,Rivera2020,BenHayun2021,CiattPRB2023}. Expanding the relativistic minimal-coupling Hamiltonian around ${\bf{P}}_0$, defining the central relativistic velocity
\begin{equation} \label{AppA_V_rel}
{\bf{V}} = \frac{{c{\bf{P}}_0 }}{{\sqrt {m^2 c^2  + P_0^2 } }} = V{\bf{u}}_V ,
\end{equation}
neglecting spin-flip and ponderomotive ${\bf{\hat A}}^2$ terms, and omitting an inconsequential constant energy offset yields the linear free-electron Hamiltonian $\hat H_{el} = {\bf{V}}\cdot{\bf{\hat P}}$ and the minimal-coupling interaction Hamiltonian $\hat H_{{\mathop{\rm int}}} = e{\bf{V}}\cdot{\bf{\hat A}}({\bf{\hat R}})$ \cite{DiGiulio2019,DiGiulio2020,Rivera2020,BenHayun2021,CiattPRB2023}. Expressing the Coulomb-gauge vector potential operator via its spectral decomposition ${\bf{\hat A}}({\bf{r}}) = \int d\omega\,\frac{1}{i\omega}{\bf{\hat E}}_\omega^\perp({\bf{r}}) + {\rm H.c.}$ \cite{Ciatt1,CiattDirect} and noting that for an aloof electron trajectory (which does not intersect $\mathscr{V}$) the longitudinal electric field operator vanishes along the electron path, one directly obtains the interaction Hamiltonian $\hat H_{{\mathop{\rm int}}}$ in the third of Eqs.~(\ref{H_MLNF_main}). 

To construct the kinematic state space of the free electron, we introduce the simultaneous eigenstates of the paraxial Hamiltonian $\hat H_{el} = {\bf{V}} \cdot {\bf{\hat P}}$ and the transverse momentum operator ${\bf{\hat P}}_\perp = {\bf{\hat P}} - ({\bf{u}}_V \cdot {\bf{\hat P}}){\bf{u}}_V$. Noting that $\hat H_{el}$ depends exclusively on the longitudinal momentum operator $\hat P_V = {\bf{u}}_V \cdot {\bf{\hat P}}$ through the relation $\hat H_{el} = V \hat P_V$, we can conveniently change variables from the standard three-dimensional momentum basis $\left|{\bf{P}}\right\rangle$ to the energy-transverse momentum basis. Defining the energy eigenvalue $E = V P_V$, the rescaled states
\begin{equation}
\left|E, {\bf{P}}_\perp\right\rangle \equiv \frac{1}{\sqrt{V}}\left|{\bf{P}}\right\rangle_{{\bf{P}} = \frac{E}{V}{\bf{u}}_V + {\bf{P}}_\perp}
\end{equation}
fulfill the eigenvalue equations $\hat H_{el}\left|E, {\bf{P}}_\perp\right\rangle = E\left|E, {\bf{P}}_\perp\right\rangle$ and ${\bf{\hat P}}_\perp\left|E, {\bf{P}}_\perp\right\rangle = {\bf{P}}_\perp\left|E, {\bf{P}}_\perp\right\rangle$. Their orthonormality, completeness, and coordinate representation follow directly from standard plane-wave properties:
\begin{eqnarray} \label{AppA_E_Pt_basis}
\left\langle {{E,{\bf{P}}_\perp }}\mathrel{\left | {\vphantom {{E,{\bf{P}}_\perp } {E',{\bf{P}}'_\perp }}}\right. \kern-\nulldelimiterspace}{{E',{\bf{P}}'_\perp }} \right\rangle &=& \delta \left( {E - E'} \right)\delta \left( {{\bf{P}}_\perp  - {\bf{P}}'_\perp } \right), \nonumber \\
\int_{ - \infty }^{ + \infty } {dE} \int {d^2 {\bf{P}}_\perp } \;\left| {E,{\bf{P}}_\perp } \right\rangle \left\langle {E,{\bf{P}}_\perp } \right| &=& \hat I_e , \nonumber \\
\left\langle {{\bf{R}}}\mathrel{\left | {\vphantom {{\bf{R}} {E,{\bf{P}}_\perp }}}\right. \kern-\nulldelimiterspace}{{E,{\bf{P}}_\perp }} \right\rangle &=& \frac{1}{{\sqrt {\left( {2\pi \hbar } \right)^3 V} }}e^{i\frac{E}{{\hbar V}}R_V } e^{\frac{i}{\hbar }{\bf{P}}_\perp  \cdot {\bf{R}}_\perp } ,
\end{eqnarray}
where $\hat I_e$ is the electron identity operator while $R_V = {\bf{u}}_V \cdot {\bf{R}}$ and ${\bf{R}}_\perp = {\bf{R}} - R_V{\bf{u}}_V$ denote the longitudinal and transverse components of the position basis eigenvalues.

The initial electron state at $t \to -\infty$ is the normalized pure wave packet 
\begin{equation}
\left|\psi\right\rangle = \int_{-\infty}^{+\infty} dE \int d^2{\bf{P}}_\perp\,\chi^{(\Delta E)}(E - E_0)\bar\phi({\bf{P}}_\perp)\left|E, {\bf{P}}_\perp\right\rangle,
\end{equation}
where $E_0 = V P_0$ is the central energy, and $\int d^2{\bf{P}}_\perp\,|\bar\phi({\bf{P}}_\perp)|^2 = 1$. The longitudinal spectral profile $\chi^{(\eta)}(\zeta)$ (with energy spread parameter $\eta = \Delta E$) is defined explicitly as
\begin{equation} \label{AppA_chi_def}
\chi ^{\left( \eta  \right)} \left( \zeta  \right) = \frac{1}{{\sqrt {\pi \eta } }}\frac{{\sin \left( {\zeta /\eta } \right)}}{{\zeta /\eta }}.
\end{equation}
Crucially, the square of this spectral amplitude defines a regularized Dirac delta sequence, which is well-known to converge to the exact Dirac delta function $\delta(\zeta)$ in the distributional sense in the monoenergetic limit ($\eta \to 0^+$):
\begin{equation} \label{AppA_chi_sq_limit}
\left[ {\chi ^{\left( \eta  \right)} \left( \zeta  \right)} \right]^2  = \delta ^{\left( \eta  \right)} \left( \zeta  \right) \mathop  \to \limits_{\eta  \to 0^ +} \delta \left( \zeta  \right).
\end{equation}
In the coordinate representation, the initial electron wave function 
\begin{equation}
\psi({\bf{R}}) = \left\langle {\bf{R}}|\psi \right\rangle = \varphi(R_V)\phi({\bf{R}}_\perp)
\end{equation}
factorizes into the transverse spatial profile $\phi({\bf{R}}_\perp)$ and the longitudinal wave train $\varphi(R_V)$ of spatial coherence length $\ell \equiv \frac{2\hbar V}{\Delta E}$:
\begin{align} \label{AppA_psi_coord}
\phi \left( {{\bf{R}}_\perp } \right) &= \frac{1}{{2\pi \hbar }}\int {d^2 {\bf{P}}_\perp } \;e^{\frac{i}{\hbar }{\bf{P}}_\perp  \cdot {\bf{R}}_\perp } \bar \phi \left( {{\bf{P}}_\perp } \right), \nonumber \\
\varphi \left( {R_V } \right) &= \frac{1}{{\sqrt {2\pi \hbar V} }}\int_{ - \infty }^{ + \infty } {dE} \;e^{i\frac{E}{{\hbar V}}R_V } \chi ^{\left( {\Delta E} \right)} \left( {E - E_0 } \right)  = H \left( {\frac{\ell }{2} - \left| {R_V } \right|} \right)\frac{1}{{\sqrt \ell  }}e^{i\frac{{E_0 }}{{\hbar V}}R_V } ,
\end{align}
where $H(x)$ is the Heaviside step function. Note that taking the limit $\ell \to \infty$ (i.e., $\Delta E = \frac{2\hbar V}{\ell} \to 0^+$) corresponds to the monochromatic (monoenergetic) electron limit, in which the longitudinal wave packet approaches a plane wave of well-defined initial energy $E_0$. Finally, the coordinate matrix elements of the initial electron density operator $\hat\rho_{el} = \left|\psi\right\rangle\left\langle\psi\right|$ are given by
\begin{equation} \label{AppA_rho_el_matrix}
\rho _{el} \left( {{\bf{R}}_1 ,{\bf{R}}_2 } \right) = \left\langle {{\bf{R}}_1 }\mathrel{\left | {\vphantom {{\bf{R}}_1  \psi }}\right. \kern-\nulldelimiterspace}{\psi } \right\rangle \left\langle {\psi }\mathrel{\left | {\vphantom {\psi  {{\bf{R}}_2 }}}\right. \kern-\nulldelimiterspace}{{{\bf{R}}_2 }} \right\rangle  = \frac{1}{\ell }H \left( {\frac{\ell }{2} - \left| {R_{1V} } \right|} \right) H \left( {\frac{\ell }{2} - \left| {R_{2V} } \right|} \right)e^{i\frac{{E_0 }}{{\hbar V}}\left( {R_{1V}  - R_{2V} } \right)} \phi \left( {{\bf{R}}_{1\perp} } \right)\phi ^* \left( {{\bf{R}}_{2\perp} } \right).
\end{equation}

\section{Transverse-momentum-resolved energy exchange probability}

In this Appendix, we provide the complete mathematical derivation of the exact transverse-momentum-resolved energy exchange probability $\frac{d\mathscr{P}}{dE d^2{\bf{P}}_\perp}$ presented in the main text. The derivation is structured into four sequential steps:
\begin{enumerate}[(1)]
\item In Sec.~B\,1, we solve the interaction-picture dynamics of the coupled electron-field-matter system. Exploiting the exact truncation of the Magnus expansion at second order, we derive the time-evolution operator in terms of a radiation-reaction phase and multimodal Weyl displacement operators.
\item In Sec.~B\,2, we evaluate this time-evolution operator in the asymptotic post-interaction limit ($t \to +\infty$). By mapping the temporal integrations onto longitudinal spatial coordinates, we establish the exact factorized form of the total density operator $\hat{\rho}_I(\infty)$ prior to the environmental trace.
\item In Sec.~B\,3, we trace out the macroscopic electromagnetic field degrees of freedom to obtain the asymptotic electron reduced density operator $\hat{\rho}_{elI}(\infty)$. This step relies on a normal-ordered evaluation for the scattering modes and a thermal Wick expansion for the internal material modes, culminating in the exact recombination of the macroscopic exponents to demonstrate the fundamental zero-point energy-gain cancellation.
\item Finally, in Sec.~B\,4, we project the resulting reduced density operator onto the continuous kinematic electron basis $\left|E, {\bf{P}}_\perp\right\rangle$ to derive the final master formula for the joint probability distribution $\frac{d\mathscr{P}}{dE d^2{\bf{P}}_\perp}$.
\end{enumerate}

\subsection{Interaction-picture evolution operator and exact Magnus expansion}

In this section, we derive the exact time-evolution operator of the coupled electron-field-matter system in the interaction picture by exploiting the bosonic algebra of the MLNF polariton operators (\ref{AppA_boson_comm}) and the fundamental integral relation  (\ref{AppA_Fund_Id}). 

Moving to the interaction picture with respect to the unperturbed Hamiltonian $\hat H_0 = \hat H_{el} + \hat H_{sem}$ via the unitary operator $\hat U_0(t) = e^{-\frac{i}{\hbar}\hat H_0 t} = e^{-\frac{i}{\hbar}{\bf{V}}\cdot{\bf{\hat P}}t}e^{-\frac{i}{\hbar}\hat H_{sem}t}$, the canonical electron position operator transforms as a uniform rectilinear translation, $e^{\frac{i}{\hbar }{\bf{V}} \cdot {\bf{\hat P}}t} {\bf{\hat R}}\,e^{ - \frac{i}{\hbar }{\bf{V}} \cdot {\bf{\hat P}}t}  = {\bf{\hat R}} + {\bf{V}}t$, whereas the polariton annihilation operators evolve harmonically as $e^{\frac{i}{\hbar}\hat H_{sem}t}{\bf{\hat g}}_{\omega s}({\bf{n}})e^{-\frac{i}{\hbar}\hat H_{sem}t} = e^{-i\omega t}{\bf{\hat g}}_{\omega s}({\bf{n}})$ and $e^{\frac{i}{\hbar}\hat H_{sem}t}{\bf{\hat f}}_{\omega\nu}({\bf{r}})e^{-\frac{i}{\hbar}\hat H_{sem}t} = e^{-i\omega t}{\bf{\hat f}}_{\omega\nu}({\bf{r}})$, yielding $e^{\frac{i}{\hbar}\hat H_{sem}t}{\bf{\hat E}}_\omega({\bf{r}})e^{-\frac{i}{\hbar}\hat H_{sem}t} = e^{-i\omega t}{\bf{\hat E}}_\omega({\bf{r}})$. Consequently, the interaction Hamiltonian $\hat H_{{\mathop{\rm int}} I}(t) = \hat U_0^\dag(t)\hat H_{{\mathop{\rm int}}}\hat U_0(t)$ takes the form
\begin{equation} \label{AppB_Hint_I}
\hat H_{{\mathop{\rm int}} I} \left( t \right) = \int {d\omega } \;\frac{{eV}}{{i\omega }}{\bf{u}}_V  \cdot \left[ {e^{ - i\omega t} {\bf{\hat E}}_\omega  \left( {{\bf{\hat R}} + {\bf{V}}t} \right) - e^{i\omega t} {\bf{\hat E}}_\omega ^\dag  \left( {{\bf{\hat R}} + {\bf{V}}t} \right)} \right].
\end{equation}
Because $\hat H_{{\mathop{\rm int}} I}(t)$ depends on the electron position operator ${\bf{\hat R}}$ but not on the conjugate momentum ${\bf{\hat P}}$, it is diagonal in the electron coordinate basis $\left|{\bf{R}}\right\rangle$ (${\bf{\hat R}}\left|{\bf{R}}\right\rangle = {\bf{R}}\left|{\bf{R}}\right\rangle$) and can be represented as
\begin{equation} \label{AppB_Hint_proj}
\hat H_{{\mathop{\rm int}} I} \left( t \right) = i\hbar \int {d^3 {\bf{R}}} \;\left| {\bf{R}} \right\rangle \left\langle {\bf{R}} \right|\hat O_{sem} \left( {{\bf{R}},t} \right),
\end{equation}
where we have introduced the field-matter generator parameterized by the classical electron coordinate ${\bf{R}}$:
\begin{equation} \label{AppB_Osem_def}
\hat O_{sem} \left( {{\bf{R}},t} \right) = \int {d\omega } \;\frac{{eV}}{{\hbar \omega }}{\bf{u}}_V  \cdot \left[ {e^{i\omega t} {\bf{\hat E}}_\omega ^\dag  \left( {{\bf{R}} + {\bf{V}}t} \right) - e^{ - i\omega t} {\bf{\hat E}}_\omega  \left( {{\bf{R}} + {\bf{V}}t} \right)} \right],
\end{equation}
which is evidently anti-Hermitian, i.e. $ \hat O_{sem}^\dag  \left( {{\bf{R}},t} \right) =  - \hat O_{sem} \left( {{\bf{R}},t} \right)$. Substituting the MLNF field expansion (\ref{AppA_E_omega_recalled}) into Eq.~(\ref{AppB_Osem_def}), the generator $\hat O_{sem}({\bf{R}},t)$ splits into the sum of two commuting anti-Hermitian operators acting on the scattering ($s$) and material ($em$) polariton sectors, respectively: 
\begin{equation} \label{AppB_Osem_split}
\hat O_{sem} \left( {{\bf{R}},t} \right) = \hat O_s \left( {{\bf{R}},t} \right) + \hat O_{em} \left( {{\bf{R}},t} \right),
\end{equation}
where
\begin{eqnarray} \label{AppB_Os_Oem_explicit}
\hat O_s \left( {{\bf{R}},t} \right) &=& \int {d\omega } \;\frac{{eV}}{{\hbar \omega }}{\bf{u}}_V  \cdot \int {do_{\bf{n}} } \left[ {e^{i\omega t} {\cal F}_{\omega s}^* \left( {\left. {{\bf{R}} + {\bf{V}}t} \right|{\bf{n}}} \right) \cdot {\bf{\hat g}}_{\omega s}^\dag  \left( {\bf{n}} \right) - e^{ - i\omega t} {\cal F}_{\omega s} \left( {\left. {{\bf{R}} + {\bf{V}}t} \right|{\bf{n}}} \right) \cdot {\bf{\hat g}}_{\omega s} \left( {\bf{n}} \right)} \right], \nonumber \\
\hat O_{em} \left( {{\bf{R}},t} \right) &=& \int {d\omega } \;\frac{{eV}}{{\hbar \omega }}{\bf{u}}_V  \cdot \int_\mathscr{V} {d^3 {\bf{r}}\,} \sum_\nu  {\left[ {e^{i\omega t} {\cal G}_{\omega \nu }^* \left( {\left. {{\bf{R}} + {\bf{V}}t} \right|{\bf{r}}} \right) \cdot {\bf{\hat f}}_{\omega \nu }^\dag  \left( {\bf{r}} \right) - e^{ - i\omega t} {\cal G}_{\omega \nu } \left( {\left. {{\bf{R}} + {\bf{V}}t} \right|{\bf{r}}} \right) \cdot {\bf{\hat f}}_{\omega \nu } \left( {\bf{r}} \right)} \right]}. 
\end{eqnarray}

We now evaluate the two-position, two-time commutator $[\hat O_{sem}({\bf{R}},t), \hat O_{sem}({\bf{R}}',t')]$. Using Eq.~(\ref{AppB_Osem_def}) together with the second of Eqs.~(\ref{AppA_E_comm}), we obtain
\begin{eqnarray} \label{AppB_Ocomm_1}
\left[ {\hat O_{sem} \left( {{\bf{R}},t} \right),\hat O_{sem} \left( {{\bf{R}}',t'} \right)} \right] &=& \int  {d\omega } \int  {d\omega '} \;\frac{{e^2 V^2 }}{{\hbar ^2 \omega \omega '}}{\bf{u}}_V  \cdot \bigg\{  - e^{i\left( {\omega t - \omega 't'} \right)} \left[ {{\bf{\hat E}}_\omega ^\dag  \left( {{\bf{R}} + {\bf{V}}t} \right),{\bf{\hat E}}_{\omega '} \left( {{\bf{R}}' + {\bf{V}}t'} \right)} \right] \nonumber \\
&& - e^{ - i\left( {\omega t - \omega 't'} \right)} \left[ {{\bf{\hat E}}_\omega  \left( {{\bf{R}} + {\bf{V}}t} \right),{\bf{\hat E}}_{\omega '}^\dag  \left( {{\bf{R}}' + {\bf{V}}t'} \right)} \right] \bigg\} \cdot {\bf{u}}_V .
\end{eqnarray}
Recalling that for any two vector operators $[{\bf{\hat A}}^\dag, {\bf{\hat B}}] = -[{\bf{\hat B}}, {\bf{\hat A}}^\dag]^T$ and invoking the first of Eqs.~(\ref{AppA_E_comm}) along with Lorentz reciprocity ${\cal G}_\omega^T({\bf{r}}'|{\bf{r}}) = {\cal G}_\omega({\bf{r}}|{\bf{r}}')$, the first commutator on the right-hand side of Eq.~(\ref{AppB_Ocomm_1}) evaluates to
\begin{equation} \label{AppB_E_Edag_comm_T}
\left[ {{\bf{\hat E}}_\omega ^\dag  \left( {{\bf{R}} + {\bf{V}}t} \right),{\bf{\hat E}}_{\omega '} \left( {{\bf{R}}' + {\bf{V}}t'} \right)} \right] =  - \delta \left( {\omega  - \omega '} \right)\frac{{\hbar k_\omega ^2 }}{{\pi \varepsilon _0 }}{\mathop{\rm Im}\nolimits} \left[ {{\cal G}_\omega  \left( {\left. {{\bf{R}} + {\bf{V}}t} \right|{\bf{R}}' + {\bf{V}}t'} \right)} \right].
\end{equation}
Inserting Eq.~(\ref{AppB_E_Edag_comm_T}) and the first of Eqs.~(\ref{AppA_E_comm}) into Eq.~(\ref{AppB_Ocomm_1}), integrating over $\omega'$, and using $k_\omega = \omega/c$, we find
\begin{eqnarray} \label{AppB_Ocomm_final}
\left[ {\hat O_{sem} \left( {{\bf{R}},t} \right),\hat O_{sem} \left( {{\bf{R}}',t'} \right)} \right] &=& \left[ {\hat O_s \left( {{\bf{R}},t} \right),\hat O_s \left( {{\bf{R}}',t'} \right)} \right] + \left[ {\hat O_{em} \left( {{\bf{R}},t} \right),\hat O_{em} \left( {{\bf{R}}',t'} \right)} \right] \nonumber \\
&=& \frac{{2ie^2 V^2 }}{{\pi \varepsilon _0 \hbar c^2 }}\int {d\omega } \;\sin \left[ {\omega \left( {t - t'} \right)} \right]{\bf{u}}_V  \cdot {\mathop{\rm Im}\nolimits} \left[ {{\cal G}_\omega  \left( {\left. {{\bf{R}} + {\bf{V}}t} \right|{\bf{R}}' + {\bf{V}}t'} \right)} \right] \cdot {\bf{u}}_V .
\end{eqnarray}
Crucially, Eq.~(\ref{AppB_Ocomm_final}) shows that the two-time commutator $[\hat O_{sem}({\bf{R}},t), \hat O_{sem}({\bf{R}}',t')]$ is a purely imaginary $c$-number, which implies the identical vanishing of all third- and higher-order nested commutators:
\begin{equation} \label{AppB_nested_zero}
\left[ {\hat O_{sem} \left( {{\bf{R}}'',t''} \right),\left[ {\hat O_{sem} \left( {{\bf{R}},t} \right),\hat O_{sem} \left( {{\bf{R}}',t'} \right)} \right]} \right] = 0.
\end{equation}

The interaction-picture time-evolution operator $\hat U_I(t) = {\cal T}\exp\left[-\frac{i}{\hbar}\int_{-\infty}^t d\tau\,\hat H_{{\mathop{\rm int}} I}(\tau)\right]$ (where ${\cal T}$ denotes chronological time ordering) inherits the diagonal structure of Eq.~(\ref{AppB_Hint_proj}) in the electron coordinate basis:
\begin{equation} \label{AppB_UI_proj}
\hat U_I \left( t \right) = \int {d^3 {\bf{R}}} \;\left| {\bf{R}} \right\rangle \left\langle {\bf{R}} \right|\hat U_{sem} \left( {{\bf{R}},t} \right), \quad \quad \hat U_{sem} \left( {{\bf{R}},t} \right) = {\cal T}\exp \left[ {\int_{ - \infty }^t {d\tau } \;\hat O_{sem} \left( {{\bf{R}},\tau } \right)} \right].
\end{equation}
By virtue of Eq.~(\ref{AppB_nested_zero}), the Magnus expansion \cite{Magnus1954} for the time-ordered exponential $\hat U_{sem}({\bf{R}},t)$ truncates exactly at second order, yielding the closed-form unitary operator
\begin{align} \label{AppB_Usem_Magnus}
\hat U_{sem} \left( {{\bf{R}},t} \right) &= \exp \left\{ {\int_{ - \infty }^t {d\tau } \;\hat O_{sem} \left( {{\bf{R}},\tau } \right) + \frac{1}{2}\int_{ - \infty }^t {d\tau } \int_{ - \infty }^\tau  {d\tau '} \;\left[ {\hat O_{sem} \left( {{\bf{R}},\tau } \right),\hat O_{sem} \left( {{\bf{R}},\tau '} \right)} \right]} \right\} \nonumber \\
&= e^{i\Phi \left( {{\bf{R}},t} \right)} e^{\hat K_{sem} \left( {{\bf{R}},t} \right)} .
\end{align}
In Eq.~(\ref{AppB_Usem_Magnus}), $\Phi({\bf{R}},t)$ is the real-valued electromagnetic self-interaction (radiation-reaction) phase,
\begin{eqnarray} \label{AppB_Phi_def}
\Phi \left( {{\bf{R}},t} \right) &\equiv& \frac{1}{{2i}}\int_{ - \infty }^t {d\tau } \int_{ - \infty }^\tau  {d\tau '} \;\left[ {\hat O_{sem} \left( {{\bf{R}},\tau } \right),\hat O_{sem} \left( {{\bf{R}},\tau '} \right)} \right] \nonumber \\
&=& \frac{{e^2 V^2 }}{{\pi \varepsilon _0 \hbar c^2 }}\int_{ - \infty }^t {d\tau } \int_{ - \infty }^\tau  {d\tau '} \int {d\omega } \;\sin \left[ {\omega \left( {\tau  - \tau '} \right)} \right]{\bf{u}}_V  \cdot {\mathop{\rm Im}\nolimits} \left[ {{\cal G}_\omega  \left( {\left. {{\bf{R}} + {\bf{V}}\tau } \right|{\bf{R}} + {\bf{V}}\tau '} \right)} \right] \cdot {\bf{u}}_V ,
\end{eqnarray}
while $\hat K_{sem}({\bf{R}},t) \equiv \int_{-\infty}^t d\tau\,\hat O_{sem}({\bf{R}},\tau)$ is the anti-Hermitian multipolariton generator. Owing to the commutativity between $\hat O_s$ and $\hat O_{em}$ in Eq.~(\ref{AppB_Osem_split}), this generator strictly separates into scattering and material contributions:
\begin{equation} \label{AppB_Ksem_factor}
\hat K_{sem} \left( {{\bf{R}},t} \right) = \hat K_s \left( {{\bf{R}},t} \right) + \hat K_{em} \left( {{\bf{R}},t} \right),
\end{equation}
where their explicit integral forms read
\begin{eqnarray} \label{AppB_Ks_Kem_Weyl}
\hat K_s \left( {{\bf{R}},t} \right) &=& \int {d\omega } \;\frac{{eV}}{{\hbar \omega }}{\bf{u}}_V  \cdot \int_{ - \infty }^t {d\tau } \int {do_{\bf{n}} } \left[ {e^{i\omega \tau } {\cal F}_{\omega s}^* \left( {\left. {{\bf{R}} + {\bf{V}}\tau } \right|{\bf{n}}} \right) \cdot {\bf{\hat g}}_{\omega s}^\dag  \left( {\bf{n}} \right) - e^{ - i\omega \tau } {\cal F}_{\omega s} \left( {\left. {{\bf{R}} + {\bf{V}}\tau } \right|{\bf{n}}} \right) \cdot {\bf{\hat g}}_{\omega s} \left( {\bf{n}} \right)} \right], \nonumber \\
\hat K_{em} \left( {{\bf{R}},t} \right) &=& \int {d\omega } \;\frac{{eV}}{{\hbar \omega }}{\bf{u}}_V  \cdot \int_{ - \infty }^t {d\tau } \int_\mathscr{V} {d^3 {\bf{r}}\,} \sum_\nu  {\left[ {e^{i\omega \tau } {\cal G}_{\omega \nu }^* \left( {\left. {{\bf{R}} + {\bf{V}}\tau } \right|{\bf{r}}} \right) \cdot {\bf{\hat f}}_{\omega \nu }^\dag  \left( {\bf{r}} \right) - e^{ - i\omega \tau } {\cal G}_{\omega \nu } \left( {\left. {{\bf{R}} + {\bf{V}}\tau } \right|{\bf{r}}} \right) \cdot {\bf{\hat f}}_{\omega \nu } \left( {\bf{r}} \right)} \right]} . \nonumber \\
\end{eqnarray}
To summarize, the complete interaction-picture time-evolution operator reads
\begin{equation} \label{AppB_UI_master_summary}
\hat U_I \left( t \right) = \int {d^3 {\bf{R}}} \;\left| {\bf{R}} \right\rangle \left\langle {\bf{R}} \right|e^{i\Phi \left( {{\bf{R}},t} \right)} e^{\hat K_s \left( {{\bf{R}},t} \right)} e^{\hat K_{em} \left( {{\bf{R}},t} \right)},
\end{equation}
where the radiation-reaction phase $\Phi({\bf{R}},t)$ and the multipolariton generators $\hat K_s({\bf{R}},t)$ and $\hat K_{em}({\bf{R}},t)$ are explicitly given by
\begin{eqnarray} \label{AppB_UI_components}
\Phi \left( {{\bf{R}},t} \right) &=& \frac{{e^2 V^2 }}{{\pi \varepsilon _0 \hbar c^2 }}\int {d\omega } \int_{ - \infty }^t {d\tau } \int_{ - \infty }^\tau  {d\tau '} \;\sin \left[ {\omega \left( {\tau - \tau '} \right)} \right]{\bf{u}}_V  \cdot {\mathop{\rm Im}\nolimits} \left[ {{\cal G}_\omega  \left( {\left. {{\bf{R}} + {\bf{V}}\tau } \right|{\bf{R}} + {\bf{V}}\tau '} \right)} \right] \cdot {\bf{u}}_V , \nonumber \\ 
\hat K_s \left( {{\bf{R}},t} \right) &=& \int {d\omega } \;\frac{{eV}}{{\hbar \omega }}{\bf{u}}_V  \cdot \int_{ - \infty }^t {d\tau } \int {do_{\bf{n}} } \left[ {e^{i\omega \tau } {\cal F}_{\omega s}^* \left( {\left. {{\bf{R}} + {\bf{V}}\tau } \right|{\bf{n}}} \right) \cdot {\bf{\hat g}}_{\omega s}^\dag  \left( {\bf{n}} \right) - e^{ - i\omega \tau } {\cal F}_{\omega s} \left( {\left. {{\bf{R}} + {\bf{V}}\tau } \right|{\bf{n}}} \right) \cdot {\bf{\hat g}}_{\omega s} \left( {\bf{n}} \right)} \right], \nonumber \\ 
\hat K_{em} \left( {{\bf{R}},t} \right) &=& \int {d\omega } \;\frac{{eV}}{{\hbar \omega }}{\bf{u}}_V  \cdot \int_{ - \infty }^t {d\tau } \int_\mathscr{V} {d^3 {\bf{r}}\,} \sum\limits_\nu  {} \left[ {e^{i\omega \tau } {\cal G}_{\omega \nu }^* \left( {\left. {{\bf{R}} + {\bf{V}}\tau } \right|{\bf{r}}} \right) \cdot {\bf{\hat f}}_{\omega \nu }^\dag  \left( {\bf{r}} \right) - e^{ - i\omega \tau } {\cal G}_{\omega \nu } \left( {\left. {{\bf{R}} + {\bf{V}}\tau } \right|{\bf{r}}} \right) \cdot {\bf{\hat f}}_{\omega \nu } \left( {\bf{r}} \right)} \right]. \nonumber \\
\end{eqnarray}

\subsection{Asymptotic density operator $\hat{\rho}_{I}(\infty)$}

In this section, we derive the total interaction-picture density operator in the post-interaction limit, $\hat{\rho}_{I}(\infty)$. As a necessary prelude, we first evaluate the asymptotic limit $t \to +\infty$ of the exact time-evolution operator given in Eq.~(\ref{AppB_UI_master_summary}) of Appendix~B\,1.

We begin by evaluating the limit $t \to +\infty$ in the phase $\Phi({\bf{R}},t)$ and in the multipolariton generators $\hat K_{s} ( {{\bf{R}},t} )$ and $\hat K_{em} ( {{\bf{R}},t} )$ defined in Eq.~(\ref{AppB_UI_components}). Decomposing the electron coordinate as ${\bf{R}} = {\bf{R}}_\perp + R_V{\bf{u}}_V$, the classical trajectory at time $\tau$ reads ${\bf{R}} + {\bf{V}}\tau = {\bf{R}}_\perp + (R_V + V\tau){\bf{u}}_V$. In the asymptotic limit, the time integration domain $\tau \in (-\infty, +\infty)$ is invariant under translations. By performing the change of longitudinal variable $z = R_V + V\tau \in (-\infty, +\infty)$ (with $d\tau = dz/V$ and $\tau - \tau' = (z - z')/V$), the dependence on the longitudinal coordinate $R_V$ identically vanishes from the integration limits. Consequently, the asymptotic self-interaction phase is $\Phi \left( {{\bf{R}} ,\infty } \right) = \Phi \left( {{\bf{R}}_\perp } \right)$ where
\begin{equation} \label{AppB_Phi_Rt}
\Phi \left( {{\bf{R}}_\perp } \right) = \frac{{e^2 }}{{\pi \varepsilon _0 \hbar c^2 }}\int_{ - \infty }^{ + \infty } {dz} \int_{ - \infty }^z {dz'} \int {d\omega } \;\sin \left[ {\frac{\omega }{V}\left( {z - z'} \right)} \right]{\bf{u}}_V  \cdot {\mathop{\rm Im}\nolimits} \left[ {{\cal G}_\omega  \left( {\left. {{\bf{R}}_\perp  + z{\bf{u}}_V } \right|{\bf{R}}_\perp  + z'{\bf{u}}_V } \right)} \right] \cdot {\bf{u}}_V . 
\end{equation}
Similarly, applying the change of variable $z = R_V + V\tau$ to the time integrals in the multipolariton generators (using $e^{\pm i\omega\tau} = e^{\mp i\frac{\omega}{V}R_V}e^{\pm i\frac{\omega}{V}z}$), the dependence on the longitudinal coordinate factorizes exactly into a plane-wave phase. The asymptotic commuting anti-Hermitian exponents $\hat K_s({\bf{R}},\infty) \equiv \hat K_s({\bf{R}})$ and $\hat K_{em}({\bf{R}},\infty) \equiv \hat K_{em}({\bf{R}})$ thus read explicitly:
\begin{eqnarray} \label{AppB_Ks_Kem_defs}
\hat K_s \left( {\bf{R}} \right) &=& \int {d\omega } \;\frac{e}{{\hbar \omega }}\int {do_{\bf{n}} } \;\int_{ - \infty }^{ + \infty } {dz} \; {\bf{u}}_V  \cdot  \left[ {e^{ - i\frac{\omega }{V}\left( {R_V - z} \right)} {\cal F}_{\omega s}^* \left( {\left. {{\bf{R}}_\perp  + z{\bf{u}}_V } \right|{\bf{n}}} \right) \cdot {\bf{\hat g}}_{\omega s}^\dag  \left( {\bf{n}} \right)} \right. \nonumber \\
&& \quad \quad \quad \quad \quad \quad \quad \quad \quad \quad \;\; \left. { - e^{i\frac{\omega }{V}\left( {R_V - z} \right)} {\cal F}_{\omega s} \left( {\left. {{\bf{R}}_\perp  + z{\bf{u}}_V } \right|{\bf{n}}} \right) \cdot {\bf{\hat g}}_{\omega s} \left( {\bf{n}} \right)} \right], \nonumber \\
\hat K_{em} \left( {\bf{R}} \right) &=& \int {d\omega } \;\frac{e}{{\hbar \omega }}\int_\mathscr{V} {d^3 {\bf{r}}\,} \sum\limits_\nu  {\int_{ - \infty }^{ + \infty } {dz} \;{\bf{u}}_V  \cdot \left[ {e^{ - i\frac{\omega }{V}\left( {R_V - z} \right)} {\cal G}_{\omega \nu }^* \left( {\left. {{\bf{R}}_\perp  + z{\bf{u}}_V } \right|{\bf{r}}} \right) \cdot {\bf{\hat f}}_{\omega \nu }^\dag  \left( {\bf{r}} \right)} \right.} \nonumber \\
&& \quad \quad \quad \quad \quad \quad \quad \quad \quad \quad \quad \quad \; \left. { - e^{i\frac{\omega }{V}\left( {R_V - z} \right)} {\cal G}_{\omega \nu } \left( {\left. {{\bf{R}}_\perp  + z{\bf{u}}_V } \right|{\bf{r}}} \right) \cdot {\bf{\hat f}}_{\omega \nu } \left( {\bf{r}} \right)} \right].
\end{eqnarray}
Consequently, evaluating the time-evolution operator of Eq.~(\ref{AppB_UI_master_summary}) in the asymptotic limit $t \to +\infty$, the exact nonperturbative post-interaction evolution operator of the coupled electron-field-matter system is given by
\begin{equation} \label{AppB_UI_infty}
\hat U_I \left( \infty  \right) = \int {d^3 {\bf{R}}} \;\left| {\bf{R}} \right\rangle \left\langle {\bf{R}} \right|e^{i\Phi \left( {{\bf{R}}_\perp  } \right)} e^{\hat K_s \left( {\bf{R}} \right)} e^{\hat K_{em} \left( {\bf{R}} \right)} .
\end{equation}

We now evaluate the asymptotic density operator $\hat{\rho}_I(\infty) = \hat U_I(\infty)\hat{\rho}_I(-\infty)\hat U_I^\dag(\infty)$, where the initial state strictly factorizes as $\hat{\rho}_I(-\infty) = \hat{\rho}_{el} \hat{\rho}_s \hat{\rho}_{em}$ according to Eq.~(\ref{Initial_State_main}). Expanding the initial electron density operator $\hat{\rho}_{el} = \left|\psi\right\rangle\left\langle\psi\right|$ in the coordinate basis via Eq.~(\ref{AppA_rho_el_matrix}) and applying Eq.~(\ref{AppB_UI_infty}), we have
\begin{equation} \label{AppB_rho_I_infty}
\hat \rho _I \left( \infty  \right) = \int {d^3 {\bf{R}}_1 } \int {d^3 {\bf{R}}_2 } \;\rho _{el} \left( {{\bf{R}}_1 ,{\bf{R}}_2 } \right)e^{i\left[ {\Phi \left( {{\bf{R}}_{1\perp} } \right) - \Phi \left( {{\bf{R}}_{2\perp} } \right)} \right]} \left| {{\bf{R}}_1 } \right\rangle \left\langle {{\bf{R}}_2 } \right|e^{\hat K_s \left( {{\bf{R}}_1 } \right)} \hat \rho _s e^{ - \hat K_s \left( {{\bf{R}}_2 } \right)} e^{\hat K_{em} \left( {{\bf{R}}_1 } \right)} \hat \rho _{em} e^{ - \hat K_{em} \left( {{\bf{R}}_2 } \right)} .
\end{equation}

\subsection{Asymptotic electron reduced density operator $\hat{\rho}_{elI}(\infty)$}

Having established the total asymptotic density operator $\hat{\rho}_I(\infty)$, we now evaluate the closed-form expression for the asymptotic electron reduced density operator $\hat{\rho}_{elI}(\infty) = {\rm Tr}_{sem} \left[ \hat{\rho}_I(\infty) \right]$. Taking the partial trace ${\rm Tr}_{sem} = {\rm Tr}_s{\rm Tr}_{em}$ over the polariton Fock space of Eq.~(\ref{AppB_rho_I_infty}) and using the cyclic property of the trace yields
\begin{eqnarray} \label{AppB_rho_elI_step1}
\hat \rho _{elI} \left( \infty  \right) &=& \int {d^3 {\bf{R}}_1 } \int {d^3 {\bf{R}}_2 } \;\rho _{el} \left( {{\bf{R}}_1 ,{\bf{R}}_2 } \right)e^{i\left[ {\Phi \left( {{\bf{R}}_{1\perp} } \right) - \Phi \left( {{\bf{R}}_{2\perp} } \right)} \right]} \nonumber \\
&& \times {\rm Tr}_s \left[ {\hat \rho _s e^{ - \hat K_s \left( {{\bf{R}}_2 } \right)} e^{\hat K_s \left( {{\bf{R}}_1 } \right)} } \right] {\rm Tr}_{em} \left[ {\hat \rho _{em} e^{ - \hat K_{em} \left( {{\bf{R}}_2 } \right)} e^{\hat K_{em} \left( {{\bf{R}}_1 } \right)} } \right] \left| {{\bf{R}}_1 } \right\rangle \left\langle {{\bf{R}}_2 } \right|.
\end{eqnarray}
We can now apply the Baker-Campbell-Hausdorff (BCH) identity $e^{-\hat K_2}e^{\hat K_1} = e^{\hat K_1 - \hat K_2}e^{\frac{1}{2}[\hat K_1, \hat K_2]}$ independently to both the scattering ($s$) and material ($em$) exponentiated operators evaluated at ${\bf{R}}_1$ and ${\bf{R}}_2$. Accordingly, the operator products in Eq.~(\ref{AppB_rho_elI_step1}) recombine as
\begin{equation} \label{AppB_Ddag_D_BCH}
e^{ - \hat K_\alpha  \left( {{\bf{R}}_2 } \right)} e^{\hat K_\alpha  \left( {{\bf{R}}_1 } \right)} = \exp \left\{ {\frac{1}{2}\left[ {\hat K_\alpha  \left( {{\bf{R}}_1 } \right),\hat K_\alpha  \left( {{\bf{R}}_2 } \right)} \right]} \right\} \exp \left[ {\hat K_\alpha  \left( {{\bf{R}}_1 } \right) - \hat K_\alpha  \left( {{\bf{R}}_2 } \right)} \right], \quad \left( {\alpha  = s,em} \right),
\end{equation}
where the $c$-number commutators evaluated directly from the definitions (\ref{AppB_Ks_Kem_defs}) via the bosonic commutation relations (\ref{AppA_boson_comm}) yield
\begin{eqnarray} \label{AppB_K_comm_evaluated}
\frac{1}{2}\left[ {\hat K_s \left( {{\bf{R}}_1 } \right),\hat K_s \left( {{\bf{R}}_2 } \right)} \right] &=&  - i{\mathop{\rm Im}\nolimits} \int {d\omega } \;e^{i\frac{\omega }{V}\left( {R_{1V}  - R_{2V} } \right)} \Gamma _{\omega s} \left( {{\bf{R}}_{1 \perp } ,{\bf{R}}_{2 \perp } } \right), \nonumber \\
\frac{1}{2}\left[ {\hat K_{em} \left( {{\bf{R}}_1 } \right),\hat K_{em} \left( {{\bf{R}}_2 } \right)} \right] &=&  - i{\mathop{\rm Im}\nolimits} \int {d\omega } \;e^{i\frac{\omega }{V}\left( {R_{1V}  - R_{2V} } \right)} \Gamma _{\omega em} \left( {{\bf{R}}_{1 \perp } ,{\bf{R}}_{2 \perp } } \right).
\end{eqnarray}
Here, we have explicitly defined the Hermitian transverse spectral correlation kernels for the scattering ($s$) and material ($em$) channels in terms of their classical spatial integrals:
\begin{eqnarray} \label{AppB_Gamma_s_em}
\Gamma _{\omega s} \left( {{\bf{R}}_{1\perp} ,{\bf{R}}_{2\perp} } \right) &\equiv& \left( {\frac{e}{{\hbar \omega }}} \right)^2 \int_{ - \infty }^{ + \infty } {dz} \int_{ - \infty }^{ + \infty } {dz'} \;e^{ - i\frac{\omega }{V}\left( {z  - z'} \right)} \nonumber \\
&& \times {\bf{u}}_V  \cdot \left[ \int {do_{\bf{n}}}  {{\cal F}_{\omega s} \left( {\left. {{\bf{R}}_{1 \perp }  + z{\bf{u}}_V } \right|{\bf{n}}} \right) \cdot {\cal F}_{\omega s}^{T*} \left( {\left. {{\bf{R}}_{2 \perp }  + z'{\bf{u}}_V } \right|{\bf{n}}} \right)} \right] \cdot {\bf{u}}_V , \nonumber \\
\Gamma _{\omega em} \left( {{\bf{R}}_{1\perp} ,{\bf{R}}_{2\perp} } \right) &\equiv& \left( {\frac{e}{{\hbar \omega }}} \right)^2 \int_{ - \infty }^{ + \infty } {dz} \int_{ - \infty }^{ + \infty } {dz'} \;e^{ - i\frac{\omega }{V}\left( {z  - z'} \right)} \nonumber \\
&& \times   {{\bf{u}}_V  \cdot \left[ \int_\mathscr{V} {d^3 {\bf{r}}\,} \sum\limits_\nu {{\cal G}_{\omega \nu } \left( {\left. {{\bf{R}}_{1 \perp }  + z{\bf{u}}_V } \right|{\bf{r}}} \right) \cdot {\cal G}_{\omega \nu }^{T*} \left( {\left. {{\bf{R}}_{2 \perp }  + z'{\bf{u}}_V } \right|{\bf{r}}} \right)} \right] \cdot {\bf{u}}_V } .
\end{eqnarray}
These kernels satisfy $\Gamma_{\omega\alpha}^*({\bf{R}}_{1\perp}, {\bf{R}}_{2\perp}) = \Gamma_{\omega\alpha}({\bf{R}}_{2\perp}, {\bf{R}}_{1\perp})$ and $\Gamma_{\omega\alpha}({\bf{R}}_\perp, {\bf{R}}_\perp) \ge 0$. By virtue of the fundamental integral identity (\ref{AppA_Fund_Id}), their sum exactly reconstructs the total macroscopic correlation kernel governed by ${\rm Im}[{\cal G}_\omega]$:
\begin{eqnarray} \label{AppB_Gamma_tot}
 \Gamma _{\omega s} \left( {{\bf{R}}_{1\perp} ,{\bf{R}}_{2\perp} } \right) + \Gamma _{\omega em} \left( {{\bf{R}}_{1\perp} ,{\bf{R}}_{2\perp} } \right) &=& \frac{{e^2 }}{{\pi \varepsilon _0 \hbar c^2 }}\int_{ - \infty }^{ + \infty } {dz } \int_{ - \infty }^{ + \infty } {dz' } \;e^{ - i\frac{\omega }{V}\left( {z  - z' } \right)} \nonumber \\
&&  \times  {\bf{u}}_V  \cdot {\mathop{\rm Im}\nolimits} \left[ {{\cal G}_\omega  \left( {\left. {{\bf{R}}_{1\perp}  + z {\bf{u}}_V } \right|{\bf{R}}_{2\perp}  + z' {\bf{u}}_V } \right)} \right] \cdot {\bf{u}}_V . 
\end{eqnarray}

Substituting Eq.~(\ref{AppB_Ddag_D_BCH}) into Eq.~(\ref{AppB_rho_elI_step1}) and using the evaluated commutators of Eq.~(\ref{AppB_K_comm_evaluated}), the asymptotic electron reduced density operator rigorously factorizes over the separate scattering and material density operators:
\begin{equation} \label{AppB_rho_elI_step2}
\hat \rho _{elI} \left( \infty  \right) = \int {d^3 {\bf{R}}_1 } \int {d^3 {\bf{R}}_2 } \;\rho _{el} \left( {{\bf{R}}_1 ,{\bf{R}}_2 } \right)e^{i\Theta \left( {{\bf{R}}_1 ,{\bf{R}}_2 } \right)} {\rm Tr}_s \left[ {\hat \rho _s e^{\hat K_s \left( {{\bf{R}}_1 } \right) - \hat K_s \left( {{\bf{R}}_2 } \right)} } \right]{\rm Tr}_{em} \left[ {\hat \rho _{em} e^{\hat K_{em} \left( {{\bf{R}}_1 } \right) - \hat K_{em} \left( {{\bf{R}}_2 } \right)} } \right]\left| {{\bf{R}}_1 } \right\rangle \left\langle {{\bf{R}}_2 } \right|,
\end{equation}
where the total two-path radiation-reaction phase $\Theta({\bf{R}}_1, {\bf{R}}_2)$, emerging from combining the elastic vacuum phase $e^{i[\Phi({\bf{R}}_{1\perp}) - \Phi({\bf{R}}_{2\perp})]}$ with the scalar factors generated by the BCH expansion, is explicitly given by
\begin{equation} \label{AppB_Theta_infty}
\Theta \left( {{\bf{R}}_1 ,{\bf{R}}_2 } \right) = \Phi \left( {{\bf{R}}_{1\perp} } \right) - \Phi \left( {{\bf{R}}_{2\perp} } \right) - {\mathop{\rm Im}\nolimits} \int {d\omega } \;e^{i\frac{\omega }{V}\left( {R_{1V}  - R_{2V} } \right)} \left[ {\Gamma _{\omega s} \left( {{\bf{R}}_{1\perp} ,{\bf{R}}_{2\perp} } \right) + \Gamma _{\omega em} \left( {{\bf{R}}_{1\perp} ,{\bf{R}}_{2\perp} } \right)} \right].
\end{equation}

We next evaluate the scattering trace ${\rm Tr}_s \{ \hat{\rho}_s \exp [ \hat K_s({\bf{R}}_1) - \hat K_s({\bf{R}}_2) ] \}$ in Eq.~(\ref{AppB_rho_elI_step2}). To this end, we explicitly separate the vacuum fluctuations of the scattering modes (which are strictly present even for the vacuum field state $\hat{\rho}_s = \left|0_s\right\rangle\left\langle 0_s\right|$) from the stimulated transitions driven by an externally prepared incident radiation ($\hat{\rho}_s \neq \left|0_s\right\rangle\left\langle 0_s\right|$). Specifically, we define the classical two-path scattering difference vector 
\begin{eqnarray} \label{AppB_C_vector_def}
{\bf{C}}_\omega  \left( {{\bf{R}}_1 ,{\bf{R}}_2 \left| {\bf{n}} \right.} \right) &\equiv& \frac{e}{{\hbar \omega }}\int_{ - \infty }^{ + \infty } {dz} \;e^{ - i\frac{\omega }{V}z} {\bf{u}}_V  \cdot \left[ {e^{i\frac{\omega }{V}R_{1V} } {\cal F}_{\omega s} \left( {\left. {{\bf{R}}_{1\perp}  + z{\bf{u}}_V } \right|{\bf{n}}} \right) - e^{i\frac{\omega }{V}R_{2V} } {\cal F}_{\omega s} \left( {\left. {{\bf{R}}_{2\perp}  + z{\bf{u}}_V } \right|{\bf{n}}} \right)} \right], 
\end{eqnarray}
allowing the operator exponent to be compactly rewritten as
\begin{equation} \label{AppB_Delta_Ks_compact}
\hat K_s \left( {{\bf{R}}_1 } \right) - \hat K_s \left( {{\bf{R}}_2 } \right) = \int {d\omega } \int {do_{\bf{n}} } \left[ {{\bf{C}}_\omega ^* \left( {{\bf{R}}_1 ,{\bf{R}}_2 \left| {\bf{n}} \right.} \right) \cdot {\bf{\hat g}}_{\omega s}^\dag  \left( {\bf{n}} \right) - {\bf{C}}_\omega  \left( {{\bf{R}}_1 ,{\bf{R}}_2 \left| {\bf{n}} \right.} \right) \cdot {\bf{\hat g}}_{\omega s} \left( {\bf{n}} \right)} \right].
\end{equation}
By invoking the BCH identity in the form $e^{\hat A + \hat B} = e^{\hat A} e^{\hat B} e^{-\frac{1}{2}[\hat A, \hat B]}$ for operators whose commutator $[\hat A, \hat B]$ is a $c$-number, we cast the exponential operator into normal order:
\begin{eqnarray} \label{AppB_Ks_BCH_normal}
\exp \left[ \hat K_s \left( {{\bf{R}}_1 } \right) - \hat K_s \left( {{\bf{R}}_2 } \right) \right] &=& \exp \left[ \int {d\omega } \int {do_{\bf{n}} } \;{\bf{C}}_\omega ^* \left( {{\bf{R}}_1 ,{\bf{R}}_2 \left| {\bf{n}} \right.} \right) \cdot {\bf{\hat g}}_{\omega s}^\dag  \left( {\bf{n}} \right) \right] \nonumber \\
&& \times \exp \left[  - \int {d\omega } \int {do_{\bf{n}} } \;{\bf{C}}_\omega  \left( {{\bf{R}}_1 ,{\bf{R}}_2 \left| {\bf{n}} \right.} \right) \cdot {\bf{\hat g}}_{\omega s} \left( {\bf{n}} \right) \right] \nonumber \\
&& \times \exp \left[  - \frac{1}{2}\int {d\omega } \int {do_{\bf{n}} } \left| {{\bf{C}}_\omega  \left( {{\bf{R}}_1 ,{\bf{R}}_2 \left| {\bf{n}} \right.} \right)} \right|^2 \right] .
\end{eqnarray}
The evaluation of the $c$-number scalar factor $|{\bf{C}}_\omega|^2$ directly reconstructs the trace of the scattering kernel $\Gamma_{\omega s}$. Consequently, the scattering trace rigorously factorizes as
\begin{equation} \label{AppB_chi_s_normal_order}
{\rm Tr}_s \left[ \hat{\rho}_s e^{ \hat K_s({\bf{R}}_1) - \hat K_s({\bf{R}}_2) } \right] = e^{\Xi _s \left( {{\bf{R}}_1 ,{\bf{R}}_2 } \right)} \Upsilon _s \left( {{\bf{R}}_1 ,{\bf{R}}_2 } \right),
\end{equation}
where $\Xi_s({\bf{R}}_1, {\bf{R}}_2)$ is the symmetric and non-positive vacuum radiative decoherence exponent,
\begin{eqnarray} \label{AppB_Xi_s0_infty}
\Xi _s \left( {{\bf{R}}_1 ,{\bf{R}}_2 } \right) =  - \frac{1}{2}\int {d\omega } \;\bigg\{ && \Gamma _{\omega s} \left( {{\bf{R}}_{1\perp} ,{\bf{R}}_{1\perp} } \right) + \Gamma _{\omega s} \left( {{\bf{R}}_{2\perp} ,{\bf{R}}_{2\perp} } \right) \nonumber \\
&& - 2{\mathop{\rm Re}\nolimits} \left[ {e^{i\frac{\omega }{V}\left( {R_{1V}  - R_{2V} } \right)} \Gamma _{\omega s} \left( {{\bf{R}}_{1\perp} ,{\bf{R}}_{2\perp} } \right)} \right] \bigg\},
\end{eqnarray}
and $\Upsilon_s({\bf{R}}_1, {\bf{R}}_2)$ defines the normal-ordered characteristic functional of the arbitrary incident quantum state of light $\hat{\rho}_s$:
\begin{eqnarray} \label{AppB_Upsilon_s_def}
\Upsilon _s \left( {{\bf{R}}_1 ,{\bf{R}}_2 } \right) &\equiv& {\rm Tr}_s \left\{ {\hat \rho _s \exp \left[ {\int {d\omega } \int {do_{\bf{n}} } \;{\bf{C}}_\omega ^* \left( {{\bf{R}}_1 ,{\bf{R}}_2 \left| {\bf{n}} \right.} \right) \cdot {\bf{\hat g}}_{\omega s}^\dag  \left( {\bf{n}} \right)} \right]} \right. \nonumber \\
&& \quad \quad \; \times \left. {\exp \left[ { - \int {d\omega } \int {do_{\bf{n}} } \;{\bf{C}}_\omega  \left( {{\bf{R}}_1 ,{\bf{R}}_2 \left| {\bf{n}} \right.} \right) \cdot {\bf{\hat g}}_{\omega s} \left( {\bf{n}} \right)} \right]} \right\}.
\end{eqnarray}
By virtue of the normal-ordered exponential operators, the trace trivially satisfies $\Upsilon_s({\bf{R}}_1, {\bf{R}}_2) = 1$ for ${\hat{\rho}_s = \left|0_s\right\rangle\left\langle 0_s\right|}$ identically for all paths ${\bf{R}}_1, {\bf{R}}_2$, correctly isolating the stimulated interactions from the vacuum decoherence $\Xi_s$.

Turning to  the material thermal trace ${\rm Tr}_{em} \{\hat{\rho}_{em}\exp [\hat K_{em}({\bf{R}}_1) - \hat K_{em}({\bf{R}}_2) ] \}$, because the operator $\hat K_{em}$ is linear in the bosonic operators ${\bf{\hat f}}_{\omega\nu}^\dag$ and ${\bf{\hat f}}_{\omega\nu}$, and $\hat{\rho}_{em}$ in Eq.~(\ref{AppA_rho_em}) is a centered Gaussian thermal state, Wick's theorem dictates that the cumulant expansion truncates exactly at second order:
\begin{equation} \label{AppB_Wick_em}
{\rm Tr}_{em} \left\{ \hat \rho _{em} \exp \left[ \hat K_{em} \left( {{\bf{R}}_1 } \right) - \hat K_{em} \left( {{\bf{R}}_2 } \right) \right] \right\} = \exp \left[ \frac{1}{2}{\rm Tr}_{em} \left\{ \hat \rho _{em} \left( \hat K_{em} \left( {{\bf{R}}_1 } \right) - \hat K_{em} \left( {{\bf{R}}_2 } \right) \right)^2 \right\} \right] \equiv \exp \left[ \Xi_{em} \left( {{\bf{R}}_1 ,{\bf{R}}_2 } \right) \right].
\end{equation}
Expanding the square of the difference operator $\hat K_{em}({\bf{R}}_1) - \hat K_{em}({\bf{R}}_2)$, the only non-vanishing terms under the diagonal thermal trace are the anti-normally and normally ordered products proportional to ${\bf{\hat f}}_{\omega \nu } {\bf{\hat f}}_{\omega \nu }^\dag$ and ${\bf{\hat f}}_{\omega \nu }^\dag  {\bf{\hat f}}_{\omega \nu }$. Substituting the thermal expectation values from Eq.~(\ref{AppA_thermal_averages})---which yield factors of $n_\omega(T_{em})+1$ and $n_\omega(T_{em})$, respectively, whose sum gives $2n_\omega(T_{em})+1$---and reorganizing the terms via the definition of the material kernel in Eq.~(\ref{AppB_Gamma_s_em}), we obtain the real and non-positive material decoherence exponent:
\begin{eqnarray} \label{AppB_Xi_infty}
\Xi_{em} \left( {{\bf{R}}_1 ,{\bf{R}}_2 } \right) =  - \int {d\omega } \left[ {n_\omega  \left( {T_{em} } \right) + \frac{1}{2}} \right]\bigg\{ &&\Gamma _{\omega em} \left( {{\bf{R}}_{1\perp} ,{\bf{R}}_{1\perp} } \right) + \Gamma _{\omega em} \left( {{\bf{R}}_{2\perp} ,{\bf{R}}_{2\perp} } \right)  \nonumber \\
 && - 2{\mathop{\rm Re}\nolimits} \left[ {e^{i\frac{\omega }{V}\left( {R_{1V}  - R_{2V} } \right)} \Gamma _{\omega em} \left( {{\bf{R}}_{1 \perp} ,{\bf{R}}_{2 \perp } } \right)} \right]\bigg\}. 
\end{eqnarray}

By substituting the evaluated traces into Eq.~(\ref{AppB_rho_elI_step2}), the asymptotic reduced density operator takes the exact form
\begin{equation} \label{AppB_rho_elI_simplified}
\hat \rho _{elI} \left( \infty  \right) = \int {d^3 {\bf{R}}_1 } \int {d^3 {\bf{R}}_2 } \;\rho _{el} \left( {{\bf{R}}_1 ,{\bf{R}}_2 } \right)e^{i\Theta \left( {{\bf{R}}_1 ,{\bf{R}}_2 } \right) + \Xi _s \left( {{\bf{R}}_1 ,{\bf{R}}_2 } \right) + \Xi _{em} \left( {{\bf{R}}_1 ,{\bf{R}}_2 } \right)} \Upsilon _s \left( {{\bf{R}}_1 ,{\bf{R}}_2 } \right)\left| {{\bf{R}}_1 } \right\rangle \left\langle {{\bf{R}}_2 } \right|.
\end{equation}
We are now in a position to sum the three macroscopic exponents governing the electron interaction: the two-path radiation-reaction phase $i\Theta$, the vacuum scattering decoherence $\Xi_s$, and the material decoherence $\Xi_{em}$ defined in Eqs.~(\ref{AppB_Theta_infty}), (\ref{AppB_Xi_s0_infty}), and (\ref{AppB_Xi_infty}), respectively. By grouping the autoterms and the cross-path terms, the combined exponent rigorously reduces to
\begin{equation} \label{AppB_Exponent_Sum}
i\Theta \left( {{\bf{R}}_1 ,{\bf{R}}_2 } \right) + \Xi _s \left( {{\bf{R}}_1 ,{\bf{R}}_2 } \right) + \Xi _{em}\left( {{\bf{R}}_1 ,{\bf{R}}_2 } \right) = i\left[ {\Phi \left( {{\bf{R}}_{1\perp} } \right) - \Phi \left( {{\bf{R}}_{2\perp} } \right)} \right] - \left[ {W\left( {{\bf{R}}_{1\perp} } \right) + W\left( {{\bf{R}}_{2\perp} } \right)} \right] + K\left( {{\bf{R}}_{1\perp} ,{\bf{R}}_{2\perp} ;R_{1V}  - R_{2V} } \right).
\end{equation}
Here, $W({\bf{R}}_\perp)$ is the total integrated inelastic depletion optical depth at impact parameter ${\bf{R}}_\perp$ containing all the autoterms,
\begin{equation} \label{AppB_W_def}
W\left( {{\bf{R}}_\perp  } \right) = \int {d\omega } \left\{ {\frac{1}{2}\Gamma _{\omega s} \left( {{\bf{R}}_\perp  ,{\bf{R}}_\perp  } \right) + \left[ {n_\omega  \left( {T_{em} } \right) + \frac{1}{2}} \right]\Gamma _{\omega em} \left( {{\bf{R}}_\perp  ,{\bf{R}}_\perp  } \right)} \right\},
\end{equation}
and $K({\bf{R}}_{1\perp}, {\bf{R}}_{2\perp}; Z)$ is the two-path spontaneous and thermal cumulant function governing the inelastic transitions (with longitudinal relative shift $Z = R_{1V} - R_{2V}$), 
\begin{eqnarray} \label{AppB_K_cumulant}
K\left( {{\bf{R}}_{1\perp} ,{\bf{R}}_{2\perp} ;Z} \right) &\equiv& \int {d\omega } \;\Bigg\{ e^{ - i\frac{\omega }{V}Z} \Big[ \Gamma _{\omega s}^* \left( {{\bf{R}}_{1\perp} ,{\bf{R}}_{2\perp} } \right) + \left( {n_\omega  \left( {T_{em} } \right) + 1} \right)\Gamma _{\omega em}^* \left( {{\bf{R}}_{1\perp} ,{\bf{R}}_{2\perp} } \right) \Big] \nonumber \\
&& \quad \quad \quad \quad \; + e^{i\frac{\omega }{V}Z} n_\omega  \left( {T_{em} } \right)\Gamma _{\omega em} \left( {{\bf{R}}_{1\perp} ,{\bf{R}}_{2\perp} } \right) \Bigg\}.
\end{eqnarray}

For the sake of expositional clarity, we summarize below the complete closed-form solution for the asymptotic electron reduced density operator $\hat{\rho}_{elI}(\infty)$ in the coordinate basis. By recombining the terms, the post-interaction state reads
\begin{eqnarray} \label{AppB_rho_elI_master_summary}
\hat \rho _{elI} \left( \infty  \right) &=& \int {d^3 {\bf{R}}_1 } \int {d^3 {\bf{R}}_2 } \;\rho _{el} \left( {{\bf{R}}_1 ,{\bf{R}}_2 } \right)e^{i\Phi \left( {{\bf{R}}_{1 \perp } } \right) - W\left( {{\bf{R}}_{1 \perp } } \right)} e^{ - i\Phi \left( {{\bf{R}}_{2 \perp } } \right) - W\left( {{\bf{R}}_{2 \perp } } \right)} \nonumber \\
&& \times e^{K\left( {{\bf{R}}_{1 \perp } ,{\bf{R}}_{2 \perp } ;R_{1V}  - R_{2V} } \right)} \Upsilon _s \left( {{\bf{R}}_1 ,{\bf{R}}_2 } \right)\left| {{\bf{R}}_1 } \right\rangle \left\langle {{\bf{R}}_2 } \right|,
\end{eqnarray}
where the coherent radiation-reaction phase $\Phi$ and the inelastic depletion depth $W$ depend solely on the transverse coordinates, while the spontaneous/thermal cross-cumulant $K$ depends additionally on the longitudinal relative shift $Z = R_{1V}  - R_{2V}$. They read explicitly:
\begin{eqnarray} \label{AppB_master_scalars}
\Phi \left( {{\bf{R}}_ \perp  } \right) &=& \frac{{e^2 }}{{\pi \varepsilon _0 \hbar c^2 }}\int_{ - \infty }^{ + \infty } {dz} \int_{ - \infty }^z {dz'} \int {d\omega } \;\sin \left[ {\frac{\omega }{V}\left( {z - z'} \right)} \right]{\bf{u}}_V  \cdot {\mathop{\rm Im}\nolimits} \left[ {{\cal G}_\omega  \left( {\left. {{\bf{R}}_ \perp   + z{\bf{u}}_V } \right|{\bf{R}}_ \perp   + z'{\bf{u}}_V } \right)} \right] \cdot {\bf{u}}_V , \nonumber \\ 
W\left( {{\bf{R}}_ \perp  } \right) &=& \int {d\omega } \left\{ {\frac{1}{2}\Gamma _{\omega s} \left( {{\bf{R}}_ \perp  ,{\bf{R}}_ \perp  } \right) + \left[ {n_\omega  \left( {T_{em} } \right) + \frac{1}{2}} \right]\Gamma _{\omega em} \left( {{\bf{R}}_ \perp  ,{\bf{R}}_ \perp  } \right)} \right\}, \nonumber \\ 
K\left( {{\bf{R}}_{1 \perp } ,{\bf{R}}_{2 \perp } ;Z} \right) &=& \int {d\omega } \left\{ {e^{ - i\frac{\omega }{V}Z} \left[ {\Gamma _{\omega s}^* \left( {{\bf{R}}_{1 \perp } ,{\bf{R}}_{2 \perp } } \right) + \left( {n_\omega  \left( {T_{em} } \right) + 1} \right)\Gamma _{\omega em}^* \left( {{\bf{R}}_{1 \perp } ,{\bf{R}}_{2 \perp } } \right)} \right]} \right. \nonumber \\
&& \quad \quad \quad \; \left. { + e^{i\frac{\omega }{V}Z} n_\omega  \left( {T_{em} } \right)\Gamma _{\omega em} \left( {{\bf{R}}_{1 \perp } ,{\bf{R}}_{2 \perp } } \right)} \right\} .
\end{eqnarray}
The active illumination dynamics are entirely encapsulated within the normally ordered characteristic functional $\Upsilon_s$ given by
\begin{equation} \label{AppB_master_Upsilon}
\Upsilon _s \left( {{\bf{R}}_1 ,{\bf{R}}_2 } \right) = {\rm Tr}_s \left\{ {\hat \rho _s \exp \left[ {\int {d\omega } \int {do_{\bf{n}} } \;{\bf{C}}_\omega ^* \left( {{\bf{R}}_1 ,{\bf{R}}_2 \left| {\bf{n}} \right.} \right) \cdot {\bf{\hat g}}_{\omega s}^\dag  \left( {\bf{n}} \right)} \right]\exp \left[ { - \int {d\omega } \int {do_{\bf{n}} } \;{\bf{C}}_\omega  \left( {{\bf{R}}_1 ,{\bf{R}}_2 \left| {\bf{n}} \right.} \right) \cdot {\bf{\hat g}}_{\omega s} \left( {\bf{n}} \right)} \right]} \right\},
\end{equation}
driven by the two-path scattering difference vector 
\begin{equation} \label{AppB_master_C}
{\bf{C}}_\omega  \left( {{\bf{R}}_1 ,{\bf{R}}_2 \left| {\bf{n}} \right.} \right) = \frac{e}{{\hbar \omega }}\int_{ - \infty }^{ + \infty } {dz} \;e^{ - i\frac{\omega }{V}z} {\bf{u}}_V  \cdot \left[ {e^{i\frac{\omega }{V}R_{1V} } {\cal F}_{\omega s} \left( {\left. {{\bf{R}}_{1 \perp }  + z{\bf{u}}_V } \right|{\bf{n}}} \right) - e^{i\frac{\omega }{V}R_{2V} } {\cal F}_{\omega s} \left( {\left. {{\bf{R}}_{2 \perp }  + z{\bf{u}}_V } \right|{\bf{n}}} \right)} \right].
\end{equation}
Finally, the two-point spatial correlation functions characterizing the vacuum scattering ($s$) and material ($em$) decoherence rates are explicitly given by
\begin{eqnarray} \label{AppB_master_Gamma}
\Gamma _{\omega s} \left( {{\bf{R}}_{1 \perp } ,{\bf{R}}_{2 \perp } } \right) &=& \left( {\frac{e}{{\hbar \omega }}} \right)^2 \int_{ - \infty }^{ + \infty } {dz} \int_{ - \infty }^{ + \infty } {dz'} \;e^{ - i\frac{\omega }{V}\left( {z - z'} \right)} {\bf{u}}_V  \cdot \left[ {\int {do_{\bf{n}} } \;{\cal F}_{\omega s} \left( {\left. {{\bf{R}}_{1\perp }  + z{\bf{u}}_V } \right|{\bf{n}}} \right) \cdot {\cal F}_{\omega s}^{T*} \left( {\left. {{\bf{R}}_{2 \perp }  + z'{\bf{u}}_V } \right|{\bf{n}}} \right)} \right] \cdot {\bf{u}}_V , \nonumber \\ 
\Gamma _{\omega em} \left( {{\bf{R}}_{1 \perp } ,{\bf{R}}_{2 \perp } } \right) &=& \left( {\frac{e}{{\hbar \omega }}} \right)^2 \int_{ - \infty }^{ + \infty } {dz} \int_{ - \infty }^{ + \infty } {dz'} \;e^{ - i\frac{\omega }{V}\left( {z - z'} \right)} {\bf{u}}_V  \cdot \left[ {\int_\mathscr{V} {d^3 {\bf{r}}\,} \sum\limits_\nu  {} {\cal G}_{\omega \nu } \left( {\left. {{\bf{R}}_{1 \perp }  + z{\bf{u}}_V } \right|{\bf{r}}} \right) \cdot {\cal G}_{\omega \nu }^{T*} \left( {\left. {{\bf{R}}_{2 \perp }  + z'{\bf{u}}_V } \right|{\bf{r}}} \right)} \right] \cdot {\bf{u}}_V . \nonumber \\
\end{eqnarray}

\subsection{Transverse-momentum-resolved probability distribution}

We now evaluate the joint energy and transverse-momentum exchange probability distribution 
\begin{equation}
\frac{{d\mathscr{P}}}{{dEd^2 {\bf{P}}_\perp  }} = {\rm Tr}_{el}\left[ {\hat \rho _{elI} \left( \infty  \right)\left| {E,{\bf{P}}_\perp  } \right\rangle \left\langle {E,{\bf{P}}_\perp  } \right|} \right]
\end{equation}
by projecting the asymptotic reduced density operator $\hat\rho_{elI}(\infty)$ of Eq.~(\ref{AppB_rho_elI_master_summary}) onto the kinematic basis states $\left| {E,{\bf{P}}_\perp  } \right\rangle$. Using the coordinate representation of the basis states given in Eq.~(\ref{AppA_E_Pt_basis}),
\begin{equation} \label{AppB_proj_elements}
\left\langle {{E,{\bf{P}}_\perp  }}\mathrel{\left | {\vphantom {{E,{\bf{P}}_\perp  } {{\bf{R}}_1 }}}\right. \kern-\nulldelimiterspace}{{{\bf{R}}_1 }} \right\rangle \left\langle {{{\bf{R}}_2 }}\mathrel{\left | {\vphantom {{{\bf{R}}_2 } {E,{\bf{P}}_\perp  }}}\right. \kern-\nulldelimiterspace}{{E,{\bf{P}}_\perp  }} \right\rangle  = \frac{1}{{\left( {2\pi \hbar } \right)^3 V}}e^{ - i\frac{E}{{\hbar V}}\left( {R_{1V}  - R_{2V} } \right)} e^{ - \frac{i}{\hbar }{\bf{P}}_\perp   \cdot \left( {{\bf{R}}_{1\perp}  - {\bf{R}}_{2\perp} } \right)} ,
\end{equation}
and substituting the initial electron density matrix elements $\rho_{el}({\bf R}_1, {\bf R}_2)$ from Eq.~(\ref{AppA_rho_el_matrix}) alongside the explicit interaction terms summarized in Eqs.~(\ref{AppB_master_scalars}) and (\ref{AppB_master_Upsilon}), we obtain the exact closed-form distribution:
\begin{eqnarray} \label{AppB_dP_master}
\frac{{d\mathscr{P}}}{{dEd^2 {\bf{P}}_\perp  }} &=& \frac{1}{{\left( {2\pi \hbar } \right)^3 V}}\int {d^2 {\bf{R}}_{1\perp} } \int {d^2 {\bf{R}}_{2\perp} } \;e^{ - \frac{i}{\hbar }{\bf{P}}_\perp   \cdot \left( {{\bf{R}}_{1\perp}  - {\bf{R}}_{2\perp} } \right)} \widetilde \phi \left( {{\bf{R}}_{1\perp} } \right)\widetilde \phi ^* \left( {{\bf{R}}_{2\perp} } \right) \nonumber \\
&& \times \frac{1}{\ell }\int_{ - \ell /2}^{\ell /2} {dR_{1V} } \int_{ - \ell /2}^{\ell /2} {dR_{2V} } \;e^{ - i\frac{{E - E_0 }}{{\hbar V}}\left( {R_{1V}  - R_{2V} } \right)} e^{K\left( {{\bf{R}}_{1\perp} ,{\bf{R}}_{2\perp} ;R_{1V}  - R_{2V} } \right)} \Upsilon _s \left( {{\bf{R}}_1 ,{\bf{R}}_2 } \right).
\end{eqnarray}
Here, $\widetilde \phi ({\bf{R}}_\perp)$ is the dressed transverse wave function defined as
\begin{equation} \label{AppB_dressed_phi}
\widetilde \phi \left( {{\bf{R}}_\perp  } \right) \equiv \phi \left( {{\bf{R}}_\perp  } \right)e^{i\Phi \left( {{\bf{R}}_\perp  } \right) - W\left( {{\bf{R}}_\perp  } \right)} ,
\end{equation}
where $\phi({\bf R}_\perp)$ is the initial transverse electron profile normalized as $\int d^2{\bf R}_\perp |\phi({\bf R}_\perp)|^2 = 1$.

To explicitly evaluate the double longitudinal integral over the finite square domain $[-\ell/2, \ell/2] \times [-\ell/2, \ell/2]$ in Eq.~(\ref{AppB_dP_master}), it is highly advantageous to perform a change of variables to the relative coordinate $Z$ and the center-of-mass coordinate $\bar R_V$. The linear transformation and its inverse read:
\begin{equation} \label{AppB_change_of_vars}
\begin{aligned}
 Z &= R_{1V}  - R_{2V} \\ 
 \bar R_V  &= \frac{1}{2}\left( {R_{1V}  + R_{2V} } \right) 
\end{aligned}
\quad \longleftrightarrow \quad
\begin{aligned}
 R_{1V}  &= \bar R_V  + \frac{1}{2}Z \\ 
 R_{2V}  &= \bar R_V  - \frac{1}{2}Z 
\end{aligned} \;.
\end{equation}
Because the Jacobian determinant of this transformation is exactly unity, the area element transforms as $dR_{1V} dR_{2V} = dZ d\bar R_V$. Mapping the boundaries of the original square domain onto the new coordinates yields a diamond-shaped integration region, leading to the exact integral identity for any integrable function $f$:
\begin{equation} \label{AppB_integral_identity}
\int_{ - \ell /2}^{\ell /2} {dR_{1V} } \int_{ - \ell /2}^{\ell /2} {dR_{2V} } \;f\left( {R_{1V} ,R_{2V} } \right) = \int_{ - \ell }^\ell  {dZ} \int_{ - \frac{{\left( {\ell  - \left| Z \right|} \right)}}{2}}^{\frac{{\left( {\ell  - \left| Z \right|} \right)}}{2}} {d\bar R_V } \;f\left( {\bar R_V  + \frac{1}{2}Z,\bar R_V  - \frac{1}{2}Z} \right).
\end{equation}
By separating the integrations over the relative separation $Z$ (which couples to the energy exchange and the spontaneous/thermal cumulant $K$) and the macroscopic center of mass $\bar R_V$ (which traces the temporal envelope of the electron pulse), the master probability distribution (\ref{AppB_dP_master}) can be identically recast into the more fundamental form:
\begin{eqnarray} \label{AppB_dP_master_CM}
\frac{{d\mathscr{P}}}{{dEd^2 {\bf{P}}_\perp  }} &=& \frac{1}{{\left( {2\pi \hbar } \right)^3 V}}\int {d^2 {\bf{R}}_{1\perp} } \int {d^2 {\bf{R}}_{2\perp} } \;e^{ - \frac{i}{\hbar }{\bf{P}}_\perp   \cdot \left( {{\bf{R}}_{1\perp}  - {\bf{R}}_{2\perp} } \right)} \widetilde \phi \left( {{\bf{R}}_{1\perp} } \right)\widetilde \phi ^* \left( {{\bf{R}}_{2\perp} } \right) \nonumber \\
&& \times \int_{ - \ell }^\ell  {dZ} \;e^{ - \frac{i}{{\hbar V}}\left( {E - E_0 } \right)Z} e^{K\left( {{\bf{R}}_{1\perp} ,{\bf{R}}_{2\perp} ;Z} \right)} \frac{1}{\ell }\int_{ - \frac{{\ell - \left| Z \right|}}{2}}^{\frac{{\ell - \left| Z \right|}}{2}} {d\bar R_V } \;\Upsilon _s \left( {{\bf{R}}_1 ,{\bf{R}}_2 } \right),
\end{eqnarray}
where the full 3D spatial arguments of the normally-ordered functional $\Upsilon_s$ are explicitly parametrized by $Z$ and $\bar R_V$ as
\begin{eqnarray} \label{AppB_R1_R2_parametrized}
{\bf{R}}_1  &=& {\bf{R}}_{1\perp}  + \left( {\bar R_V  + \frac{1}{2}Z} \right){\bf{u}}_V , \nonumber \\
{\bf{R}}_2  &=& {\bf{R}}_{2\perp}  + \left( {\bar R_V - \frac{1}{2}Z} \right){\bf{u}}_V .
\end{eqnarray}
This explicit decomposition proves particularly powerful when evaluating the transverse distribution for specific quantum states of light, since in the monoenergetic limit ($\ell \to \infty$) the integral over $d\bar R_V$ rigorously extracts the phase-matching conditions (energy conservation and sideband interference) intrinsically encoded within the optical characteristic functional $\Upsilon_s$.

\section{Evaluation of the characteristic functional and energy-momentum spectra for specific quantum states of light}

In this Appendix, we provide the detailed mathematical derivation of the normal-ordered characteristic functional $\Upsilon_s(\mathbf{R}_1, \mathbf{R}_2)$ defined in Eq.~(\ref{AppB_master_Upsilon}) and of the resulting transverse-momentum-resolved energy exchange distributions (\ref{AppB_dP_master_CM}) for three fundamental classes of incident radiation: (1) thermal radiation at temperature $T_s$, (2) multimode coherent states, and (3) multimode squeezed vacuum states.

\subsection{Incident thermal radiation at temperature $T_s$}

As a first fundamental case, we consider the configuration in which the incident free-space radiation (the asymptotic scattering polariton sector $s$) is in thermal equilibrium at temperature $T_s$, described by the canonical Gibbs density operator
\begin{equation} \label{AppC_rho_s_thermal}
\hat \rho _s  = \frac{{e^{ - \hat H_s /\left( {k_B T_s } \right)} }}{{{\rm Tr}_s \left[ {e^{ - \hat H_s /\left( {k_B T_s } \right)} } \right]}},
\end{equation}
where $\hat H_s$ is the free-scattering Hamiltonian defined in the first of Eqs.~(\ref{AppA_H_sectors}). Using the bosonic commutation relation for $\hat{\mathbf{g}}_{\omega s}(\mathbf{n})$ in the first of Eqs.~(\ref{AppA_boson_comm}), the non-vanishing one- and two-point thermal expectation values over $\hat\rho_s$ are given by
\begin{eqnarray} \label{AppC_thermal_s_averages}
{\rm Tr}_s \left[ {\hat \rho _s \hat{\mathbf{g}}_{\omega s}^\dag  \left( \mathbf{n} \right)\hat{\mathbf{g}}_{\omega 's} \left( \mathbf{n}' \right)} \right] &=& n_\omega  \left( T_s \right)\delta \left( \omega  - \omega ' \right)\delta \left( o_\mathbf{n}  - o_{\mathbf{n}'}  \right){\cal I}_\mathbf{n} , \nonumber \\
{\rm Tr}_s \left[ {\hat \rho _s \hat{\mathbf{g}}_{\omega s} \left( \mathbf{n} \right)\hat{\mathbf{g}}_{\omega 's}^\dag  \left( \mathbf{n}' \right)} \right] &=& \left[ {n_\omega  \left( T_s \right) + 1} \right]\delta \left( \omega  - \omega ' \right)\delta \left( o_\mathbf{n}  - o_{\mathbf{n}'}  \right){\cal I}_\mathbf{n} ,
\end{eqnarray}
with all other one- and two-point averages vanishing, where $n_\omega(T_s) = \left[e^{\hbar\omega/(k_B T_s)} - 1\right]^{-1}$ is the Bose-Einstein thermal distribution at temperature $T_s$.

To evaluate the normal-ordered characteristic functional $\Upsilon_s(\mathbf{R}_1, \mathbf{R}_2)$  defined in Eq.~(\ref{AppB_master_Upsilon}), we first recombine the normal-ordered product of exponentials into a single exponential via the Baker-Campbell-Hausdorff identity $e^{\hat A}e^{\hat B} = e^{\hat A + \hat B}e^{\frac{1}{2}[\hat A, \hat B]}$:
\begin{eqnarray} \label{AppC_Upsilon_T_step1}
\Upsilon _s \left( \mathbf{R}_1 ,\mathbf{R}_2 \right) &=& \exp \left[ {\frac{1}{2}\int {d\omega } \int {do_\mathbf{n} } \; \left| {\mathbf{C}_\omega  \left( {{\bf{R}}_1 ,{\bf{R}}_2 \left| {\bf{n}} \right.} \right)} \right|^2 } \right] \nonumber \\
&& \times {\rm Tr}_s \left\{ {\hat \rho _s \exp \left\{ {\int {d\omega } \int {do_\mathbf{n} } \;\left[ {\mathbf{C}_\omega ^* \left( {{\bf{R}}_1 ,{\bf{R}}_2 \left| {\bf{n}} \right.} \right) \cdot \hat{\mathbf{g}}_{\omega s}^\dag  \left( \mathbf{n} \right) - \mathbf{C}_\omega  \left( {{\bf{R}}_1 ,{\bf{R}}_2 \left| {\bf{n}} \right.} \right) \cdot \hat{\mathbf{g}}_{\omega s} \left( \mathbf{n} \right)} \right]} \right\}} \right\}.
\end{eqnarray}
Because the operator in the exponent of Eq.~(\ref{AppC_Upsilon_T_step1}) is linear in the bosonic operators $\hat{\mathbf{g}}_{\omega s}^\dag(\mathbf{n})$ and $\hat{\mathbf{g}}_{\omega s}(\mathbf{n})$ and the thermal state $\hat\rho_s$ in Eq.~(\ref{AppC_rho_s_thermal}) is a centered Gaussian state, Wick's theorem dictates that the cumulant expansion of the trace truncates exactly at second order:
\begin{eqnarray} \label{AppC_Upsilon_T_Wick}
&& {\rm Tr}_s \left\{ {\hat \rho _s \exp \left\{ {\int {d\omega } \int {do_\mathbf{n} } \;\left[ {\mathbf{C}_\omega ^* \left( {{\bf{R}}_1 ,{\bf{R}}_2 \left| {\bf{n}} \right.} \right) \cdot \hat{\mathbf{g}}_{\omega s}^\dag  \left( \mathbf{n} \right) - \mathbf{C}_\omega  \left( {{\bf{R}}_1 ,{\bf{R}}_2 \left| {\bf{n}} \right.} \right) \cdot \hat{\mathbf{g}}_{\omega s} \left( \mathbf{n} \right)} \right]} \right\}} \right\} \nonumber \\
&& \quad = \exp \left\{ {\frac{1}{2}{\rm Tr}_s \left\{ {\hat \rho _s \left\{ {\int {d\omega } \int {do_\mathbf{n} } \;\left[ {\mathbf{C}_\omega ^* \left( {{\bf{R}}_1 ,{\bf{R}}_2 \left| {\bf{n}} \right.} \right) \cdot \hat{\mathbf{g}}_{\omega s}^\dag  \left( \mathbf{n} \right) - \mathbf{C}_\omega  \left( {{\bf{R}}_1 ,{\bf{R}}_2 \left| {\bf{n}} \right.} \right) \cdot \hat{\mathbf{g}}_{\omega s} \left( \mathbf{n} \right)} \right]} \right\}^2 } \right\}} \right\} \nonumber \\
&& \quad = \exp \left\{ { - \frac{1}{2}\int {d\omega } \;\left[ {2n_\omega  \left( T_s \right) + 1} \right]\int {do_\mathbf{n} } \;\left| {\mathbf{C}_\omega  \left( {{\bf{R}}_1 ,{\bf{R}}_2 \left| {\bf{n}} \right.} \right)} \right|^2 } \right\},
\end{eqnarray}
where in the last step we have expanded the square and applied the thermal expectation values (\ref{AppC_thermal_s_averages}) together with the transversality relation $\mathbf{C}_\omega\left( {{\bf{R}}_1 ,{\bf{R}}_2 \left| {\bf{n}} \right.} \right)\cdot{\cal I}_\mathbf{n} = \mathbf{C}_\omega\left( {{\bf{R}}_1 ,{\bf{R}}_2 \left| {\bf{n}} \right.} \right)$. Substituting Eq.~(\ref{AppC_Upsilon_T_Wick}) into Eq.~(\ref{AppC_Upsilon_T_step1}), the positive vacuum prefactor $+\frac{1}{2}\int d\omega \int do_\mathbf{n}\,|\mathbf{C}_\omega|^2$ coming from the BCH recombination exactly cancels the zero-point contribution $-\frac{1}{2}\int d\omega \int do_\mathbf{n}\,|\mathbf{C}_\omega|^2$ in Eq.~(\ref{AppC_Upsilon_T_Wick}). Recalling the definition of the scattering correlation kernel $\Gamma_{\omega s}(\mathbf{R}_{1\perp}, \mathbf{R}_{2\perp})$ in the first of Eqs.~(\ref{AppB_master_Gamma}), the characteristic functional reduces to
\begin{eqnarray} \label{AppC_Upsilon_T_explicit}
\Upsilon _s \left( \mathbf{R}_1 ,\mathbf{R}_2 \right) &=& \exp \left[ { - \int {d\omega } \;n_\omega  \left( T_s \right)\int {do_\mathbf{n} } \;\left| \mathbf{C}_\omega  \left( {{\bf{R}}_1 ,{\bf{R}}_2 \left| {\bf{n}} \right.} \right) \right|^2 } \right] \nonumber \\
&=& \exp \left\{  - \int {d\omega } \;n_\omega  \left( T_s \right) \left\{ \Gamma _{\omega s} \left( \mathbf{R}_{1\perp} ,\mathbf{R}_{1\perp} \right) + \Gamma _{\omega s} \left( \mathbf{R}_{2\perp} ,\mathbf{R}_{2\perp} \right) - 2{\mathop{\rm Re}\nolimits} \left[ {e^{i\frac{\omega }{V}\left( R_{1V}  - R_{2V} \right)} \Gamma _{\omega s} \left( \mathbf{R}_{1\perp} ,\mathbf{R}_{2\perp} \right)} \right] \right\} \right\}. \nonumber \\
\end{eqnarray}

We now insert Eq.~(\ref{AppC_Upsilon_T_explicit}) into the center-of-mass master formula (\ref{AppB_dP_master_CM}). Because the thermal state (\ref{AppC_rho_s_thermal}) is stationary, $\Upsilon_s(\mathbf{R}_1, \mathbf{R}_2)$ depends on the longitudinal coordinates exclusively through the relative separation $Z = R_{1V} - R_{2V}$ and is independent of the center-of-mass coordinate $\bar R_V$. Consequently, $\Upsilon_s$ factors out of the $\bar R_V$ integral in Eq.~(\ref{AppB_dP_master_CM}), which evaluates directly to $\frac{1}{\ell}\int_{-(\ell - |Z|)/2}^{(\ell - |Z|)/2} d\bar R_V = 1 - |Z|/\ell$. Writing $2{\mathop{\rm Re}\nolimits}\left[e^{i\frac{\omega}{V}Z}\Gamma_{\omega s}(\mathbf{R}_{1\perp}, \mathbf{R}_{2\perp})\right] = e^{-i\frac{\omega}{V}Z}\Gamma_{\omega s}^*(\mathbf{R}_{1\perp}, \mathbf{R}_{2\perp}) + e^{i\frac{\omega}{V}Z}\Gamma_{\omega s}(\mathbf{R}_{1\perp}, \mathbf{R}_{2\perp})$, we see that $\Upsilon_s(\mathbf{R}_1, \mathbf{R}_2)$ combines with $W(\mathbf{R}_\perp)$ in Eq.~(\ref{AppB_W_def}) and $K(\mathbf{R}_{1\perp}, \mathbf{R}_{2\perp}; Z)$ in Eq.~(\ref{AppB_K_cumulant}) to place the scattering and material sectors on a completely symmetric footing at their respective temperatures $T_s$ and $T_{em}$. Specifically, evaluating this expression in the monoenergetic electron limit ($\ell \to \infty$), the triangular envelope $1 - |Z|/\ell \to 1$ and the longitudinal integration bounds naturally extend to infinity. The transverse-momentum-resolved energy exchange distribution thus reduces to the exact form
\begin{eqnarray} \label{AppC_dP_T_final}
\frac{{d\mathscr{P}}}{{dEd^2 \mathbf{P}_\perp  }} &=& \frac{1}{{\left( {2\pi \hbar } \right)^3 V}}\int {d^2 \mathbf{R}_{1\perp} } \int {d^2 \mathbf{R}_{2\perp} } \;\widetilde \phi _T \left( \mathbf{R}_{1\perp} \right)\widetilde \phi _T^* \left( \mathbf{R}_{2\perp} \right)e^{ - \frac{i}{\hbar }\mathbf{P}_\perp   \cdot \left( \mathbf{R}_{1\perp}  - \mathbf{R}_{2\perp} \right)} \nonumber \\
&& \times \int_{ - \infty }^{ + \infty } {dZ} \;e^{ - i\frac{{E - E_0 }}{{\hbar V}}Z} e^{K_T \left( \mathbf{R}_{1\perp} ,\mathbf{R}_{2\perp} ;Z \right)} ,
\end{eqnarray}
where, in accordance with Eq.~(\ref{AppB_dressed_phi}), the dressed transverse wave function $\widetilde\phi_T(\mathbf{R}_\perp)$, the optical depth $W_T(\mathbf{R}_\perp)$, and the two-path cumulant $K_T(\mathbf{R}_{1\perp}, \mathbf{R}_{2\perp}; Z)$ read
\begin{eqnarray} \label{AppC_W_K_T_defs}
\widetilde \phi _T \left( \mathbf{R}_\perp  \right) &=& \phi \left( \mathbf{R}_\perp  \right)e^{i\Phi \left( \mathbf{R}_\perp  \right) - W_T \left( \mathbf{R}_\perp  \right)} , \nonumber \\
W_T \left( \mathbf{R}_\perp  \right) &=& \int {d\omega } \;\left\{ {\left[ {n_\omega  \left( T_s \right) + \frac{1}{2}} \right]\Gamma _{\omega s} \left( \mathbf{R}_\perp  ,\mathbf{R}_\perp  \right) + \left[ {n_\omega  \left( T_{em} \right) + \frac{1}{2}} \right]\Gamma _{\omega em} \left( \mathbf{R}_\perp  ,\mathbf{R}_\perp  \right)} \right\}, \nonumber \\
K_T \left( \mathbf{R}_{1\perp} ,\mathbf{R}_{2\perp} ;Z \right) &=& \int {d\omega } \;\Bigg\{ e^{ - i\frac{\omega }{V}Z} \Big[ {\left( n_\omega  \left( T_s \right) + 1 \right)\Gamma _{\omega s}^* \left( \mathbf{R}_{1\perp} ,\mathbf{R}_{2\perp} \right) + \left( n_\omega  \left( T_{em} \right) + 1 \right)\Gamma _{\omega em}^* \left( \mathbf{R}_{1\perp} ,\mathbf{R}_{2\perp} \right)} \Big] \nonumber \\
&& \quad \quad \; + e^{i\frac{\omega }{V}Z} \Big[ {n_\omega  \left( T_s \right)\Gamma _{\omega s} \left( \mathbf{R}_{1\perp} ,\mathbf{R}_{2\perp} \right) + n_\omega  \left( T_{em} \right)\Gamma _{\omega em} \left( \mathbf{R}_{1\perp} ,\mathbf{R}_{2\perp} \right)} \Big] \Bigg\}.
\end{eqnarray}

\subsection{Multimode coherent state}

Although the formalism fully accommodates ultrashort optical pulses, for simplicity we restrict our analysis here to the case of continuous-wave (monochromatic) coherent illumination at a fixed frequency $\bar{\omega}$. The incident radiation is prepared in a pure monochromatic coherent state $\hat\rho_s = \left| \psi_s \right\rangle \left\langle \psi_s \right|$, generated by a displacement operator acting on the scattering vacuum:
\begin{equation} \label{AppC_coh_state}
\left| {\psi _s } \right\rangle  = \exp \left\{ {\int {do_\mathbf{n} } \;\left[ {\bm{\alpha } \left( \mathbf{n} \right) \cdot \hat{\mathbf{g}}_{\bar \omega s}^\dag  \left( \mathbf{n} \right) - \bm{\alpha }^* \left( \mathbf{n} \right) \cdot \hat{\mathbf{g}}_{\bar \omega s} \left( \mathbf{n} \right)} \right]} \right\}\left| {0_s } \right\rangle ,
\end{equation}
where $\bm{\alpha}(\mathbf{n})$ is the transverse angular amplitude profile ($\mathbf{n}\cdot\bm{\alpha}(\mathbf{n}) = 0$). This state is the right eigenstate of the scattering annihilation operator, satisfying $\hat{\mathbf{g}}_{\omega s}(\mathbf{n}) \left| \psi_s \right\rangle = \bm{\alpha}(\mathbf{n})\delta(\omega - \bar{\omega}) \left| \psi_s \right\rangle$. Evaluating the expectation value of the positive-frequency macroscopic electric field operator $\mathbf{\hat E}_\omega(\mathbf{r})$ from Eq.~(\ref{E_omega_main}) over the uncorrelated state $\hat\rho_s \hat\rho_{em}$ directly yields the classical electric field $\mathbf{E}^{{\rm cla}}_{\bar \omega}(\mathbf{r})\delta(\omega - \bar{\omega})$, where
\begin{equation} \label{AppC_E_cla}
\mathbf{E}_{\bar \omega }^{{\rm cla}} \left( \mathbf{r} \right) = \int {do_\mathbf{n} } \;{\cal F}_{\bar \omega s} \left( \left. \mathbf{r} \right|\mathbf{n} \right) \cdot \bm{\alpha } \left( \mathbf{n} \right) ,
\end{equation}
whose profile naturally comprises both the incident free-space wave and the exact scattered near and far fields dressing the arbitrary nanostructure.

To evaluate the normal-ordered characteristic functional $\Upsilon_s(\mathbf{R}_1, \mathbf{R}_2)$ defined in Eq.~(\ref{AppB_Upsilon_s_def}), the creation and annihilation operators are directly replaced by their classical eigenvalues, eliminating the frequency integration:
\begin{equation} \label{AppC_Upsilon_coh_step1}
\Upsilon _s \left( \mathbf{R}_1 ,\mathbf{R}_2 \right) = \exp \left\{ {\int {do_\mathbf{n} } \;\left[ {\mathbf{C}_{\bar \omega }^* \left( {{\bf{R}}_1 ,{\bf{R}}_2 \left| {\bf{n}} \right.} \right) \cdot \bm{\alpha }^* \left( \mathbf{n} \right) - \mathbf{C}_{\bar \omega } \left( {{\bf{R}}_1 ,{\bf{R}}_2 \left| {\bf{n}} \right.} \right) \cdot \bm{\alpha } \left( \mathbf{n} \right)} \right]} \right\}.
\end{equation}
Substituting the two-path scattering difference vector $\mathbf{C}_{\bar\omega}\left( {{\bf{R}}_1 ,{\bf{R}}_2 \left| {\bf{n}} \right.} \right)$ defined in Eq.~(\ref{AppB_C_vector_def}) into the functional, we recognize the emergence of the classical electric field along the paraxial electron trajectory. To conform with the standard theoretical convention for semiclassical PINEM interactions, we define the complex spectral coupling parameter $\beta(\mathbf{R}_\perp)$ at impact parameter $\mathbf{R}_\perp$ as
\begin{equation} \label{AppC_beta_def}
\beta \left( \mathbf{R}_\perp  \right) \equiv \frac{e}{{\hbar \bar \omega }}\int_{ - \infty }^{ + \infty } {dz} \;e^{ - i\frac{{\bar \omega }}{V}z} \mathbf{u}_V  \cdot \mathbf{E}_{\bar \omega }^{{\rm cla}} \left( \mathbf{R}_\perp   + z\mathbf{u}_V  \right).
\end{equation}
With this definition, the integral over the scattering modes yields $-\int do_\mathbf{n}\,\mathbf{C}_{\bar\omega}\left( {{\bf{R}}_1 ,{\bf{R}}_2 \left| {\bf{n}} \right.} \right)\cdot\bm{\alpha}(\mathbf{n}) = -e^{i\frac{\bar\omega}{V}R_{1V}}\beta(\mathbf{R}_{1\perp}) + e^{i\frac{\bar\omega}{V}R_{2V}}\beta(\mathbf{R}_{2\perp})$. Equation~(\ref{AppC_Upsilon_coh_step1}) then rigorously factorizes into single-path phase modulations:
\begin{eqnarray} \label{AppC_Upsilon_factor}
\Upsilon _s \left( \mathbf{R}_1 ,\mathbf{R}_2 \right) &=& \exp \left[ { -\beta \left( \mathbf{R}_{1\perp}  \right)e^{i\frac{{\bar \omega }}{V}R_{1V} }  + \beta ^* \left( \mathbf{R}_{1\perp}  \right)e^{ - i\frac{{\bar \omega }}{V}R_{1V} } } \right] \nonumber \\
&\times&  \exp \left[ {   {\beta \left( \mathbf{R}_{2\perp}  \right)e^{i\frac{{\bar \omega }}{V}R_{2V} }  - \beta ^* \left( \mathbf{R}_{2\perp}  \right)e^{ - i\frac{{\bar \omega }}{V}R_{2V} } } } \right].
\end{eqnarray}
Expressing the exponents in terms of sine functions and utilizing the Jacobi-Anger identity for the Bessel functions of the first kind $J_m$, namely $e^{iz\sin \phi }  = \sum_{n =  - \infty }^{ + \infty } {J_n \left( z \right)e^{in\phi } }$, the functional $\Upsilon_s$ evaluates explicitly to a double harmonic series of discrete multiphoton exchanges:
\begin{eqnarray} \label{AppC_Jacobi_Anger_Double}
\Upsilon _s \left( \mathbf{R}_1 ,\mathbf{R}_2 \right) &=& \sum\limits_{m_1  =  - \infty }^{ + \infty } {\sum\limits_{m_2  =  - \infty }^{ + \infty } {J_{m_1 } \left( 2\left| \beta \left( \mathbf{R}_{1\perp}  \right) \right| \right)J_{m_2 } \left( 2\left| \beta \left( \mathbf{R}_{2\perp}  \right) \right| \right)} } \nonumber \\
&& \times e^{ - im_1 \arg \left[ {\beta \left( \mathbf{R}_{1\perp}  \right)} \right]} e^{im_2 \arg \left[ {\beta \left( \mathbf{R}_{2\perp}  \right)} \right]} e^{ - i\frac{{\bar \omega }}{V}\left( {m_1 R_{1V}  - m_2 R_{2V} } \right)} ,
\end{eqnarray}
where the reversal of the summation index $m_1 \to -m_1$ in the exponents naturally arises from the property $J_m \left( { - z} \right) = J_{ - m} \left( z \right)$ of the Bessel functions of the first kind.

Because the coherent field establishes a macroscopic phase grating that breaks continuous longitudinal translation invariance, $\Upsilon_s(\mathbf{R}_1, \mathbf{R}_2)$ depends separately on $R_{1V}$ and $R_{2V}$. Expressing $R_{1V}$ and $R_{2V}$ in terms of the relative coordinate $Z$ and the center-of-mass coordinate $\bar R_V$ according to Eqs.~(\ref{AppB_change_of_vars}), the longitudinal phase factor in Eq.~(\ref{AppC_Jacobi_Anger_Double}) separates into relative and center-of-mass oscillations as
\begin{equation} \label{AppC_coh_phase_sep}
e^{ - i\frac{{\bar \omega }}{V}\left( {m_1 R_{1V}  - m_2 R_{2V} } \right)}  = e^{ - i\frac{{m_1  + m_2 }}{2}\frac{{\bar \omega }}{V}Z} e^{ - i\left( {m_1  - m_2 } \right)\frac{{\bar \omega }}{V}\bar R_V } .
\end{equation}
Inserting Eqs.~(\ref{AppC_Jacobi_Anger_Double}) and (\ref{AppC_coh_phase_sep}) into the center-of-mass master formula (\ref{AppB_dP_master_CM}), the probability distribution takes the form
\begin{eqnarray} \label{AppC_dP_coh_CM}
 \frac{{d\mathscr{P}}}{{dEd^2 {\bf{P}}_ \perp  }} &=& \frac{1}{{\left( {2\pi \hbar } \right)^3 V}}\sum\limits_{m_1  =  - \infty }^{ + \infty } {\sum\limits_{m_2  =  - \infty }^{ + \infty } {} } \int {d^2 {\bf{R}}_{1 \perp } } \int {d^2 {\bf{R}}_{2 \perp } } e^{ - \frac{i}{\hbar }{\bf{P}}_ \perp   \cdot \left( {{\bf{R}}_{1 \perp }  - {\bf{R}}_{2 \perp } } \right)}  \nonumber \\ 
 &\times& \left\{ {\widetilde \phi \left( {{\bf{R}}_{1 \perp } } \right)J_{m_1 } \left( {2\left| {\beta \left( {{\bf{R}}_{1 \perp } } \right)} \right|} \right)e^{ - im_1 \arg \left[ {\beta \left( {{\bf{R}}_{1 \perp } } \right)} \right]} } \right\}\left\{ {\widetilde \phi \left( {{\bf{R}}_{2 \perp } } \right)J_{m_2 } \left( {2\left| {\beta \left( {{\bf{R}}_{2 \perp } } \right)} \right|} \right)e^{ - im_2 \arg \left[ {\beta \left( {{\bf{R}}_{2 \perp } } \right)} \right]} } \right\}^*  \nonumber \\ 
 &\times& \int_{ - \ell }^\ell  {dZ} \;e^{ - i\frac{{E - E_0  + \frac{{m_1  + m_2 }}{2}\hbar \bar \omega }}{{\hbar V}}Z} e^{K\left( {{\bf{R}}_{1 \perp } ,{\bf{R}}_{2 \perp } ;Z} \right)} \frac{1}{\ell }\int_{ - \left( {\ell  - \left| Z \right|} \right)/2}^{\left( {\ell  - \left| Z \right|} \right)/2} {d\bar R_V } e^{ - i\left( {m_1  - m_2 } \right)\frac{{\bar \omega }}{V}\bar R_V } .
\end{eqnarray}
We now evaluate the asymptotic behavior of the integral over $\bar R_V$ in the monoenergetic electron limit, wherein the initial longitudinal coherence length is macroscopically large ($\ell \to \infty$). For any off-diagonal term ($m_1 \neq m_2$), the integral of the complex exponential evaluates to a bounded sine function, which vanishes identically when multiplied by the overall $1/\ell$ prefactor as $\ell \to \infty$. Conversely, for $m_1 = m_2$, the exponential is unity and the integration yields $\ell - |Z|$, so that dividing by $\ell$ gives $1 - |Z|/\ell \to 1$. This establishes the exact Kronecker-delta relation:
\begin{equation} \label{AppC_barRV_limit}
\lim_{\ell  \to \infty } \frac{1}{\ell }\int_{ - \frac{{\ell  - \left| Z \right|}}{2}}^{\frac{{\ell  - \left| Z \right|}}{2}} {d\bar R_V \;e^{ - i\left( {m_1  - m_2 } \right)\frac{{\bar \omega }}{V}\bar R_V } }  = \delta _{m_1 ,m_2 } .
\end{equation}
Thus, the macroscopic spatial extent of the paraxial electron strictly enforces the energetic phase-matching condition $m_1 = m_2 \equiv m$, restoring effective longitudinal translation invariance over the electron envelope. Taking the limit $\ell \to \infty$ in Eq.~(\ref{AppC_dP_coh_CM}), the integration bounds for $Z$ extend to infinity, and relabeling $m \to -m$ (noting that the factors $(-1)^m$ arising from $J_{-m}(x) = (-1)^m J_m(x)$ in the $\mathbf{R}_{1\perp}$ and $\mathbf{R}_{2\perp}$ amplitudes mutually cancel as $(-1)^{2m} = 1$), the exact closed-form distribution takes the form of a generalized diagonal discrete sideband expansion:
\begin{eqnarray} \label{AppC_dP_coh_diag}
\frac{{d\mathscr{P}}}{{dEd^2 \mathbf{P}_\perp  }} &=& \frac{1}{{\left( {2\pi \hbar } \right)^3 V}}\sum\limits_{m =  - \infty }^{ + \infty } {\int {d^2 \mathbf{R}_{1\perp} } \int {d^2 \mathbf{R}_{2\perp} } \;e^{ - \frac{i}{\hbar }\mathbf{P}_\perp   \cdot \left( \mathbf{R}_{1\perp}  - \mathbf{R}_{2\perp} \right)} \widetilde \phi _m \left( \mathbf{R}_{1\perp} \right)\widetilde \phi _m^* \left( \mathbf{R}_{2\perp} \right) } \nonumber \\
&& \times \int_{ - \infty }^{ + \infty } {dZ} \;e^{ - i\frac{{E - E_0  - m\hbar \bar \omega }}{{\hbar V}}Z} e^{K\left( \mathbf{R}_{1\perp} ,\mathbf{R}_{2\perp} ;Z \right)} ,
\end{eqnarray}
with the conventionally defined dressed transverse amplitude of order $m$
\begin{equation} \label{AppC_phi_m_def}
\widetilde \phi_m(\mathbf{R}_\perp) \equiv \widetilde \phi(\mathbf{R}_\perp)J_m(2|\beta(\mathbf{R}_\perp)|)e^{im\arg[\beta(\mathbf{R}_\perp)]}.
\end{equation}

\subsection{Multimode squeezed vacuum state}

As in the coherent-state analysis of Appendix~C\,2, although the general formalism natively encompasses broadband ultrashort squeezed pulses, for simplicity we restrict our investigation here to quasi-monochromatic continuous-wave squeezed vacuum illumination centered at frequency $\bar{\omega}$ within a narrow spectral bandwidth $\Delta\omega \ll \bar{\omega}$. The incident radiation is prepared in a pure multimode squeezed vacuum state $\hat\rho_s = \left| \psi_s \right\rangle \left\langle \psi_s \right|$, generated by the unitary squeezing operator $\hat S_s$ acting on the free-space scattering vacuum over the spectral interval $\Delta\omega$:
\begin{equation} \label{AppC_sq_state}
\left| {\psi _s } \right\rangle  = \hat S_s \left| {0_s } \right\rangle  = \exp \left\{ {\frac{1}{2}\int_{\Delta \omega } {d\omega } \int {do_\mathbf{n} } \;\left[ {\zeta ^* \left( \mathbf{n} \right)\hat{\mathbf{g}}_{\omega s} \left( \mathbf{n} \right) \cdot \hat{\mathbf{g}}_{\omega s} \left( \mathbf{n} \right) - \zeta \left( \mathbf{n} \right)\hat{\mathbf{g}}_{\omega s}^\dag  \left( \mathbf{n} \right) \cdot \hat{\mathbf{g}}_{\omega s}^\dag  \left( \mathbf{n} \right)} \right]} \right\}\left| {0_s } \right\rangle ,
\end{equation}
where $\zeta(\mathbf{n}) = r(\mathbf{n})e^{i\theta(\mathbf{n})}$ is the dimensionless complex angular squeezing parameter around frequency $\bar{\omega}$, with amplitude $r(\mathbf{n}) \ge 0$ and phase $\theta(\mathbf{n})$. Under this unitary transformation, all scattering modes outside the bandwidth $\Delta\omega$ remain unperturbed in their vacuum state, whereas the driven annihilation and creation operators within $\Delta\omega$ transform according to the exact Bogoliubov relations:
\begin{eqnarray} \label{AppC_Bogoliubov}
\hat S_s^\dag  \hat{\mathbf{g}}_{\omega s} \left( \mathbf{n} \right)\hat S_s &=& \hat{\mathbf{g}}_{\omega s} \left( \mathbf{n} \right)\cosh \left[ {r\left( \mathbf{n} \right)} \right] - \hat{\mathbf{g}}_{\omega s}^\dag  \left( \mathbf{n} \right)e^{i\theta \left( \mathbf{n} \right)} \sinh \left[ {r\left( \mathbf{n} \right)} \right], \nonumber \\
\hat S_s^\dag  \hat{\mathbf{g}}_{\omega s}^\dag  \left( \mathbf{n} \right)\hat S_s &=& \hat{\mathbf{g}}_{\omega s}^\dag  \left( \mathbf{n} \right)\cosh \left[ {r\left( \mathbf{n} \right)} \right] - \hat{\mathbf{g}}_{\omega s} \left( \mathbf{n} \right)e^{ - i\theta \left( \mathbf{n} \right)} \sinh \left[ {r\left( \mathbf{n} \right)} \right].
\end{eqnarray}
To evaluate the normal-ordered characteristic functional $\Upsilon_s(\mathbf{R}_1, \mathbf{R}_2)$ defined in Eq.~(\ref{AppB_master_Upsilon}), we observe that, by virtue of normal ordering, all vacuum modes outside $\Delta\omega$ yield a unit trace, restricting the frequency integration exclusively to the narrow band $\Delta\omega$ where $\mathbf{C}_\omega(\mathbf{R}_1, \mathbf{R}_2|\mathbf{n}) \simeq \mathbf{C}_{\bar\omega}(\mathbf{R}_1, \mathbf{R}_2|\mathbf{n})$. Recombining the normal-ordered exponential product over $\Delta\omega$ into the unitary multimode displacement operator
\begin{equation}
\hat D_s \left( \mathbf{C}_{\bar \omega } \right) = \exp \left\{ {\int_{\Delta \omega } {d\omega } \int {do_\mathbf{n} } \;\left[ {\mathbf{C}_{\bar \omega }^* \left( {\mathbf{R}_1 ,\mathbf{R}_2 \left| \mathbf{n} \right.} \right) \cdot \hat{\mathbf{g}}_{\omega s}^\dag  \left( \mathbf{n} \right) - \mathbf{C}_{\bar \omega } \left( {\mathbf{R}_1 ,\mathbf{R}_2 \left| \mathbf{n} \right.} \right) \cdot \hat{\mathbf{g}}_{\omega s} \left( \mathbf{n} \right)} \right]} \right\}
\end{equation}
via the Baker-Campbell-Hausdorff identity, we obtain
\begin{equation} \label{AppC_chi_sq_step1}
\Upsilon _s \left( \mathbf{R}_1 ,\mathbf{R}_2 \right) = \exp \left( {\frac{{\Delta \omega }}{2}\int {do_\mathbf{n} } \;\left| {\mathbf{C}_{\bar \omega } \left( {\mathbf{R}_1 ,\mathbf{R}_2 \left| \mathbf{n} \right.} \right)} \right|^2 } \right)\left\langle {0_s } \right|\hat S_s^\dag  \hat D_s \left( \mathbf{C}_{\bar \omega } \right)\hat S_s \left| {0_s } \right\rangle .
\end{equation}
Applying the Bogoliubov transformation (\ref{AppC_Bogoliubov}) directly to the argument of the displacement operator gives $\hat S_s^\dag \hat D_s(\mathbf{C}_{\bar\omega}) \hat S_s = \hat D_s(\mathbf{K}_{\bar\omega})$, where the transformed complex vector $\mathbf{K}_{\bar\omega}(\mathbf{R}_1, \mathbf{R}_2|\mathbf{n})$ reads
\begin{equation} \label{AppC_K_vector}
\mathbf{K}_{\bar \omega } \left( {\mathbf{R}_1 ,\mathbf{R}_2 \left| \mathbf{n} \right.} \right) = \mathbf{C}_{\bar \omega } \left( {\mathbf{R}_1 ,\mathbf{R}_2 \left| \mathbf{n} \right.} \right)\cosh \left[ {r\left( \mathbf{n} \right)} \right] + \mathbf{C}_{\bar \omega }^* \left( {\mathbf{R}_1 ,\mathbf{R}_2 \left| \mathbf{n} \right.} \right)e^{ - i\theta \left( \mathbf{n} \right)} \sinh \left[ {r\left( \mathbf{n} \right)} \right].
\end{equation}
Evaluating the standard vacuum expectation value $\langle 0_s | \hat D_s(\mathbf{K}_{\bar\omega}) | 0_s \rangle = \exp(-\frac{\Delta\omega}{2} \int do_\mathbf{n} |\mathbf{K}_{\bar\omega}|^2)$, the squared modulus of the transformed vector in Eq.~(\ref{AppC_K_vector}) expands directly to
\begin{equation} \label{AppC_K_sq}
\left| {\mathbf{K}_{\bar \omega } } \right|^2  = \left| {\mathbf{C}_{\bar \omega } } \right|^2 \left( {1 + 2\sinh ^2 \left[ {r\left( \mathbf{n} \right)} \right]} \right) + 2\sinh \left[ {r\left( \mathbf{n} \right)} \right]\cosh \left[ {r\left( \mathbf{n} \right)} \right]{\mathop{\rm Re}\nolimits} \left[ {e^{ - i\theta \left( \mathbf{n} \right)} \mathbf{C}_{\bar \omega }  \cdot \mathbf{C}_{\bar \omega } } \right],
\end{equation}
where the identity $\cosh^2[r(\mathbf{n})] = 1 + \sinh^2[r(\mathbf{n})]$ has been applied. 

Substituting Eq.~(\ref{AppC_K_sq}) back into Eq.~(\ref{AppC_chi_sq_step1}), the total exponent of the characteristic functional becomes
\begin{eqnarray} \label{AppC_Exponent_comp}
&& \frac{{\Delta \omega }}{2}\int {do_\mathbf{n} } \;\left| {\mathbf{C}_{\bar \omega } } \right|^2 - \frac{{\Delta \omega }}{2}\int {do_\mathbf{n} } \;\left| {\mathbf{K}_{\bar \omega } } \right|^2  \nonumber \\
&& \quad = \Delta \omega \int {do_\mathbf{n} } \;\left\{ {\frac{1}{2}\left| {\mathbf{C}_{\bar \omega } } \right|^2 - \frac{1}{2}\left| {\mathbf{C}_{\bar \omega } } \right|^2 \left( {1 + 2\sinh ^2 \left[ {r\left( \mathbf{n} \right)} \right]} \right) - \sinh \left[ {r\left( \mathbf{n} \right)} \right]\cosh \left[ {r\left( \mathbf{n} \right)} \right]{\mathop{\rm Re}\nolimits} \left[ {e^{ - i\theta \left( \mathbf{n} \right)} \mathbf{C}_{\bar \omega }  \cdot \mathbf{C}_{\bar \omega } } \right]} \right\} \nonumber \\
&& \quad =  - \Delta \omega \int {do_\mathbf{n} } \;\sinh \left[ {r\left( \mathbf{n} \right)} \right]\left\{ {\sinh \left[ {r\left( \mathbf{n} \right)} \right]\left| {\mathbf{C}_{\bar \omega } } \right|^2  + \cosh \left[ {r\left( \mathbf{n} \right)} \right]{\mathop{\rm Re}\nolimits} \left[ {e^{ - i\theta \left( \mathbf{n} \right)} \mathbf{C}_{\bar \omega }  \cdot \mathbf{C}_{\bar \omega } } \right]} \right\}.
\end{eqnarray}
As clearly shown, the positive vacuum prefactor $+\frac{1}{2}|\mathbf{C}_{\bar\omega}|^2$ coming from the BCH recombination exactly cancels the negative unit scalar $-\frac{1}{2}|\mathbf{C}_{\bar\omega}|^2$ originating from the zero-point expansion of $|\mathbf{K}_{\bar\omega}|^2$. The exact characteristic functional for the quasi-monochromatic multimode squeezed vacuum state thus reduces to the closed-form Gaussian expression:
\begin{eqnarray} \label{AppC_Upsilon_sq_final}
\Upsilon _s \left( {\mathbf{R}_1 ,\mathbf{R}_2 } \right)  &=& \exp \left\{ { - \Delta \omega \int {do_{\mathbf{n}} } \sinh ^2 \left[ {r\left( {\mathbf{n}} \right)} \right]\left| {\mathbf{C}_{\bar \omega } \left( {\mathbf{R}_1 ,\mathbf{R}_2 \left| \mathbf{n} \right.} \right)} \right|^2 } \right\} \nonumber \\ 
&\times& \exp \left\{ { - \Delta \omega \int {do_{\mathbf{n}} } \sinh \left[ {r\left( {\mathbf{n}} \right)} \right]\cosh \left[ {r\left( {\mathbf{n}} \right)} \right]{\mathop{\rm Re}\nolimits} \left[ {e^{ - i\theta \left( {\mathbf{n}} \right)} \mathbf{C}_{\bar \omega } \left( {\mathbf{R}_1 ,\mathbf{R}_2 \left| \mathbf{n} \right.} \right) \cdot \mathbf{C}_{\bar \omega } \left( {\mathbf{R}_1 ,\mathbf{R}_2 \left| \mathbf{n} \right.} \right)} \right]} \right\}. 
\end{eqnarray}
In complete analogy with the coherent coupling parameter $\beta(\mathbf{R}_\perp)$ of Eq.~(\ref{AppC_beta_def}), we introduce the dimensionless directional coupling vector $\bm{\gamma}(\mathbf{R}_\perp|\mathbf{n})$ at impact parameter $\mathbf{R}_\perp$:
\begin{equation} \label{AppC_gamma_vec_def}
\bm{\gamma } \left( {\left. {\mathbf{R}_\perp  } \right|\mathbf{n}} \right) \equiv \sqrt {\Delta \omega } \,\frac{e}{{\hbar \bar \omega }}\int_{ - \infty }^{ + \infty } {dz} \;e^{ - i\frac{{\bar \omega }}{V}z} \mathbf{u}_V  \cdot {\cal F}_{\bar \omega s} \left( {\left. {\mathbf{R}_\perp   + z\mathbf{u}_V } \right|\mathbf{n}} \right),
\end{equation}
in terms of which the two-path scattering difference vector $\mathbf{C}_{\bar\omega}\left( {\mathbf{R}_1 ,\mathbf{R}_2 \left| \mathbf{n} \right.} \right)$ of Eq.~(\ref{AppB_C_vector_def}) satisfies $\sqrt{\Delta\omega}\,\mathbf{C}_{\bar\omega}\left( {\mathbf{R}_1 ,\mathbf{R}_2 \left| \mathbf{n} \right.} \right) = e^{i\frac{\bar\omega}{V}R_{1V}}\bm{\gamma}(\mathbf{R}_{1\perp}|\mathbf{n}) - e^{i\frac{\bar\omega}{V}R_{2V}}\bm{\gamma}(\mathbf{R}_{2\perp}|\mathbf{n})$. Substituting this decomposition into Eq.~(\ref{AppC_Upsilon_sq_final}), the angular integration over the scattering directions $\mathbf{n}$ is entirely encapsulated into two dimensionless macroscopic squeezing correlation kernels: the normal (phase-insensitive) squeezing correlation kernel $\Lambda_{ij}^N$ originating from the first exponent in Eq.~(\ref{AppC_Upsilon_sq_final}), and the anomalous (phase-sensitive) two-photon squeezing kernel $\Lambda_{ij}^A$ originating from the second exponent, defined respectively as
\begin{eqnarray} \label{AppC_Gamma_Lambda_sq_defs}
\Lambda_{ij}^N &\equiv& \int {do_\mathbf{n} } \;\sinh ^2 \left[ {r\left( \mathbf{n} \right)} \right]\bm{\gamma } \left( {\left. {\mathbf{R}_{i\perp} } \right|\mathbf{n}} \right) \cdot \bm{\gamma }^* \left( {\left. {\mathbf{R}_{j\perp} } \right|\mathbf{n}} \right), \nonumber \\
\Lambda_{ij}^A &\equiv& \int {do_\mathbf{n} } \;\sinh \left[ {r\left( \mathbf{n} \right)} \right]\cosh \left[ {r\left( \mathbf{n} \right)} \right]e^{ - i\theta \left( \mathbf{n} \right)} \bm{\gamma } \left( {\left. {\mathbf{R}_{i\perp} } \right|\mathbf{n}} \right) \cdot \bm{\gamma } \left( {\left. {\mathbf{R}_{j\perp} } \right|\mathbf{n}} \right).
\end{eqnarray}
Here, $\Lambda_{12}^N$ is Hermitian ($\Lambda_{12}^{N*} = \Lambda_{21}^N$) with non-negative diagonal elements $\Lambda_{ii}^N \ge 0$, whereas $\Lambda_{12}^A$ is complex-symmetric under path exchange ($\Lambda_{21}^A = \Lambda_{12}^A$) and involves the unconjugated inner product $\bm{\gamma}(\mathbf{R}_{i\perp}|\mathbf{n})\cdot\bm{\gamma}(\mathbf{R}_{j\perp}|\mathbf{n})$. In terms of the kernels (\ref{AppC_Gamma_Lambda_sq_defs}), the characteristic functional (\ref{AppC_Upsilon_sq_final}) factorizes into the product of a phase-insensitive normal contribution and an anomalous phase-sensitive contribution:
\begin{eqnarray} \label{AppC_Upsilon_sq_factorized}
\Upsilon _s \left( {\mathbf{R}_1 ,\mathbf{R}_2 } \right) &=& \exp \left\{ { - \Lambda_{11}^N - \Lambda_{22}^N + 2{\mathop{\rm Re}\nolimits} \left[ {e^{i\frac{{\bar \omega }}{V}\left( {R_{1V}  - R_{2V} } \right)} \Lambda_{12}^N} \right]} \right\} \nonumber \\ 
&\times& \exp \left\{ { - {\mathop{\rm Re}\nolimits} \left[ {e^{i2\frac{{\bar \omega }}{V}R_{1V} } \Lambda_{11}^A + e^{i2\frac{{\bar \omega }}{V}R_{2V} } \Lambda_{22}^A - 2e^{i\frac{{\bar \omega }}{V}\left( {R_{1V}  + R_{2V} } \right)} \Lambda_{12}^A} \right]} \right\}. 
\end{eqnarray}
We now insert the characteristic functional (\ref{AppC_Upsilon_sq_factorized}) into the center-of-mass master formula (\ref{AppB_dP_master_CM}) to evaluate the transverse-momentum-resolved probability distribution. In the monoenergetic electron limit ($\ell \to \infty$), we first expand the exponential of each of the four real harmonics in Eq.~(\ref{AppC_Upsilon_sq_factorized}) via the generating function of the modified Bessel functions of the first kind $I_n$, namely $e^{z\cos\phi} = \sum_{n = -\infty}^{+\infty} I_n(z)e^{in\phi}$. The characteristic functional expands into the multi-harmonic series
\begin{eqnarray} \label{AppC_Upsilon_sq_Bessel_exp}
\Upsilon _s \left( {\mathbf{R}_1 ,\mathbf{R}_2 } \right) &=& e^{ - \Lambda_{11}^N - \Lambda_{22}^N} \sum\limits_{n_{12} =  - \infty }^{ + \infty } \sum\limits_{k_{11}  =  - \infty }^{ + \infty } \sum\limits_{k_{22}  =  - \infty }^{ + \infty } \sum\limits_{k_{12} =  - \infty }^{ + \infty } \left( { - 1} \right)^{k_{11}  + k_{22} }   I_{n_{12}} \left( {2\left| {\Lambda_{12}^N} \right|} \right) I_{k_{11} } \left( {\left| {\Lambda_{11}^A} \right|} \right)I_{k_{22} } \left( {\left| {\Lambda_{22}^A} \right|} \right)I_{k_{12}} \left( {2\left| {\Lambda_{12}^A} \right|} \right) \nonumber \\
&\times&  e^{i\left[ {n_{12}\arg \left( {\Lambda _{12}^N } \right) + k_{11} \arg \left( {\Lambda _{11}^A } \right) + k_{22} \arg \left( {\Lambda _{22}^A } \right) + k_{12}\arg \left( {\Lambda _{12}^A } \right)} \right]} 
e^{i\frac{{\bar \omega }}{V}\left[ {n_{12}\left( {R_{1V}  - R_{2V} } \right) + 2k_{11} R_{1V}  + 2k_{22} R_{2V}  + k_{12}\left( {R_{1V}  + R_{2V} } \right)} \right]} . \nonumber \\
\end{eqnarray}
Expressing the longitudinal coordinates $R_{1V}$ and $R_{2V}$ in terms of the relative coordinate $Z$ and the center-of-mass coordinate $\bar R_V$ according to Eqs.~(\ref{AppB_change_of_vars}) and (\ref{AppB_R1_R2_parametrized}), the longitudinal phase factor in Eq.~(\ref{AppC_Upsilon_sq_Bessel_exp}) exactly separates into relative and center-of-mass oscillations:
\begin{equation} \label{AppC_Phase_Separation}
e^{i\frac{{\bar \omega }}{V}\left[ {n_{12}\left( {R_{1V}  - R_{2V} } \right) + 2k_{11} R_{1V}  + 2k_{22} R_{2V}  + k_{12}\left( {R_{1V}  + R_{2V} } \right)} \right]}  = e^{i\left( {n_{12} + k_{11}  - k_{22} } \right)\frac{{\bar \omega }}{V}Z} e^{2i\left( {k_{11}  + k_{22}  + k_{12}} \right)\frac{{\bar \omega }}{V}\bar R_V } .
\end{equation}
Inserting Eqs.~(\ref{AppC_Upsilon_sq_Bessel_exp}) and (\ref{AppC_Phase_Separation}) into the master formula (\ref{AppB_dP_master_CM}) and performing the integration over the center-of-mass coordinate $\bar R_V$ in the monoenergetic limit $\ell \to \infty$ by invoking the exact relation (\ref{AppC_barRV_limit}), the $\bar R_V$ integral yields the Kronecker delta $\delta_{k_{12}, -(k_{11} + k_{22})}$. Setting $k_{12} = -(k_{11} + k_{22})$ and grouping all contributions sharing the same net longitudinal harmonic index $m = n_{12} + k_{11} - k_{22}$, the asymptotic energy and transverse-momentum probability distribution takes the exact discrete sideband representation:
\begin{eqnarray} \label{AppC_dP_sq_final}
\frac{{d\mathscr{P}}}{{dEd^2 \mathbf{P}_\perp  }} &=& \frac{1}{{\left( {2\pi \hbar } \right)^3 V}}\sum\limits_{m =  - \infty }^{ + \infty } {\int {d^2 \mathbf{R}_{1\perp} } \int {d^2 \mathbf{R}_{2\perp} } \;e^{ - \frac{i}{\hbar }\mathbf{P}_\perp   \cdot \left( \mathbf{R}_{1\perp}  - \mathbf{R}_{2\perp} \right)} \widetilde \phi _{sq} \left( \mathbf{R}_{1\perp} \right)\widetilde \phi _{sq}^* \left( \mathbf{R}_{2\perp} \right)M_m \left( \mathbf{R}_{1\perp} ,\mathbf{R}_{2\perp} \right) } \nonumber \\
&& \times \int_{ - \infty }^{ + \infty } {dZ} \;e^{ - i\frac{{E - E_0  - m\hbar \bar \omega }}{{\hbar V}}Z} e^{K\left( \mathbf{R}_{1\perp} ,\mathbf{R}_{2\perp} ;Z \right)} ,
\end{eqnarray}
where the squeeze-dressed transverse amplitude becomes
\begin{equation} \label{AppC_W_sq_def}
\widetilde\phi_{sq}({\bf{R}}_{i\perp}) = \phi({\bf{R}}_{i\perp})e^{i\Phi({\bf{R}}_{i\perp}) - W({\bf{R}}_{i\perp}) - \Lambda_{ii}^N},
\end{equation}
and the two-point transverse correlation weight $M_m(\mathbf{R}_{1\perp}, \mathbf{R}_{2\perp})$ of the $m$-th squeezed sideband reads explicitly
\begin{eqnarray} \label{AppC_Wm_sq_def}
M_m \left( {\mathbf{R}_{1\perp} ,\mathbf{R}_{2\perp} } \right) &=& \sum\limits_{k_{11}  =  - \infty }^{+\infty}  {\sum\limits_{k_{22}  =  - \infty }^{+\infty}  {\left( { - 1} \right)^{k_{11}  + k_{22} } } }  I_{m - k_{11}  + k_{22} } \left( {2\left| {\Lambda_{12}^N} \right|} \right)I_{k_{11} } \left( {\left| {\Lambda_{11}^A} \right|} \right)I_{k_{22} } \left( {\left| {\Lambda_{22}^A} \right|} \right)I_{k_{11}  + k_{22} } \left( {2\left| {\Lambda_{12}^A} \right|} \right) \nonumber \\ 
 &\times& e^{i\left\{ {\left( {m - k_{11}  + k_{22} } \right)\arg \left[ {\Lambda_{12}^N} \right] + k_{11} \arg \left[ {\Lambda_{11}^A} \right] + k_{22} \arg \left[ {\Lambda_{22}^A} \right] - \left( {k_{11}  + k_{22} } \right)\arg \left[ {\Lambda_{12}^A} \right]} \right\}} .
\end{eqnarray}

\end{document}